# Seeking Stability by being Lazy and Shallow

Lazy and shallow instantiation is user friendly

Gert-Jan Bottu[*]
KU Leuven
Belgium
gertjan.bottu@kuleuven.be

Richard A. Eisenberg
Tweag
Paris, France
rae@richarde.dev

**Abstract**

Designing a language feature often requires a choice between several, similarly expressive possibilities. Given that user studies are generally impractical, we propose using *stability* as a way of making such decisions. Stability is a measure of whether the meaning of a program alters under small, seemingly innocuous changes in the code (e.g., inlining).

Directly motivated by a need to pin down a feature in GHC/Haskell, we apply this notion of stability to analyse four approaches to the instantiation of polymorphic types, concluding that the most stable approach is lazy (instantiate a polytype only when absolutely necessary) and shallow (instantiate only top-level type variables, not variables that appear after explicit arguments).

## 1 Introduction

Programmers naturally wish to get the greatest possible utility from their work. They thus embrace *polymorphism*: the idea that one function can work with potentially many types. A simple example is *const* :: ∀ *a b*. *a* → *b* → *a*, which returns its first argument, ignoring its second. The question then becomes: what concrete types should *const* work with at a given call site? For example, if we say *const True* 'x', then a compiler needs to figure out that *a* should become *Bool* and *b* should become *Char*. The process of taking a type variable and substituting in a concrete type is called *instantiation*. Choosing a correct instantiation is important; for *const*, the choice of *a* ↦ *Bool* means that the return type of *const True* 'x' is *Bool*. A context expecting a different type would lead to a type error.

[*]This work was partially completed while Bottu was an intern at Tweag.

In the above example, the choices for *a* and *b* in the type of *const* were inferred. Haskell, among other languages, also gives programmers the opportunity to *specify* the instantiation for these arguments [Eisenberg et al. 2016]. For example, we might say *const* @*Bool* @*Char True* 'x' (choosing the instantiations for both *a* and *b*) or *const* @*Bool True* 'x' (still allowing inference for *b*). However, once we start allowing user-directed instantiation, many thorny design issues arise. For example, will **let** *f* = *const* **in** *f* @*Bool True* 'x' be accepted?

Our concerns are rooted in concrete design questions in Haskell, as embodied by the Glasgow Haskell Compiler (GHC). Specifically, as Haskell increasingly has features in support of type-level programming, how should its instantiation behave? Should instantiating a type like *Int* → ∀ *a*. *a* → *a* yield *Int* → *α* → *α* (where *α* is a unification variable), or should instantiation stop at the regular argument of type *Int*? This is a question of the *depth* of instantiation. Suppose now *f* :: *Int* → ∀ *a*. *a* → *a*. Should *f* 5 have type ∀ *a*. *a* → *a* or *α* → *α*? This is a question of the *eagerness* of instantiation. As we explore in Section 3, these questions have real impact on our users.

Unlike much type-system research, our goal is *not* simply to make a type-safe and expressive language. Type-safe instantiation is well understood [e.g., Damas and Milner 1982; Reynolds 1974]. Instead, we wish to examine the *usability* of a design around instantiation. Unfortunately, proper scientific studies around usability are essentially intractable, as we would need pools of comparable experts in several designs executing a common task. Instead of usability, then, we draw a fresh focus to a property we name *stability*.

Intuitively, a language is *stable* if small, seemingly-innocuous changes to the source code of a program do not cause large changes to the program's behaviour; we expand on this definition in Section 3. We use stability as our metric for evaluating instantiation schemes in GHC.

Our contributions are as follows:

- The introduction of stability properties relevant for examining instantiation in Haskell, along with examples of how these properties affect programmer experience. (Section 3)
- A family of type systems, based on the bidirectional type-checking algorithm implemented in GHC [Eisenberg et al. 2016; Peyton Jones et al. 2007; Serrano et al. 2020]. It is parameterised over the flavour of type instantiation. (Section 4)
- An analysis of how different choices of instantiation flavour either respect or do not respect the similarities we identify. We conclude that lazy, shallow instantiation is the most stable. (Section 5; proofs in Appendix E)

Though we apply stability as the mechanism of studying instantiation within Haskell, we believe our approach is more widely applicable, both to other user-facing design questions within Haskell and in the context of other languages.

The appendices mentioned in the text can be found in the extended version at http://arxiv.org/abs/2106.14938.

## 2  Background

This section describes instantiation in GHC today and sets our baseline understanding for the remainder of the paper.

### 2.1  Instantiation in GHC

Visible type application and variable specificity are fixed attributes of the designs we are considering.

***Visible type application.*** Since GHC 8.0, Haskell has supported visible instantiation of type variables, based on the order in which those variables occur [Eisenberg et al. 2016]. Given $const :: \forall \, a \, b. \, a \to b \to a$, we can write $const \; @Int \; @Bool$, which instantiates the type variables, giving us an expression of type $Int \to Bool \to Int$. If a user wants to visibly instantiate a later type parameter (say, $b$) without choosing an earlier one, they can write $@\_$ to skip a parameter. The expression $const \; @\_ \; @Bool$ has type $\alpha \to Bool \to \alpha$, for any type $\alpha$.

***Specificity.*** Eisenberg et al. [2016, Section 3.1] introduce the notion of type variable *specificity*. The key idea is that quantified type variables are either written by the user (these are called *specified*) or invented by the compiler (these are called *inferred*). A specified variable is available for explicit instantiation using, e.g., $@Int$; an inferred variable may not be explicitly instantiated.

Following GHC, we use braces to denote inferred variables. Thus, if we have the Haskell program

$id_1 :: a \to a$

$id_1 \; x = x$

$id_2 \; x = x$

then we would write that $id_1 :: \forall \, a. \, a \to a$ (with a specified $a$) and $id_2 :: \forall \, \{a\}. \, a \to a$ (with an inferred $a$). Accordingly, $id_1 \; @Int$ is a function of type $Int \to Int$, while $id_2 \; @Int$ is a type error.

### 2.2  Deep vs. Shallow Instantiation

The first aspect of instantiation we seek to vary is its *depth*, which type variables get instantiated. Concretely, shallow instantiation affects only the type variables bound before any explicit arguments. Deep instantiation, on the other hand, also instantiates all variables bound after any number of explicit arguments. For example, consider a function $f :: \forall \, a. \, a \to (\forall \, b. \, b \to b) \to \forall \, c. \, c \to c$. A shallow instantiation of $f$'s type instantiates only $a$, whereas deep instantiation also affects $c$, despite $c$'s deep binding site. Neither instantiation flavour touches $b$ however, as $b$ is not an argument of $f$.

Versions of GHC up to 8.10 perform deep instantiation, as originally introduced by Peyton Jones et al. [2007], but GHC 9.0 changes this design, as proposed by Peyton Jones [2019] and inspired by Serrano et al. [2020]. In this paper, we study this change through the lens of stability.

### 2.3  Eager vs. Lazy Instantiation

Our work also studies the *eagerness* of instantiation, which determines the location in the code where instantiation happens. Eager instantiation immediately instantiates a polymorphic type variable as soon as it is mentioned. In contrast, lazy instantiation holds off instantiation as long as possible until instantiation is necessary in order to, say, allow a variable to be applied to an argument.

For example, consider these functions:

$pair :: \forall \, a. \, a \to \forall \, b. \, b \to (a, b)$

$pair \; x \; y = (x, y)$

$myPairX \; x = pair \; x$

What type do we expect to infer for $myPairX$? With eager instantiation, the type of a polymorphic expression is instantiated as soon as it occurs. Thus, $pair \; x$ will have a type $\beta \to (\alpha, \beta)$, assuming we have guessed $x :: \alpha$. (We use Greek letters to denote unification variables.) With neither $\alpha$ nor $\beta$ constrained, we will generalise both, and infer $\forall \, \{a\} \, \{b\}. \, a \to b \to (a, b)$ for $myPairX$. Crucially, this type is *different* than the type of $pair$.

Let us now replay this process with lazy instantiation. The variable $pair$ has type $\forall \, a. \, a \to \forall \, b. \, b \to (a, b)$. In order to apply $pair$ of that type to $x$, we must instantiate the first quantified type variable $a$ to a fresh unification variable $\alpha$, yielding the type $\alpha \to \forall \, b. \, b \to (\alpha, b)$. This is indeed a function type, so we can consume the argument $x$, yielding $pair \; x :: \forall \, b. \, b \to (\alpha, b)$. We have now type-checked the expression *pair x*, and thus we take the parameter *x* into account and generalise this type to produce the inferred type *myPairX* :: ∀ {*a*}. *a* → ∀ *b*. *b* → (*a*, *b*). This is the same as the type given for *pair*, modulo the specificity of *a*.

As we have seen, thus, the choice of eager or lazy instantiation can change the inferred types of definitions. In a language that allows visible instantiation of type variables, the difference between these types is user-visible. With lazy instantiation, *myPairX True* @*Char* 'x' is accepted, whereas with eager instantiation, it would be rejected.

## 3  Stability

We have described stability as a measure of how small transformations—call them *similarities*—in user-written code might drastically change the behaviour of a program. This section lays out the specific similarities we will consider with respect to our instantiation flavours. There are naturally *many* transformations one might think of applying to a source program. We have chosen ones that relate best to instantiation; others (e.g. does a function behave differently in curried form as opposed to uncurried form?) do not distinguish among our flavours and are thus less interesting in our concrete context. We include examples demonstrating each of these, showing how instantiation can become muddled. While these examples are described in terms of types inferred for definitions that have no top-level signature, many of the examples can easily be adapted to include a signature. After presenting our formal model of Haskell instantiation, we check our instantiation flavours against these similarities in Section 5, with proofs in Appendix E.

We must first define what we mean by the "behaviour" of a program. We consider two different notions of behaviour, both the *compile time* semantics of a program (that is, whether the program is accepted and what types are assigned to its variables) and its *runtime* semantics (that is, what the program executes to, assuming it is still well typed). We write, for example, $\overset{C+R}{\Longleftrightarrow}$ to denote a similarity that we expect to respect both compile and runtime semantics, whereas $\overset{R}{\Longleftrightarrow}$ is one that we expect only to respect runtime semantics, but may change compile time semantics. Similarly, $\overset{C+R}{\Longrightarrow}$ denotes a one-directional similarity that we expect to respect both compile and runtime semantics.

**Similarity 1: Let-Inlining and Extraction**

A key concern for us is around let-inlining and -extraction. That is, if we bind an expression to a new variable and use that variable instead of the original expression, does our program change meaning? Or if we inline a definition, does our program change meaning? These notions are captured in Similarity 1:[1]

$$\text{let } x = e_1 \text{ in } e_2 \overset{C+R}{\Longleftrightarrow} [e_1/x] \, e_2$$

**Example 1: myId.**  The Haskell standard library defines *id* :: ∀ *a*. *a* → *a* as the identity function. Suppose we made a synonym of this (using the implicit top-level **let** of Haskell files), with the following:

*myId* = *id*

Note that there is no type signature. Even in this simple example, our choice of instantiation eagerness changes the type we infer:

| *myId* | eager | lazy |
|---|---|---|
| deep or shallow | ∀ {*a*}. *a* → *a* | ∀ *a*. *a* → *a* |

Under eager instantiation, the mention of *id* is immediately instantiated, and thus we must re-generalise in order to get a polymorphic type for *myId*. Generalising always produces inferred variables, and so the inferred type for *myId* starts with ∀ {*a*}, meaning that *myId* cannot be a drop-in replacement for *id*, which might be used with explicit type instantiation. On the other hand, lazy instantiation faithfully replicates the type of *id* and uses it as the type of *myId*.

**Example 2: myPair.**  This problem gets even worse if the original function has a non-prenex type, like our *pair*, above. Our definition is now:

*myPair* = *pair*

With this example, both design axes around instantiation matter:

| *myPair* | eager | lazy |
|---|---|---|
| deep | ∀ {*a*} {*b*}. *a* → *b* → (*a*, *b*) | ∀ *a*. *a* → ∀ *b*. *b* → (*a*, *b*) |
| shallow | ∀ {*a*}. *a* → ∀ *b*. *b* → (*a*, *b*) | ∀ *a*. *a* → ∀ *b*. *b* → (*a*, *b*) |

All we want is to define a simple synonym, and yet reasoning about the types requires us to consider both depth and eagerness of instantiation.

**Example 3: myPairX.**  The *myPairX* example above acquires a new entanglement once we account for specificity. We define *myPairX* with this:

*myPairX x* = *pair x*

We infer these types:

| *myPairX* | eager | lazy |
|---|---|---|
| deep or shallow | ∀ {*a*} {*b*}. *a* → *b* → (*a*, *b*) | ∀ {*a*}. *a* → ∀ *b*. *b* → (*a*, *b*) |

Unsurprisingly, the generalised variables end up as inferred, instead of specified.

---

[1] A language with a strict let construct will observe a runtime difference between a let binding and its expansion, but this similarity would still hold with respect to type-checking.

**Similarity 2: Signature Property**

The second similarity annotates a let binding with the inferred type $\sigma$ of the bound expression $e_1$. We expect this similarity to be one-directional, as dropping a type annotation may indeed change the compile time semantics of a program, as we hope programmers expect.

$$\overline{f\,\overline{\pi_i} = e_i}^{\,i} \stackrel{C+R}{\Longrightarrow} f : \sigma; \overline{f\,\overline{\pi_i} = e_i}^{\,i}, \text{ where } \sigma \text{ is the inferred type of } f$$

***Example 4: infer.*** Though not yet implemented, we consider a version of Haskell that includes the ability to abstract over type variables, the subject of an approved proposal for GHC [Eisenberg 2018]. With this addition, we can imagine writing *infer*:

*infer* = $\lambda$ @$a$ ($x$ :: $a$) $\rightarrow$ $x$

We would infer these types:

| *infer* | eager | lazy |
|---|---|---|
| deep or shallow | $\forall$ {$a$}. $a \rightarrow a$ | $\forall$ $a$. $a \rightarrow a$ |

Note that the eager variant will infer a type containing an inferred quantified variable {$a$}. this is because the expression $\lambda$ @$a$ ($x$ :: $a$) $\rightarrow$ $x$ is instantly instantiated; it is then **let**-generalised to get the type in the table above.

If we change our program to include these types as annotations, the eager type, with its inferred variable, will be rejected. The problem is that we cannot check an abstraction $\lambda$ @$a$ $\rightarrow$ … against an expected type $\forall$ {$a$}. …: the whole point of having an inferred specificity is to prevent such behaviour, as an inferred variable should not correspond to either abstractions or applications in the term.

**Similarity 3: Type Signatures**

Changing a type signature should not affect runtime semantics—except in the case of type classes (or other feature that interrupts parametricity). Because our paper elides type classes, we can state this similarity quite generally; more fleshed-out settings would require a caveat around the lack of type-class constraints.

$$f : \sigma_1; \overline{f\,\overline{\pi_i} = e_i}^{\,i} \stackrel{R}{\Longleftrightarrow} f : \sigma_2; \overline{f\,\overline{\pi_i} = e_i}^{\,i}$$

***Example 5: swizzle.*** Suppose we have this function defined[2]:

*undef* :: $\forall$ $a$. *Int* $\rightarrow$ $a$ $\rightarrow$ $a$
*undef* = *undefined*

Now, we write a synonym but with a slightly different type:

*swizzle* :: *Int* $\rightarrow$ $\forall$ $a$. $a$ $\rightarrow$ $a$
*swizzle* = *undef*

Shockingly, *undef* and *swizzle* have different runtime behaviour: forcing *undef* diverges (unsurprisingly), but forcing *swizzle* has no effect. The reason is that the definition of *swizzle* is not as simple as it looks. In the System-F-based core language used within GHC, we have *swizzle* = $\lambda$($n$ :: *Int*) $\rightarrow$ $\Lambda$($a$ :: *Type*) $\rightarrow$ *undef* @$a$ $n$. Accordingly, *swizzle* is a function, which is already a value[3].

---

[2]This example is inspired by Peyton Jones [2019].

Under shallow instantiation, *swizzle* would simply be rejected, as its type is different than *undef*'s. The only way *swizzle* can be accepted is if it is deeply skolemised (see *Application* in Section 4), a necessary consequence of deep instantiation.

| *swizzle* | eager or lazy |
|---|---|
| deep | converges |
| shallow | rejected |

**Similarity 4: Pattern-Inlining and Extraction**

The fourth similarity represents changing variable patterns (written to the left of the = in a function definition) into $\lambda$-binders (written on the right of the =), and vice versa. Here, we assume the patterns $\overline{\pi}$ contain only (expression and type) variables. The three-place *wrap* relation is unsurprising. It denotes that wrapping the patterns $\overline{\pi}$ around the expression $e_1$ in lambda binders results in $e_1'$. Its definition can be found in Appendix C.

$$\text{let } x\,\overline{\pi} = e_1 \text{ in } e_2 \stackrel{C+R}{\Longleftrightarrow} \text{let } x = e_1' \text{ in } e_2, \text{ where } wrap\,(\overline{\pi}; e_1 \sim e_1')$$

***Example 6: infer2, again.*** Returning to the *infer* example, we might imagine moving the abstraction to the left of the =, yielding:

*infer2* @$a$ ($x$ :: $a$) = $x$

Under all instantiation schemes, *infer2* will be assigned the type $\forall$ $a$. $a \rightarrow a$. Accordingly, under eager instantiation, the choice of whether to bind the variables before the = or afterwards matters.

**Similarity 5: Single vs. Multiple Equations**

Our language model includes the ability to define a function by specifying multiple equations. The type inference algorithm in GHC differentiates between single and multiple equation declarations (see Section 5), and we do not want this distinction to affect types. While normally new equations for a function would vary the patterns compared to existing equations, we simply repeat the existing equation twice; after all, the particular choice of (well-typed) pattern should not affect compile time semantics at all.

$$f\,\overline{\pi} = e \stackrel{C}{\Longleftrightarrow} f\,\overline{\pi} = e, f\,\overline{\pi} = e$$

***Example 7: unitId1 and unitId2.*** Consider these two definitions:

*unitId1* () = *id*

---

[3]Similarly to *swizzle*, the definition of *undef* gets translated into $\Lambda$($a$ :: *Type*) $\rightarrow$ *undefined* @(*Int* $\rightarrow$ $a$ $\rightarrow$ $a$). However, this is not a value as GHC evaluates underneath the $\Lambda$ binder. The evaluation relation can be found in Appendix D.

*unitId2* () = *id*
*unitId2* () = *id*

Both of these functions ignore their input and return the polymorphic identity function. Let us look at their types:

|  |  | eager | lazy |
|---|---|---|---|
| *unitId1* | deep or shallow | $\forall \{a\}. () \to a \to a$ | $() \to \forall a. a \to a$ |
| *unitId2* | deep or shallow | $\forall \{a\}. () \to a \to a$ | $\forall \{a\}. () \to a \to a$ |

The lazy case for *UnitId1* is the odd one out: we see that the definition of *unitId1* has type $\forall\ a.\ a \to a$, do not instantiate it, and then prepend the () parameter. In the eager case, we see that both definitions instantiate *id* and then re-generalise.

However, the most interesting case is the treatment of *unitId2* under lazy instantiation. The reason the type of *unitId2* here differs from that of *unitId1* is that the pattern-match forces the instantiation of *id*. As each branch of a multiple-branch pattern-match must result in the same type, we have to seek the most general type that is still less general than each branch's type. Pattern matching thus performs an instantiation step (regardless of eagerness), in order to find this common type.

In the scenario of *unitId2*, however, this causes trouble: the match instantiates *id*, and then the type of *unitId2* is re-generalised. This causes *unitId2* to have a different inferred type than *unitId1*, leading to an instability.

**Similarity 6: $\eta$-Expansion**

And lastly, we want $\eta$-expansion not to affect types. (This change *can* reasonably affect runtime behaviour, so we would never want to assert that $\eta$-expansion maintains runtime semantics.)

$$e \overset{C}{\Longleftrightarrow} \lambda x.e\ x, \text{ where } e \text{ has a function type}$$

***Example 8: eta.*** Consider these two definitions, where *id* :: $\forall\ a.\ a \to a$:

*noEta* = *id*
*eta*   = $\lambda x \to$ *id x*

The two right-hand sides should have identical meaning, as *eta* is simply the $\eta$-expansion of *noEta*. Yet, under lazy instantiation, these two will have different types:

|  |  | eager | lazy |
|---|---|---|---|
| *noEta* | deep or shallow | $\forall \{a\}. a \to a$ | $\forall a. a \to a$ |
| *eta* | deep or shallow | $\forall \{a\}. a \to a$ | $\forall \{a\}. a \to a$ |

The problem is that the $\eta$-expansion instantiates the occurrence of *id* in *eta*, despite the lazy instantiation strategy. Under eager instantiation, the instantiation happens regardless.

### 3.1 Stability

The examples in this section show that the choice of instantiation scheme matters—and that no individual choice is clearly the best. To summarise, each of our possible schemes runs into trouble with some example; this table lists the numbers of the examples that witness a problem:

|  | eager | lazy |
|---|---|---|
| deep | 1, 2, 3, 4, 5, 6 | 5, 7, 8 |
| shallow | 1, 2, 3, 4, 6 | 7, 8 |

At this point, the best choice is unclear. Indeed, these examples are essentially where we started our exploration of this issue—with failures in each quadrant of this table, how should we design instantiation in GHC?

To understand this better, Section 4 presents a formalisation of GHC's type-checking algorithm, parameterised over the choice of depth and eagerness. Section 5 then presents properties derived from the similarities of this section and checks which variants of our type system uphold which properties. The conclusion becomes clear: lazy, shallow instantiation respects the most similarities.

We now fix the definition of stability we will work toward in this paper:

**Definition** (Stability). *A language is considered stable when all of the program similarities above are respected.*

We note here that the idea of judging a language by its robustness in the face of small transformations is not new; see, for example, Le Botlan and Rémy [2009] or Schrijvers et al. [2019], who also consider a similar property. However, we believe ours is the first work to focus on it as the primary criterion of evaluation.

Our goal in this paper is not to eliminate instability, which would likely be too limiting, leaving us potentially with either the Hindley-Milner implicit type system or a System F explicit one. Both are unsatisfactory. Instead, our goal is to make the consideration of stability a key guiding principle in language design. The rest of this paper uses the lens of stability to examine design choices around ordered explicit type instantiation. We hope that this treatment serves as an exemplar for other language design tasks and provides a way to translate vague notions of an "intuitive" design into concrete language properties that can be proved or disproved. Furthermore, we believe that instantiation is an interesting subject of study, as any language with polymorphism must consider these issues, making them less esoteric than they might first appear.

## 4 The Mixed Polymorphic $\lambda$-Calculus

In order to assess the stability of our different designs, this section develops a polymorphic, stratified $\lambda$-calculus with both implicit and explicit polymorphism. We call it the Mixed Polymorphic $\lambda$-calculus, or MPLC. Our formalisation (based on Eisenberg et al. [2016] and Serrano et al. [2020]) features

$$
\begin{array}{rcll}
\delta & ::= & \mathcal{S} \mid \mathcal{D} & \textit{Depth} \\
\epsilon & ::= & \mathcal{E} \mid \mathcal{L} & \textit{Eagerness} \\
\tau & ::= & a \mid \tau_1 \to \tau_2 \mid T\,\overline{\tau} & \textit{Monotype} \\
\rho & ::= & \tau \mid \sigma \to \phi^\delta & \textit{Instantiated type} \\
\sigma & ::= & \rho \mid \forall \overline{a}.\sigma \mid \forall \{\overline{a}\}.\sigma \mid \sigma_1 \to \sigma_2 & \textit{Type scheme} \\
\phi^\delta & ::= & \rho \quad (\delta = \mathcal{D}) & \textit{Instantiated result} \\
 & \mid & \sigma \quad (\delta = \mathcal{S}) & \\
\eta^\epsilon & ::= & \rho \quad (\epsilon = \mathcal{E}) & \textit{Synthesised type} \\
 & \mid & \sigma \quad (\epsilon = \mathcal{L}) &
\end{array}
$$

$$
\begin{array}{rcll}
e & ::= & h\,\overline{arg} \mid \lambda x.e \mid \Lambda a.e \mid \mathrm{let}\,decl\,\mathrm{in}\,e & \textit{Expression} \\
h & ::= & x \mid K \mid e : \sigma \mid e & \textit{Application head} \\
arg & ::= & e \mid @\sigma & \textit{Application argument} \\
decl & ::= & x : \sigma; \overline{x\,\overline{\pi_i} = e_i}^i \mid \overline{x\,\overline{\pi_i} = e_i}^i & \textit{Declaration} \\
\pi & ::= & x \mid K\,\overline{\pi} \mid @\sigma & \textit{Pattern} \\
\Sigma & ::= & \cdot \mid \Sigma, T\,\overline{a} \mid \Sigma, K : \overline{a}; \overline{\sigma}; T & \textit{Static context} \\
\Gamma, \Delta & ::= & \Sigma \mid \Gamma, x : \sigma \mid \Gamma, a & \textit{Context} \\
\psi & ::= & \tau \mid @a & \textit{Arg. descriptor}
\end{array}
$$

**Figure 1.** Mixed Polymorphic $\lambda$-Calculus (MPLC) Syntax

explicit type instantiation and abstraction, as well as type variable specificity. In order to support visible type application, even when instantiating eagerly, we must consider all the arguments to a function before doing our instantiation, lest some arguments be type arguments. Furthermore, type signatures are allowed in the calculus, and the bidirectional type system [Pierce and Turner 2000] permits higher-rank [Odersky and Läufer 1996] functions. Some other features, such as local let declarations defining functions with multiple equations, are added to support some of the similarities we wish to study.

We have built this system to support flexibility in both of our axes of instantiation scheme design. That is, the calculus is parameterised over choices of instantiation depth and eagerness. In this way, our calculus is essentially a *family* of type systems: choose your design, and you can instantiate our rules accordingly.

### 4.1 Syntax

The syntax for MPLC is shown in Figure 1. We define two meta parameters $\delta$ and $\epsilon$ denoting the depth and eagerness of instantiation respectively. In the remainder of this paper, grammar and relations which are affected by one of these parameters will be annotated as such. A good example of this are types $\phi^\delta$ and $\eta^\epsilon$, as explained below.

Keeping all the moving pieces straight can be challenging. We thus offer some mnemonics to help the reader: In the remainder of the paper, aspects pertaining to **e**ager instantiation are highlighted in emerald, while **l**azy features are highlighted in lavender. Similarly, instantiation under the **s**hallow scheme is drawn using a **s**triped line, as in $\Gamma \vdash \sigma \xdashrightarrow{inst\,\mathcal{S}} \rho$.

***Types.*** Our presentation of the MPLC contains several different type categories, used to constrain type inference. Monotypes $\tau$ represent simple ground types without any polymorphism, while type schemes $\sigma$ can be polymorphic, including under arrows. In contrast, instantiated types $\rho$ cannot have any top-level polymorphism. However, depending on the depth $\delta$ of instantiation, a $\rho$-type may or may not feature nested foralls on the right of function arrows.

This dependency on the depth $\delta$ of type instantiation is denoted using an instantiated result type $\phi^\delta$ on the right of the function arrow. Instantiating shallowly, $\phi^\mathcal{S}$ is a type scheme $\sigma$, but deep instantiation sees $\phi^\mathcal{D}$ as an instantiated type $\rho$. This makes sense: $Int \to \forall a.a \to a$ *is* a fully instantiated type under shallow instantiation, but not under deep. We also have synthesised types $\eta^\epsilon$ to denote the output of the type synthesis judgement $\Gamma \vdash e \Rightarrow \eta^\epsilon$, which infers a type from an expression. The shape of this type depends on the eagerness $\epsilon$ of type instantiation: under lazy instantiation ($\mathcal{L}$), inference can produce full type schemes $\sigma$; but under eager instantiation ($\mathcal{E}$), synthesised types $\eta^\epsilon$ are always instantiated types $\rho$: any top-level quantified variable would have been instantiated away.

Finally, an argument descriptor $\psi$ represents a type synthesised from analysing a function argument pattern. Descriptors are assembled into type schemes $\sigma$ with the *type* ($\overline{\psi}; \sigma_0 \sim \sigma$) judgement, in Figure 5.

***Expressions.*** Expressions $e$ are mostly standard; we explain the less common forms here.

As inspired by Serrano et al. [2020], applications are modelled as applying a head $h$ to a (maximally long) list of arguments $\overline{arg}$. The main idea is that under eager instantiation, type instantiation for the head is postponed until it has been applied to its arguments. A head $h$ is thus defined to be either a variable $x$, a data constructor $K$, an annotated expression $e : \sigma$ or a simple expression $e$. This last form will not be typed with a type scheme under eager instantiation—that is, we will not be able to use explicit instantiation—but is required to enable application of a lambda expression. As we feature both term and type application, an argument $arg$ is defined to be either an expression $e$ or a type argument $@\sigma$.

Our syntax additionally includes explicit abstractions over type variables, written with $\Lambda$. Though the design of this feature (inspired by Eisenberg et al. [2018, Appendix B]) is straightforward in our system, its inclusion drives some of the challenge of maintaining stability.

Lastly, let-expressions are modelled on the syntax of Haskell. These contain a single (non-recursive) declaration *decl*, which may optionally have a type signature $x : \sigma$, followed by the definition $\overline{x\ \overline{\pi}_i = e_i}^i$. The patterns $\overline{\pi}$ on the left of the equals sign can each be either a simple variable $x$, type $@\sigma$ or a saturated data constructor $K\ \overline{\pi}$.

| | | |
|---|---|---|
| Fig. 2 | $\Gamma \vdash e \Rightarrow \eta^\epsilon$ | Synthesise type $\eta^\epsilon$ for $e$ |
| Fig. 2 | $\Gamma \vdash e \Leftarrow \sigma$ | Check $e$ against type $\sigma$ |
| Fig. 2 | $\Gamma \vdash^H h \Rightarrow \sigma$ | Synthesise type $\sigma$ for head $h$ |
| Fig. 2 | $\Gamma \vdash^A \overline{arg} \Leftarrow \sigma \Rightarrow \sigma'$ | Check $\overline{arg}$ against $\sigma$, resulting in type $\sigma'$ |
| Fig. 3 | $\Gamma \vdash decl \Rightarrow \Gamma'$ | Extend context with a decl. |
| Fig. 4 | $\Gamma \vdash^P \overline{\pi} \Rightarrow \overline{\psi}; \Delta$ | Synthesise types $\overline{\psi}$ for patterns $\overline{\pi}$, binding context $\Delta$ |
| Fig. 4 | $\Gamma \vdash^P \overline{\pi} \Leftarrow \sigma \Rightarrow \sigma'; \Delta$ | Check $\overline{\pi}$ against $\sigma$, with residual type $\sigma'$, binding $\Delta$ |
| Fig. 5 | $\Gamma \vdash \sigma \xrightarrow{inst\ \delta} \rho$ | Instantiate $\sigma$ to $\rho$ |
| Fig. 5 | $\Gamma \vdash \sigma \xrightarrow{skol\ \delta} \rho; \Gamma'$ | Skolemise $\sigma$ to $\rho$, binding $\Gamma'$ |
| App. C | $binders^\delta(\sigma) = \overline{a}; \rho$ | Extract type var. binders $\overline{a}$ and residual $\rho$ from $\sigma$ |
| App. C | $wrap\ (\overline{\pi}; e_1 \sim e_2)$ | Bind patterns $\overline{\pi}$ in $e_1$ to get $e_2$ |

**Table 1.** Relation Overview

***Contexts.*** Typing contexts $\Gamma$ are entirely standard, storing both the term variables $x$ with their types and the type variables $a$ in scope; these type variables may appear in both terms (as the calculus features explicit type application) and types. The type constructors and data constructors are stored in a static context $\Sigma$, which forms the basis of typing contexts $\Gamma$. This static context contains the data type definitions by storing both type constructors $T\ \overline{a}$ and data constructors $K\ :\ \overline{a}; \overline{\sigma}; T$. Data constructor types contain the list of quantified variables $\overline{a}$, the argument types $\overline{\sigma}$, and the resulting type $T$; when $K\ :\ \overline{a}; \overline{\sigma}; T$, then the use of $K$ in an expression would have type $\forall \overline{a}.\overline{\sigma} \to T\ \overline{a}$, abusing syntax slightly to write a list of types $\overline{\sigma}$ to the left of an arrow.

### 4.2 Type system overview

Table 1 provides a high-level overview of the different typing judgements for the MPLC. The detailed rules can be found in Figures 2–5. The starting place to understand our rules is in Figure 2. These judgements implement a bidirectional type system, fairly standard with the exception of their treatment of a list of arguments all at once[4].

Understanding this aspect of the system hinges on rule TM-INFAPP, which synthesises the type of the head $h$ and uses its type to check the arguments $\overline{arg}$. The argument-checking judgement $\Gamma \vdash^A \overline{arg} \Leftarrow \sigma \Rightarrow \sigma'$ (inspired by Dunfield and Krishnaswami [2013]) uses the function's type $\sigma$ to learn what type is expected of each argument; after checking all arguments, the judgement produces a residual type $\sigma'$. The judgement's rules

---

[4]This is a well-known technique to reduce the number of traversals through the applications, known as *spine form* [Cervesato and Pfenning 2003].

$\boxed{\Gamma \vdash^H h \Rightarrow \sigma}$ *(Head Type Synthesis)*

H-VAR
$$\frac{x : \sigma \in \Gamma}{\Gamma \vdash^H x \Rightarrow \sigma}$$

H-CON
$$\frac{K : \overline{a}; \overline{\sigma}; T \in \Gamma}{\Gamma \vdash^H K \Rightarrow \forall \overline{a}.\overline{\sigma} \to T\ \overline{a}}$$

H-ANN
$$\frac{\Gamma \vdash e \Leftarrow \sigma}{\Gamma \vdash^H e : \sigma \Rightarrow \sigma}$$

H-INF
$$\frac{\Gamma \vdash e \Rightarrow \eta^\epsilon}{\Gamma \vdash^H e \Rightarrow \eta^\epsilon}$$

$\boxed{\Gamma \vdash e \Rightarrow \eta^\epsilon}$ *(Term Type Synthesis)*

TM-INFAPP
$$\frac{\Gamma \vdash^H h \Rightarrow \sigma \quad \Gamma \vdash^A \overline{arg} \Leftarrow \sigma \Rightarrow \sigma' \quad \Gamma \vdash \sigma' \xrightarrow{inst\ \delta} \eta^\epsilon}{\Gamma \vdash h\ \overline{arg} \Rightarrow \eta^\epsilon}$$

TM-INFABS
$$\frac{\Gamma, x : \tau_1 \vdash e \Rightarrow \eta_2^\epsilon}{\Gamma \vdash \lambda x.e \Rightarrow \tau_1 \to \eta_2^\epsilon}$$

TM-INFLET
$$\frac{\Gamma \vdash decl \Rightarrow \Gamma' \quad \Gamma' \vdash e \Rightarrow \eta^\epsilon}{\Gamma \vdash \text{let } decl \text{ in } e \Rightarrow \eta^\epsilon}$$

TM-INFTYABS
$$\frac{\Gamma, a \vdash e \Rightarrow \eta_1^\epsilon \quad \Gamma \vdash \forall a.\eta_1^\epsilon \xrightarrow{inst\ \delta} \eta_2^\epsilon}{\Gamma \vdash \Lambda a.e \Rightarrow \eta_2^\epsilon}$$

$\boxed{\Gamma \vdash e \Leftarrow \sigma}$ *(Term Type Checking)*

TM-CHECKABS
$$\frac{\Gamma \vdash \sigma \xrightarrow{skol\ S} \sigma_1 \to \sigma_2; \Gamma_1 \quad \Gamma_1, x : \sigma_1 \vdash e \Leftarrow \sigma_2}{\Gamma \vdash \lambda x.e \Leftarrow \sigma}$$

TM-CHECKLET
$$\frac{\Gamma \vdash decl \Rightarrow \Gamma' \quad \Gamma' \vdash e \Leftarrow \sigma}{\Gamma \vdash \text{let } decl \text{ in } e \Leftarrow \sigma}$$

TM-CHECKINF
$$\frac{\Gamma \vdash \sigma \xrightarrow{skol\ \delta} \rho; \Gamma_1 \quad \Gamma_1 \vdash e \Rightarrow \eta^\epsilon \quad \Gamma_1 \vdash \eta^\epsilon \xrightarrow{inst\ \delta} \rho \quad e \neq \lambda, \Lambda, \text{let}}{\Gamma \vdash e \Leftarrow \sigma}$$

TM-CHECKTYABS
$$\frac{\sigma = \forall \{\overline{a}\}.\forall a.\sigma' \quad \Gamma, \overline{a}, a \vdash e \Leftarrow \sigma'}{\Gamma \vdash \Lambda a.e \Leftarrow \sigma}$$

$\boxed{\Gamma \vdash^A \overline{arg} \Leftarrow \sigma \Rightarrow \sigma'}$ *(Argument Type Checking)*

ARG-EMPTY
$$\overline{\Gamma \vdash^A \cdot \Leftarrow \sigma \Rightarrow \sigma}$$

ARG-APP
$$\frac{\Gamma \vdash e \Leftarrow \sigma_1 \quad \Gamma \vdash^A \overline{arg} \Leftarrow \sigma_2 \Rightarrow \sigma'}{\Gamma \vdash^A e, \overline{arg} \Leftarrow \sigma_1 \to \sigma_2 \Rightarrow \sigma'}$$

ARG-INST
$$\frac{\Gamma \vdash^A e, \overline{arg} \Leftarrow [\tau_1/a]\ \sigma_2 \Rightarrow \sigma_3}{\Gamma \vdash^A e, \overline{arg} \Leftarrow \forall a.\sigma_2 \Rightarrow \sigma_3}$$

ARG-TYAPP
$$\frac{\Gamma \vdash^A \overline{arg} \Leftarrow [\sigma_1/a]\ \sigma_2 \Rightarrow \sigma_3}{\Gamma \vdash^A @\sigma_1, \overline{arg} \Leftarrow \forall a.\sigma_2 \Rightarrow \sigma_3}$$

ARG-INFINST
$$\frac{\Gamma \vdash^A \overline{arg} \Leftarrow [\tau_1/a]\ \sigma_2 \Rightarrow \sigma_3}{\Gamma \vdash^A \overline{arg} \Leftarrow \forall \{a\}.\sigma_2 \Rightarrow \sigma_3}$$

**Figure 2.** Term Typing for Mixed Polymorphic $\lambda$-Calculus

$\boxed{\Gamma \vdash decl \Rightarrow \Gamma'}$ *(Declaration Checking)*

**Decl-NoAnnSingle**
$$\frac{\Gamma \vdash^P \overline{\pi} \Rightarrow \overline{\psi}; \Delta \quad \Gamma, \Delta \vdash e \Rightarrow \eta^\epsilon \quad type\,(\overline{\psi}; \eta^\epsilon \sim \sigma) \quad \overline{a} = f_v(\sigma) \setminus dom\,(\Gamma)}{\Gamma \vdash x\,\overline{\pi} = e \Rightarrow \Gamma, x : \forall\,\overline{\{a\}}.\sigma}$$

**Decl-NoAnnMulti**
$$\frac{\begin{array}{c} i > 1 \quad \overline{\Gamma \vdash^P \overline{\pi_i} \Rightarrow \overline{\psi}; \Delta_i}^i \\ \overline{\Gamma, \Delta_i \vdash e_i \Rightarrow \eta_i^\epsilon}^i \\ \overline{\Gamma, \Delta_i \vdash \eta_i^\epsilon \xrightarrow{inst\,\delta} \rho}^i \quad type\,(\overline{\psi}; \rho \sim \sigma) \\ \overline{a} = f_v(\sigma) \setminus dom\,(\Gamma) \quad \sigma' = \forall\,\overline{\{a\}}.\sigma \end{array}}{\Gamma \vdash \overline{x\,\overline{\pi_i} = e_i}^i \Rightarrow \Gamma, x : \sigma'}$$

**Decl-Ann**
$$\frac{\overline{\Gamma \vdash^P \overline{\pi_i} \Leftarrow \sigma \Rightarrow \sigma_i'; \Delta_i}^i \quad \overline{\Gamma, \Delta_i \vdash e_i \Leftarrow \sigma_i'}^i}{\Gamma \vdash x : \sigma; \overline{x\,\overline{\pi_i} = e_i}^i \Rightarrow \Gamma, x : \sigma}$$

**Figure 3.** Declaration Typing for Mixed Polymorphic $\lambda$-Calculus

walk down the list, checking term arguments (rule Arg-App), implicitly instantiating specified variables where necessary (rule Arg-Inst, which spots a term-level argument $e$ but does not consume it), uses type arguments for instantiation (rule Arg-TyApp), and eagerly instantiates inferred type arguments (rule Arg-InfInst).

Our type system also includes let-declarations, which allow for the definition of functions, with or without type signatures, and supporting multiple equations defined by pattern-matching. Checking declarations and dealing with patterns is accomplished by the judgements in Figures 3 and 4, respectively, although the details may be skipped on a first reading: we include these rules for completeness and as the basis of our stability-oriented evaluation (Section 5). These rules do not directly offer insight into our treatment of instantiation.

Instead, the interesting aspects of our formulation are in the instantiation and skolemisation judgements.

### 4.3 Instantiation and Skolemisation

When we are type-checking the application of a polymorphic function, we must *instantiate* its type variables: this changes a function $id :: \forall\,a.\,a \to a$ into $id :: \tau \to \tau$, where $\tau$ is any monotype. On the other hand, when we are type-checking the body of a polymorphic definition, we must *skolemise* its type variables: this changes a definition $(\lambda x \to x) :: \forall\,a.\,a \to a$ so that we assign $x$ to have type $a$, where $a$ is a *skolem constant*—a fresh type, unequal to any other.

$\boxed{\Gamma \vdash^P \overline{\pi} \Rightarrow \overline{\psi}; \Delta}$ *(Pattern Synthesis)*

**Pat-InfEmpty**
$$\frac{}{\Gamma \vdash^P \cdot \Rightarrow \cdot; \cdot}$$

**Pat-InfVar**
$$\frac{\Gamma, x : \tau_1 \vdash^P \overline{\pi} \Rightarrow \overline{\psi}; \Delta}{\Gamma \vdash^P x, \overline{\pi} \Rightarrow \tau_1, \overline{\psi}; x : \tau_1, \Delta}$$

**Pat-InfCon**
$$\frac{\begin{array}{c} K : \overline{a_0}; \overline{\sigma_0}; T \in \Gamma \\ \Gamma \vdash^P \overline{\pi} \Leftarrow [\overline{\sigma_1}, \overline{\tau_0}/\overline{a_0}]\,(\overline{\sigma_0} \to T\,\overline{a_0}) \Rightarrow T\,\overline{\tau}; \Delta_1 \\ \Gamma, \Delta_1 \vdash^P \overline{\pi}' \Rightarrow \overline{\psi}; \Delta_2 \end{array}}{\Gamma \vdash^P (K\,@\overline{\sigma_1}\,\overline{\pi}), \overline{\pi}' \Rightarrow T\,\overline{\tau}, \overline{\psi}; \Delta_1, \Delta_2}$$

**Pat-InfTyVar**
$$\frac{\Gamma, a \vdash^P \overline{\pi} \Rightarrow \overline{\psi}; \Delta}{\Gamma \vdash^P @a, \overline{\pi} \Rightarrow @a, \overline{\psi}; a, \Delta}$$

$\boxed{\Gamma \vdash^P \overline{\pi} \Leftarrow \sigma \Rightarrow \sigma'; \Delta}$ *(Pattern Checking)*

**Pat-CheckEmpty**
$$\frac{}{\Gamma \vdash^P \cdot \Leftarrow \sigma \Rightarrow \sigma; \cdot}$$

**Pat-CheckVar**
$$\frac{\Gamma, x : \sigma_1 \vdash^P \overline{\pi} \Leftarrow \sigma_2 \Rightarrow \sigma'; \Delta}{\Gamma \vdash^P x, \overline{\pi} \Leftarrow \sigma_1 \to \sigma_2 \Rightarrow \sigma'; x : \sigma_1, \Delta}$$

**Pat-CheckCon**
$$\frac{\begin{array}{c} K : \overline{a_0}; \overline{\sigma_0}; T \in \Gamma \quad \Gamma \vdash \sigma_1 \xrightarrow{inst\,\delta} \rho_1 \\ \Gamma \vdash^P \overline{\pi} \Leftarrow [\overline{\sigma_1}, \overline{\tau_0}/\overline{a_0}]\,(\overline{\sigma_0} \to T\,\overline{a_0}) \Rightarrow \rho_1; \Delta_1 \\ \Gamma, \Delta_1 \vdash^P \overline{\pi}' \Leftarrow \sigma_2 \Rightarrow \sigma_2'; \Delta_2 \end{array}}{\Gamma \vdash^P (K\,@\overline{\sigma_1}\,\overline{\pi}), \overline{\pi}' \Leftarrow \sigma_1 \to \sigma_2 \Rightarrow \sigma_2'; \Delta_1, \Delta_2}$$

**Pat-CheckForall**
$$\frac{\Gamma, a \vdash^P \overline{\pi} \Leftarrow \sigma \Rightarrow \sigma'; \Delta \quad \overline{\pi} \neq \cdot \text{ and } \overline{\pi} \neq @\sigma, \overline{\pi}'}{\Gamma \vdash^P \overline{\pi} \Leftarrow \forall\,a.\sigma \Rightarrow \sigma'; a, \Delta}$$

**Pat-CheckTyVar**
$$\frac{\Gamma, a \vdash^P \overline{\pi} \Leftarrow [a/b]\,\sigma_1 \Rightarrow \sigma_2; \Delta}{\Gamma \vdash^P @a, \overline{\pi} \Leftarrow \forall\,b.\sigma_1 \Rightarrow \sigma_2; a, \Delta}$$

**Pat-CheckInfForall**
$$\frac{\Gamma, a \vdash^P \overline{\pi} \Leftarrow \sigma \Rightarrow \sigma'; \Delta \quad \overline{\pi} \neq \cdot}{\Gamma \vdash^P \overline{\pi} \Leftarrow \forall\,\{a\}.\sigma \Rightarrow \sigma'; a, \Delta}$$

**Figure 4.** Pattern Typing for Mixed Polymorphic $\lambda$-Calculus

These constants are bound in the context returned from the skolemisation judgement.

Naturally, the behaviour of both instantiation and skolemisation depend on the instantiation depth; see Figure 5. Both rule Inst-Inst and rule Skol-Skol use the *binders* helper

$$\boxed{\Gamma \vdash \sigma \xrightarrow{inst\ \delta} \rho} \qquad \text{(Type instantiation)}$$

$$\text{Inst-Inst} \quad \frac{binders^\delta(\sigma) = \overline{a}; \rho}{\Gamma \vdash \sigma \xrightarrow{inst\ \delta} [\overline{\tau}/\overline{a}]\,\rho}$$

$$\boxed{\Gamma \vdash \sigma \xrightarrow{skol\ \delta} \rho; \Gamma'} \qquad \text{(Type skolemisation)}$$

$$\text{Skol-Skol} \quad \frac{binders^\delta(\sigma) = \overline{a}; \rho}{\Gamma \vdash \sigma \xrightarrow{skol\ \delta} \rho; \Gamma, \overline{a}}$$

$$\boxed{type\ (\overline{\psi}; \sigma \sim \sigma')} \qquad \text{(Telescope Type Construction)}$$

$$\text{Type-Empty} \quad \frac{}{type\ (\cdot; \sigma \sim \sigma)} \qquad \text{Type-Var} \quad \frac{type\ (\overline{\psi}; \sigma_2 \sim \sigma_2')}{type\ (\tau_1, \overline{\psi}; \sigma_2 \sim \tau_1 \to \sigma_2')}$$

$$\text{Type-TyVar} \quad \frac{type\ (\overline{\psi}; \sigma \sim \sigma')}{type\ (@a, \overline{\psi}; \sigma \sim \forall a.\sigma')}$$

**Figure 5.** Type Instantiation and Skolemisation

function: $binders^\delta(\sigma) = \overline{a}; \rho$ extracts out bound type variables $\overline{a}$ and a residual type $\rho$ from a polytype $\sigma$. The depth, though, is key: the shallow ($\mathcal{S}$) version of our type system, $binders$ gathers only type variables bound at the top, while the deep ($\mathcal{D}$) version looks to the right past arrows. As examples, we have $binders^\mathcal{S}(\forall a.a \to \forall b.b \to b) = a; a \to \forall b.b \to b$ and $binders^\mathcal{D}(\forall a.a \to \forall b.b \to b) = a, b; a \to b \to b$. The full definition (inspired by Peyton Jones et al. [2007, Section 4.6.2]) is in Appendix C.

Some usages of these relations happen only for certain choices of instantiation flavour. For example, see rule Tm-InfApp. We see the last premise instantiates the result of the application— but its emerald colour tells us that this instantiation happens only under the eager flavour[5]. Indeed, this particular use of instantiation is the essence of eager instantiation: even after a function has been applied to all of its arguments, the eager scheme continues to instantiate. Similarly, rule Tm-InfTyAbs instantiates eagerly in the eager flavour.

The lazy counterpart to the eager instantiation in rule Tm-InfApp is the instantiation in rule Tm-CheckInf. This rule is the catch-all case in the checking judgement, and it is used when we are checking an application against an expected type, as in the expression $f\ a\ b\ c :: T\ Int\ Bool$. In this example, if $f\ a\ b\ c$ still has a polymorphic type, then we will need to instantiate it in order to check the type against the monomorphic $T\ Int\ Bool$. This extra instantiation would always be redundant in the eager flavour (the application is instantiated eagerly when inferring its type) but is vital in the lazy flavour.

Several other rules interact with instantiation in interesting ways:

*λ-expressions.* Rule Tm-CheckAbs checks a $\lambda$-expression against an expected type $\sigma$. However, this expected type may be a polytype. We thus must first skolemise it, revealing a function type $\sigma_1 \to \sigma_2$ underneath (if this is not possible, type checking fails). In order to support explicit type abstraction inside a lambda binder $\lambda x.\Lambda a.e$, rule Tm-CheckAbs never skolemises under an arrow: note the fixed $\mathcal{S}$ visible in the rule. As an example, this is necessary in order to accept $(\lambda x\ @b\ (y :: b) \to y) :: \forall a.a \to \forall b.b \to b$, where it would be disastrous to deeply skolemise the expected type when examining the outer $\lambda$.

*Declarations without a type annotation.* Rule Decl-NoAnnMulti is used for synthesising a type for a multiple-equation function definition that is not given a type signature. When we have multiple equations for a function, we might imagine synthesising different polytypes for each equation. We could then imagine trying to find some type that each equation's type could instantiate to, while still retaining as much polymorphism as possible. This would seem to be hard for users to predict, and hard for a compiler to implement. Our type system here follows GHC in instantiating the types of all equations to be a monotype, which is then re-generalised. This extra instantiation is not necessary under eager instantiation, which is why it is coloured in lavender.

For a single equation (rule Decl-NoAnnSingle), synthesising the original polytype, without instantiation and re-generalisation is straightforward, and so that is what we do (also following GHC).

## 5 Evaluation

This section evaluates the impact of the type instantiation flavour on the stability of the programming language. To this end, we define a set of eleven properties, based on the informal definition of stability from Section 3. Every property is analysed against the four instantiation flavours, the results of which are shown in Table 2, which also references the proof appendix for each of the properties, in the column labeled App.

We do not investigate the type safety of our formalism, as the MPLC is a subset of System F. We can thus be confident that programs in our language can be assigned a sensible runtime semantics without going wrong.

---

[5]We can also spot this fact by examining the metavariables. Instantiation takes us from a $\sigma$-type to a $\rho$-type, but the result in rule Tm-InfApp is a $\eta^\epsilon$-type: a $\rho$-type in the eager flavour, but a $\sigma$-type in the lazy flavour.

| Sim. | Property | | Phase | App. | $\mathcal{E}$ $\mathcal{S}$ | $\mathcal{E}$ $\mathcal{D}$ | $\mathcal{L}$ $\mathcal{S}$ | $\mathcal{L}$ $\mathcal{D}$ |
|---|---|---|---|---|---|---|---|---|
| 1 | 1 | Let inl. | C | E.1 | ✓ | ✓ | ✓ | ✓ |
|   | 2 | Let extr. | C | E.1 | ✗ | ✗ | ✓ | ✓ |
|   | 3 |  | R | E.3 | ✓ | ✗ | ✓ | ✗ |
| 2 | 4 | Signature prop | C |  | ✗ | ✗ | ✗ | ✗ |
|   | 4b | *restricted* |  | E.4 | ✗ | ✗ | ✓ | ✓ |
|   | 5 |  | R | E.4 | ✓ | ✗ | ✓ | ✗ |
| 3 | 6 | Type sign. | R | E.4 | ✓ | ✗ | ✓ | ✗ |
| 4 | 7 | Pattern inl. | C | E.5 | ✗ | ✗ | ✓ | ✓ |
|   | 8 |  | R | E.5 | ✓ | ✗ | ✓ | ✓ |
|   | 9 | Pattern extr. | C | E.5 | ✗ | ✗ | ✓ | ✓ |
| 5 | 10 | Single/multi | C | E.6 | ✓ | ✓ | ✗ | ✗ |
| 6 | 11 | $\eta$-expansion | C |  | ✗ | ✗ | ✗ | ✗ |
|   | 11b | *restricted* |  | E.7 | ✗ | ✓ | ✗ | ✗ |

**Table 2.** Property Overview

### 5.1 Contextual Equivalence

Following the approach of GHC, rather than providing an operational semantics of our type system directly, we instead define an elaboration of the surface language presented in this paper to explicit System F, our core language. It is important to remark that elaborating deep instantiation into this core language involves semantics-changing $\eta$-expansion. This allows us to understand the behaviour of Example 5, *swizzle*, which demonstrates a change in runtime semantics arising from a type signature. This change is caused by $\eta$-expansion, observable only in the core language.

The definition of this core language and the elaboration from MPLC to core are in Appendix D. The meta variable $t$ refers to core terms, and $\rightsquigarrow$ denotes elaboration. In the core language, $\eta$-expansion is expressed through the use of an expression wrapper $\dot{t}$, an expression with a hole, which re-types the expression that gets filled in. The full details can be found in Appendix D. We now provide an intuitive definition of contextual equivalence in order to describe what it means for runtime semantics to remain unchanged.

**Definition 1** (Contextual Equivalence). *Two core expressions $t_1$ and $t_2$ are contextually equivalent, written $t_1 \simeq t_2$, if there does not exist a context that can distinguish them. That is, $t_1$ and $t_2$ behave identically in all contexts.*

Here, we understand a context to be a core expression with a hole, similar to an expression wrapper, which instantiates the free variables of the expression that gets filled in. More concretely, the expression built by inserting $t_1$ and $t_2$ to the context should either both evaluate to the same value, or both diverge. A formal definition of contextual equivalence can be found in Appendix E.2.

### 5.2 Properties

**let-*inlining and extraction*.** We begin by analysing Similarity 1, which expands to the three properties described in this subsection.

**Property 1** (Let Inlining is Type Preserving).
- $\Gamma \vdash \text{let } x = e_1 \text{ in } e_2 \Rightarrow \eta^\epsilon \supset \Gamma \vdash [e_1/x] \, e_2 \Rightarrow \eta^\epsilon$
- $\Gamma \vdash \text{let } x = e_1 \text{ in } e_2 \Leftarrow \sigma \supset \Gamma \vdash [e_1/x] \, e_2 \Leftarrow \sigma$

**Property 2** (Let Extraction is Type Preserving).
- $\Gamma \vdash [e_1/x] \, e_2 \Rightarrow \eta_2^\epsilon \wedge \Gamma \vdash e_1 \Rightarrow \eta_1^\epsilon$
  $\supset \Gamma \vdash \text{let } x = e_1 \text{ in } e_2 \Rightarrow \eta_2^\epsilon$
- $\Gamma \vdash [e_1/x] \, e_2 \Leftarrow \sigma_2 \wedge \Gamma \vdash e_1 \Rightarrow \eta_1^\epsilon$
  $\supset \Gamma \vdash \text{let } x = e_1 \text{ in } e_2 \Leftarrow \sigma_2$

**Property 3** (Let Inlining is Runtime Semantics Preserving).
- $\Gamma \vdash \text{let } x = e_1 \text{ in } e_2 \Rightarrow \eta^\epsilon \rightsquigarrow t_1$
  $\wedge \Gamma \vdash [e_1/x] \, e_2 \Rightarrow \eta^\epsilon \rightsquigarrow t_2 \supset t_1 \simeq t_2$
- $\Gamma \vdash \text{let } x = e_1 \text{ in } e_2 \Leftarrow \sigma \rightsquigarrow t_1$
  $\wedge \Gamma \vdash [e_1/x] \, e_2 \Leftarrow \sigma \rightsquigarrow t_2 \supset t_1 \simeq t_2$

As an example for why Property 2 does not hold under eager instantiation, consider $id @ Int$. Extracting the $id$ function into a new let-binder fails to type check, because $id$ will be instantiated and then re-generalised. This means that explicit type instantiation can no longer work on the extracted definition.

The runtime semantics properties (both these and later ones) struggle under deep instantiation. This is demonstrated by Example 5, *swizzle*, where we see that non-prenex quantification can cause $\eta$-expansion during elaboration and thus change runtime semantics.

**Signature Property.** Similarity 2 gives rise to these properties about signatures.

**Property 4** (Signature Property is Type Preserving).
$\Gamma \vdash \overline{x \, \overline{\pi_i} = e_i}^i \Rightarrow \Gamma' \wedge x : \sigma \in \Gamma'$
$\supset \Gamma \vdash x : \sigma; \overline{x \, \overline{\pi_i} = e_i}^i \Rightarrow \Gamma'$

As an example of how this goes wrong under eager instantiation, consider the definition $x = \Lambda a. \lambda y. (y : a)$. Annotating $x$ with its inferred type $\forall \{a\}. a \rightarrow a$ is rejected, because rule TM-CHECKTYABS requires a *specified* quantified variable, not an *inferred* one.

However, similarly to eager evaluation, even lazy instantiation needs to instantiate the types at some point. In order to type a multi-equation declaration, a single type needs to be constructed that subsumes the types of every branch. In our type system, rule DECL-NOANNMULTI simplifies this process by first instantiating every branch type (following the example set by GHC), thus breaking Property 4. We thus introduce a simplified version of this property, limited to single equation declarations. This raises a possible avenue of future work: parameterising the type system over the handling of multi-equation declarations.

**Property 4b** (Signature Property is Type Preserving (Single Equation)).
$\Gamma \vdash x\,\overline{\pi} = e \Rightarrow \Gamma' \wedge x : \sigma \in \Gamma' \supset \Gamma \vdash x : \sigma; x\,\overline{\pi} = e \Rightarrow \Gamma'$

**Property 5** (Signature Property is Runtime Semantics Preserving).
$\Gamma \vdash \overline{x\,\overline{\pi_i} = e_i}^i \Rightarrow \Gamma' \rightsquigarrow x : \sigma = t_1$
$\wedge\; \Gamma \vdash x : \sigma; \overline{x\,\overline{\pi_i} = e_i}^i \Rightarrow \Gamma' \rightsquigarrow x : \sigma = t_2 \supset t_1 \simeq t_2$

***Type Signatures.*** Similarity 3 gives rise to the following property about runtime semantics.

**Property 6** (Type Signatures are Runtime Semantics Preserving).
$\Gamma \vdash x : \sigma_1; \overline{x\,\overline{\pi_i} = e_i}^i \Rightarrow \Gamma_1 \rightsquigarrow x : \sigma_1 = t_1$
$\wedge\; \Gamma \vdash x : \sigma_2; \overline{x\,\overline{\pi_i} = e_i}^i \Rightarrow \Gamma_1 \rightsquigarrow x : \sigma_2 = t_2$
$\wedge\; \Gamma \vdash \sigma_1 \xrightarrow{inst\,\delta} \rho \rightsquigarrow \dot{t}_1 \wedge \Gamma \vdash \sigma_2 \xrightarrow{inst\,\delta} \rho \rightsquigarrow \dot{t}_2$
$\supset \dot{t}_1[t_1] \simeq \dot{t}_2[t_2]$

Consider let $x : \forall a.Int \rightarrow a \rightarrow a; x = undefined$ in $x$ 'seq' (), which diverges. Yet under deep instantiation, this version terminates: let $x : Int \rightarrow \forall a.a \rightarrow a; x = undefined$ in $x$ 'seq' (). Under shallow instantiation, the second program is rejected, because $undefined$ cannot be instantiated to the type $Int \rightarrow \forall a.a \rightarrow a$, as that would be impredicative. You can find the typing rules for $undefined$ and $seq$ in Appendix D.1.

***Pattern Inlining and Extraction.*** The properties in this section come from Similarity 4. Like in that similarity, we assume that the patterns are just variables (either implicit type variables or explicit term variables).

**Property 7** (Pattern Inlining is Type Preserving).
$\Gamma \vdash x\,\overline{\pi} = e_1 \Rightarrow \Gamma' \wedge wrap\,(\overline{\pi}; e_1 \sim e_2) \supset \Gamma \vdash x = e_2 \Rightarrow \Gamma'$

The failure of pattern inlining under eager instantiation will feel similar: if we take $id\,@a\,x = x : a$, we will infer a type $\forall a.a \rightarrow a$. Yet if we write $id = \Lambda a.\lambda x.(x : a)$, then eager instantiation will give us the different type $\forall \{a\}.a \rightarrow a$.

**Property 8** (Pattern Inlining / Extraction is Runtime Semantics Preserving).
$\Gamma \vdash x\,\overline{\pi} = e_1 \Rightarrow \Gamma' \rightsquigarrow x : \sigma = t_1 \wedge wrap\,(\overline{\pi}; e_1 \sim e_2)$
$\wedge\; \Gamma \vdash x = e_2 \Rightarrow \Gamma' \rightsquigarrow x : \sigma = t_2 \supset t_1 \simeq t_2$

**Property 9** (Pattern Extraction is Type Preserving).
$\Gamma \vdash x = e_2 \Rightarrow \Gamma' \wedge wrap\,(\overline{\pi}; e_1 \sim e_2) \supset \Gamma \vdash x\,\overline{\pi} = e_1 \Rightarrow \Gamma'$

***Single vs. multiple equations.*** Similarity 5 says that there should be no observable change between the case for a single equation and multiple (redundant) equations with the same right-hand side. That gets formulated into the following property.

**Property 10** (Single/multiple Equations is Type Preserving).
$\Gamma \vdash x\,\overline{\pi} = e \Rightarrow \Gamma, x : \sigma \supset \Gamma \vdash x\,\overline{\pi} = e, x\,\overline{\pi} = e \Rightarrow \Gamma'$

This property favours the otherwise-unloved eager flavour. Imagine $f \_ = pair$. Under eager instantiation, this definition is accepted as type synthesis produces an instantiated type. Yet if we simply duplicate this equation under lazy instantiation (realistic scenarios would vary the patterns on the left-hand side, but duplication is simpler to state and addresses the property we want), then rule DECL-NOANNMULTI will reject as it requires the type to be instantiated.

***$\eta$-expansion.*** Similarity 6 leads to the following property.

**Property 11** ($\eta$-expansion is Type Preserving).
- $\Gamma \vdash e \Rightarrow \eta^\epsilon \wedge numargs(\eta^\epsilon) = n \supset \Gamma \vdash \lambda \overline{x}^n.e\,\overline{x}^n \Rightarrow \eta^\epsilon$
- $\Gamma \vdash e \Leftarrow \sigma \wedge numargs(\rho) = n \supset \Gamma \vdash \lambda \overline{x}^n.e\,\overline{x}^n \Leftarrow \sigma$

Here, $\overline{x}^n$ represents $n$ variables. We use $numargs(\sigma)$ to count the number of explicit arguments an expression can take, possibly instantiating any intervening implicit arguments. A formal definition can be found in Figure 7 in the appendix. However, in synthesis mode this property fails for every flavour: $\eta^\epsilon$ might be a function type $\sigma_1 \rightarrow \sigma_2$ taking a type scheme $\sigma_1$ as an argument, while we only synthesise monotype arguments. We thus introduce a restricted version of Property 11, with the additional premise that $\eta^\epsilon$ can not contain any binders to the left of an arrow.

**Property 11b** ($\eta$-expansion is Type Preserving (Monotype Restriction)).
- $\Gamma \vdash e \Rightarrow \eta^\epsilon \wedge numargs(\eta^\epsilon) = n \wedge \Gamma \vdash \eta^\epsilon \xrightarrow{inst\,\delta} \tau$
  $\supset \Gamma \vdash \lambda \overline{x}^n.e\,\overline{x}^n \Rightarrow \eta^\epsilon$
- $\Gamma \vdash e \Leftarrow \sigma \wedge numargs(\rho) = n \supset \Gamma \vdash \lambda \overline{x}^n.e\,\overline{x}^n \Leftarrow \sigma$

This (restricted) property fails for all but the eager/deep flavour as $\eta$-expansion forces other flavours to instantiate arguments they otherwise would not have.

### 5.3 Conclusion

A brief inspection of Table 2 suggests how we should proceed: choose *lazy*, *shallow* instantiation. While this configuration does not respect all properties, it is the clear winner—even more so when we consider that Property 11b (one of only two that favour another mode) must be artificially restricted in order for any of our flavours to support the property.

We should note here that we authors were surprised by this result. This paper arose from the practical challenge of designing instantiation in GHC. After considerable debate among the authors of GHC, we were unable to remain comfortable with any one decision—as we see here, no choice is perfect, and so any suggestion was met with counter-examples showing how that suggestion was incorrect. Yet we had a hunch that eager instantiation was the right design. We thus formulated the similarities of Section 3 and went about building a formalisation and proving properties. Crucially, we did *not* select the similarities to favour a particular result, though we did choose to avoid reasonable similarities that would not show any difference between instantiation flavours. At an early stage of this work, we continued to believe that eager instantiation was superior. It was only through careful analysis, guided by our proofs and counter-examples, that we realised that lazy instantiation was winning. We are now convinced by our own analysis.

## 6 Instantiation in GHC

Given the connection between this work and GHC, we now turn to examine some practicalities of how lazy instantiation might impact the implementation.

### 6.1 Eagerness

GHC used eager instantiation from the beginning, echoing Damas and Milner [1982]. However, the GHC 8 series, which contains support for explicit type application, implements an uneasy truce, sometimes using lazy instantiation (as advocated by Eisenberg et al. [2016]), and sometimes eager. In contrast, GHC 9.0 uses eager instantiation everywhere. This change was made for practical reasons: eager instantiation simplifies the code somewhat. If we went back to using lazy instantiation, the recent experience in going from lazy to eager suggests we will have to combat these challenges:

***Displaying inferred types.*** The types inferred for functions are more exotic with lazy instantiation. For example, defining $f = \lambda\_ \to id$ would infer $f :: \forall \{a\}. a \to \forall b. b \to b$. These types, which could be reported by tools (including GHCi), might be confusing for users.

***Monomorphism restriction.*** Eager instantiation makes the monomorphism restriction easier to implement, because relevant constraints are instantiated.

The monomorphism restriction is a peculiarity of Haskell, introduced to avoid unexpected runtime evaluation[6]. It potentially applies whenever a variable is defined without a type annotation and without any arguments to the left of the =: such a definition is not allowed to infer a type constraint.

Eager instantiation is helpful in implementing the monomorphism restriction, as the implementation of **let**-generalisation can look for unsolved constraints and default the type if necessary. With lazy instantiation, on the other hand, we would have to infer the type and then make a check to see whether it is constrained, instantiating it if necessary. Of course, the monomorphism restriction itself introduces instability in the language (note that *plus* and (+) have different types), and so perhaps revisiting this design choice is worthwhile.

***Type application with un-annotated variables.*** For simplicity, we want all variables without type signatures not to work with explicit type instantiation. (Eisenberg et al. [2016, Section 3.1] expands on this point.) Eager instantiation accomplishes this, because variables without type signatures would get their polymorphism via re-generalisation. On the other hand, lazy instantiation would mean that some user-written variables might remain in a variable's type, like in the type of $f$, just above.

Yet even with eager instantiation, if instantiation is shallow, we can *still* get the possibility of visible type application on un-annotated variables: the specified variables might simply be hiding under a visible argument. Consider *myPair* from Example 2: under eager shallow instantiation, it gets assigned the type $\forall \{a\}. a \to \forall b. b \to (a, b)$. This allows for visible type application despite the lack of a signature: *myPair True* @*Char*.

### 6.2 Depth

From the introduction of support for higher-rank types in GHC 6.8, GHC has done deep instantiation, as outlined by Peyton Jones et al. [2007], the paper describing the higher-rank types feature. However, deep instantiation has never respected the runtime semantics of a program; Peyton Jones [2019] has the details. In addition, deep instantiation is required in order to support covariance of result types in the type subsumption judgement ([Peyton Jones et al. 2007, Figure 7]). This subsumption judgement, though, weakens the ability to do impredicative type inference, as described by Serrano et al. [2018] and Serrano et al. [2020]. GHC has thus, for GHC 9.0, changed to use shallow subsumption and shallow instantiation.

### 6.3 The situation today: Quick Look impredicativity has arrived

A recent innovation within GHC (due for release in the next version, GHC 9.2) is the implementation of the Quick Look algorithm for impredicative type inference [Serrano et al. 2020]. The design of that algorithm walks a delicate balance between expressiveness and stability. It introduces new instabilities: for example, if $f\ x\ y$ requires impredicative instantiation, (**let** *unused* = 5 **in** $f$) $x\ y$ will fail. Given that users who opt into impredicative type inference are choosing to lose stability properties, we deemed it more important to study type inference without impredicativity in analysing stability. While our formulation of the inference algorithm is easily integrated with the Quick Look algorithm, we leave an analysis of the stability of the combination as future work.

---

[6]The full description is in the Haskell Report, Section 4.5.5 [Marlow (editor) 2010].

## 7 Conclusion

This work introduces the concept of *stability* as a proxy for the usability of a language that supports both implicit and explicit arguments[7]. We believe that designers of all languages supporting this mix of features need to grapple with how to best mix these features; those other designers may wish to follow our lead in formalising the problem to seek the most stable design. While stability is uninteresting in languages featuring pure explicit or pure implicit instantiation, it turns out to be an important metric in the presence of mixed behaviour.

Other work on type systems tends to focus on other properties; there is thus little related work beyond the papers we have already cited.

We introduced a family of type systems, parameterised over the instantiation flavour, and featuring a mix of explicit and implicit behaviour; these systems are inspired by Peyton Jones et al. [2007], Eisenberg et al. [2016], and Serrano et al. [2020]. Using this family, we then evaluated the different flavours of instantiation, against a set of formal stability properties. The results are surprisingly unambiguous: (a) *lazy instantiation* achieves the highest stability for the compile time semantics, and (b) *shallow instantiation* results in the most stable runtime semantics.

## Acknowledgments

The authors thank collaborator Simon Peyton Jones for discussion and feedback, along with our anonymous reviewers. We also thank Tom Schrijvers for his support. This material is based upon work supported by the National Science Foundation under Grant No. 1704041. Any opinions, findings, and conclusions or recommendations expressed in this material are those of the author and do not necessarily reflect the views of the National Science Foundation. This work is partially supported by KU Leuven, under Project No. C14/20/079.

---

[7]Recent work by Schrijvers et al. [2019] also uses the term *stability* to analyse a language feature around implicit arguments. That work discusses the stability of type class instance selection in the presence of substitutions, a different concern than we have here.

Haskell '21, August 26–27, 2021, Virtual Event, Republic of Korea    Gert-Jan Bottu and Richard A. Eisenberg

## A  Instabilities around instantiation beyond Haskell

The concept of stability is important in languages that have a mix of implicit and explicit features—a very common combination, appearing in Coq, Agda, Idris, modern Haskell, C++, Java, C#, Scala, F#, and Rust, among others. This appendix walks through how a mixing of implicit and explicit features in Idris[8] and Agda[9] causes instability, alongside the features of Haskell we describe in the main paper. We use these languages to show how the issues we describe are likely going to arise in any language mixing implicit and explicit features—and how stability is a worthwhile metric in examining these features—not to critique these languages in particular.

### A.1  Explicit Instantiation

Our example languages feature explicit instantiation of implicit arguments, allowing the programmer to manually instantiate a polymorphic type, for example. Explicit instantiation broadly comes in two flavours: ordered or named parameters.

### A.2  Idris

Idris supports named parameters. If we define *const* : { *a*, *b* : *Type* } → *a* → *b* → *a* (this syntax is the Idris equivalent of the Haskell type ∀ *a b*. *a* → *b* → *a*), then we can write *const* { *b* = *Bool* } to instantiate only the second type parameter or *const* { *a* = *Int* } { *b* = *Bool* } to instantiate both. Order does not matter; *const* { *b* = *Bool* } { *a* = *Int* } works as well as the previous example. Named parameters may be easier to read than ordered parameters and are robust to the addition of new type variables.

Idris's approach suffers from an instability inherent with named parameters. Unlike Haskell, the order of quantified variables does not matter. Yet now, the choice of *names* of the parameters naturally *does* matter. Thus *const* : *c* → *d* → *c* (taking advantage of the possibility of omitting explicit quantification in Idris) has a different interface than *const* : *a* → *b* → *a*, despite the fact that the type variables scope over only the type signature they appear in.

### A.3  Agda

Agda accepts both ordered and named parameters. After defining *const* : { *a b* : *Set* } → *a* → *b* → *a*, we can write expressions like *const* { *Int* } (instantiating only *a*), *const* { *b* = *Bool* }, or *const* { _ } { *Bool* }. Despite using named parameters, order *does* matter: we cannot instantiate earlier parameters after later ones. Naming is useful for skipping parameters that the user does not wish to instantiate. Because Agda requires explicit quantification of variables used in types (except as allowed for in implicit generalisation, below), the ordering of variables must be fixed by the programmer. However, like Idris, Agda suffers from the fact that the choice of name of these local variables leaks to clients.

### A.4  Explicit Abstraction

***Binding implicit variables in named function definitions.*** If we sometimes want to explicitly instantiate an implicit argument, we will also sometimes want to explicitly abstract over an implicit argument. A classic example of why this is useful is in the *replicate* function for length-indexed vectors, here written in Idris:

*replicate* : { *n* : *Nat* } → *a* → *Vect n a*
*replicate* { *n* = *Z* }    _ = [ ]
*replicate* { *n* = *S* _ } *x* = *x* :: *replicate x*

Because a length-indexed vector *Vect* includes its length in its type, we need not always pass the desired length of a vector into the *replicate* function: type inference can figure it out. We thus decide here to make the *n* : *Nat* parameter to be implicit, putting it in braces. However, in the definition of *replicate*, we must pattern-match on the length to decide what to return. The solution is to use an explicit pattern, in braces, to match against the argument *n*.

---

[8]We work with Idris 2, as available from https://github.com/idris-lang/Idris2, at commit a7d5a9a7fdfbc3e7ee8995a07b90e6a454209cd8.
[9]We work with Agda 2.6.0.1.

Idris and Agda both support explicit abstraction in parallel to their support of explicit instantiation: when writing equations for a function, the user can use braces to denote the abstraction over an implicit parameter. Idris requires such parameters to be named, while Agda supports both named and ordered parameters, just as the languages do for instantiation. The challenges around stability are the same here as they are for explicit instantiation.

Haskell has no implemented feature analogous to this. Its closest support is that for scoped type variables, where a type variable introduced in a type signature becomes available in a function body. For example:

*const* :: ∀ *a b*. *a* → *b* → *a*
*const x y* = (*x* :: *a*)

The ∀ *a b* brings *a* and *b* into scope both in the type signature *and* in the function body. This feature in Haskell means that, like in Idris and Agda, changing the name of an apparently local variable in a type signature may affect code beyond that type signature. It also means that the top-level ∀ in a type signature is treated specially. For example, neither of the following examples are accepted by GHC:

*const*$_1$ :: ∀. ∀ *a b*. *a* → *b* → *a*
*const*$_1$ *x y* = (*x* :: *a*)
*const*$_2$ :: (∀ *a b*. *a* → *b* → *a*)
*const*$_2$ *x y* = (*x* :: *a*)

In *const*$_1$, the vacuous ∀. (which is, generally, allowed) stops the scoped-type variables mechanism from bringing *a* into scope; in *const*$_2$, the parentheses around the type serve the same function. Once again, we see how Haskell is unstable: programmers might reasonably think that syntax like ∀ *a b*. is shorthand for ∀ *a*. ∀ *b*. or that outermost parentheses would be redundant, yet neither of these facts is true.

***Binding implicit variables in an anonymous function.*** Sometimes, binding a type variable only in a function declaration is not expressive enough, however—we might want to do this in an anonymous function in the middle of some other expression.

Here is a (contrived) example of this in Agda, where ∋ allows for prefix type annotations:

_∋_ : (*A* : *Set*) → *A* → *A*
*A* ∋ *x* = *x*

*ChurchBool* : *Set*$_1$
*ChurchBool* = {*A* : *Set*} → *A* → *A* → *A*

*churchBoolToBit* : *ChurchBool* → ℕ
*churchBoolToBit b* = *b* 1 0

*one* : ℕ
*one* = *churchBoolToBit* (λ{*A*} *x*$_1$ *x*$_2$ → *A* ∋ *x*$_1$)

Here, we bind the implicit variable *A* in the argument to *churchBoolToBit*. (Less contrived examples are possible; see the Motivation section of Eisenberg [2018].)

Binding an implicit variable in a λ-expression is subtler than doing it in a function clause. Idris does not support this feature at all, requiring a named function to bind an implicit variable. Agda supports this feature, as written above, but with caveats: the construct only works sometimes. For example, the following is rejected:

*id* : {*A* : *Set*} → *A* → *A*
*id* = λ{*A*} *x* → *A* ∋ *x*

The fact that this example is rejected, but *id* {*A*} *x* = *A* ∋ *x* is accepted is another example of apparent instability— we might naïvely expect that writing a function with an explicit λ and using patterns to the left of an = are equivalent. Another interesting aspect of binding an implicit variable in a λ-abstraction is that the name of the variable is utterly arbitrary: instead of writing (λ{*A*} *x*$_1$ *x*$_2$ → *A* ∋ *x*$_1$), we can write (λ{*anything* = *A*} *x*$_1$ *x*$_2$ → *A* ∋ *x*$_1$). This is an attempt to use Agda's support for named implicits, but the name can be, well, *anything*. This would appear to be a concession to the fact that the proper name for this variable, *A* as written in the definition of *ChurchBool*, can be arbitrarily far away from the usage of the name, so Agda is liberal in accepting any replacement for it.

An accepted proposal [Eisenberg 2018] adds this feature to Haskell, though it has not been implemented as of this writing. That proposal describes that the feature would be available only when we are *checking* a term against a known type, taking advantage of GHC's bidirectional type system [Eisenberg et al. 2016; Peyton Jones et al. 2007]. One of the motivations that inspired this paper was to figure out whether we could relax this restriction. After all, it would seem plausible that we should accept a definition like *id* = λ @*a* (*x* :: *a*) → *a* without a type signature. (Here, the @*a* syntax binds *a* to an otherwise-implicit type argument.) It will turn out that, in the end, we can do this only when we instantiate lazily—see Section 5.

### A.5 Implicit Generalisation

All three languages support some form of implicit generalisation, despite the fact that the designers of Haskell famously declared that **let** should not be generalised [Vytiniotis et al. 2010] and that both Idris and Agda require type signatures on all declarations.

***Haskell.*** Haskell's **let**-generalisation is the most active, as type signatures are optional.[10] Suppose we have defined *const x y* = *x*, without a signature. What type do we infer? It could be ∀ *a b*. *a* → *b* → *a* or ∀ *b a*. *a* → *b* → *a*. This choice matters, because it affects the meaning of explicit type instantiations. A natural reaction is to suggest choosing the former inferred type, following the left-to-right scheme described above. However, in a language with a type system as rich as Haskell's, this guideline does not always work. Haskell supports type synonyms (which can reorder the occurrence of variables), class constraints (whose ordering is arbitrary) [Wadler and Blott 1989], functional dependencies (which mean that a type variable might be mentioned *only* in constraints and not in the main body of a type) [Jones 2000], and arbitrary type-level computation through type families [Chakravarty et al. 2005; Eisenberg et al. 2014]. With all of these features potentially in play, it is unclear how to order the type variables. Thus, in a concession to language stability, Haskell brutally forbids explicit type instantiation on any function whose type is inferred; we discuss the precise mechanism in the next section.

Since GHC 8.0, Haskell allows dependency within type signatures [Weirich et al. 2013], meaning that the straightforward left-to-right ordering of variables—even in a user-written type signature—might not be well-scoped. As a simple example, consider *tr* :: *TypeRep* (*a* :: *k*), where *TypeRep* :: ∀ *k*. *k* → *Type* allows runtime type representation and is part of GHC's standard library. A naive left-to-right extraction of type variables would yield ∀ *a k*. *TypeRep* (*a* :: *k*), which is ill-scoped when we consider that *a* depends on *k*. Instead, we must reorder to ∀ *k a*. *TypeRep* (*a* :: *k*). In order to support stability when instantiating explicitly, GHC thus defines a concrete sorting algorithm, called "ScopedSort", that reorders the variables; it has become part of GHC's user-facing specification. Any change to this algorithm may break user programs, and it is specified in GHC's user manual.

***Idris.*** Idris's support for implicit generalisation is harder to trigger; see Appendix B for an example of how to do it. The problem that arises in Idris is predictable: if the compiler performs the quantification, then it must choose the name of the quantified type variable. How will clients know what this name is, necessary in order to instantiate the parameter? They cannot. Accordingly, in order to support stability, Idris uses a special name for generalised variables: the variable name itself includes braces (for example, it might be {*k* : 265}) and thus can never be parsed[11].

***Agda.*** Recent versions of Agda support a new **variable** keyword[12]. Here is an example of it in action:

---

[10]Though not relevant for our analysis, some readers may want the details: Without any language extensions enabled, all declarations without signatures are generalised, meaning that defining *id x* = *x* will give *id* the type ∀ *a*. *a* → *a*. With the MonoLocalBinds extension enabled, which is activated by either of GADTs or TypeFamilies, local definitions that capture variables from an outer scope are not generalised—this is the effect of the dictum that **let** should not be generalised. As an example, the *g* in *f x* = **let** *g y* = (*y*, *x*) **in** (*g* 'a', *g True*) is not generalised, because its body mentions the captured *x*. Accordingly, *f* is rejected, as it uses *g* at two different types (*Char* and *Bool*). Adding a type signature to *g* can fix the problem.
[11]Idris 1 does not use an exotic name, but still prevents explicit instantiation, using a mechanism similar to Haskell's specificity mechanism.
[12]See https://agda.readthedocs.io/en/v2.6.0.1/language/generalization-of-declared-variables.html in the Agda manual for an description of the feature.

**variable**
  $A$ : Set
  $l_1$ $l_2$ : List $A$

The declaration says that an out-of-scope use of, say, $A$ is a hint to Agda to implicitly quantify over $A$ : Set. The order of declarations in a **variable** block is significant: note that $l_1$ and $l_2$ depend on $A$. However, because explicit instantiation by order is possible in Agda, we must specify the order of quantification when Agda does generalisation. Often, this order is derived directly from the **variable** block—but not always. Consider this (contrived) declaration:

*property* : *length* $l_2$ + *length* $l_1$ ≡ *length* $l_1$ + *length* $l_2$

What is the full, elaborated type of *property*? Note that the two lists $l_1$ and $l_2$ can have *different* element types $A$. The Agda manual calls this *nested* implicit generalisation, and it specifies an algorithm—similar to GHC's ScopedSort—to specify the ordering of variables. Indeed it must offer this specification, as leaving this part out would lead to instability; that is, it would lead to the inability for a client of *property* to know how to order their type instantiations.

## B  Example of Implicit Generalisation in Idris

It is easy to believe that a language that requires type signatures on all definitions will not have implicit generalisation. However, Idris does allow generalisation to creep in, with just the right definitions.

We start with this:

**data** *Proxy* : { $k$ : *Type* } → $k$ → *Type* **where**
  $P$ : *Proxy a*

The datatype *Proxy* here is polymorphic; its one explicit argument can be of any type.

Now, we define *poly*:

*poly* : *Proxy a*
*poly* = $P$

We have not given an explicit type to the type variable $a$ in *poly*'s type. Because *Proxy*'s argument can be of any type, $a$'s type is unconstrained. Idris *generalises* this type, giving *poly* the type { $k$ : *Type* } → { $a$ : $k$ } → *Proxy a*.

At a use site of *poly*, we must then distinguish between the possibility of instantiating the user-written $a$ and the possibility of instantiating the compiler-generated $k$. This is done by giving the $k$ variable an unusual name, {k:446} in our running Idris session.

## C  Type System Details

$\boxed{binders^\delta(\sigma) = \overline{a}; \rho}$ *(Binders)*

BNDR-SHALLOWINST
$$\overline{binders^S(\rho) = \cdot; \rho}$$

BNDR-SHALLOWFORALL
$$\frac{binders^S(\sigma) = \overline{b}; \rho}{binders^S(\forall \overline{a}.\sigma) = \overline{a}, \overline{b}; \rho}$$

BNDR-SHALLOWINFFORALL
$$\frac{binders^S(\sigma) = \overline{b}; \rho}{binders^S(\forall \overline{\{a\}}.\sigma) = \overline{\{a\}}, \overline{b}; \rho}$$

BNDR-DEEPMONO
$$\overline{binders^D(\tau) = \cdot; \tau}$$

BNDR-DEEPFUNCTION
$$\frac{binders^D(\sigma_2) = \overline{a}; \rho_2}{binders^D(\sigma_1 \to \sigma_2) = \overline{a}; \sigma_1 \to \rho_2}$$

BNDR-DEEPFORALL
$$\frac{binders^D(\sigma) = \overline{b}; \rho}{binders^D(\forall \overline{a}.\sigma) = \overline{a}, \overline{b}; \rho}$$

BNDR-DEEPINFFORALL
$$\frac{binders^D(\sigma) = \overline{b}; \rho}{binders^D(\forall \overline{\{a\}}.\sigma) = \overline{\{a\}}, \overline{b}; \rho}$$

$\boxed{wrap\ (\overline{\pi}; e_1 \sim e_2)}$ *(Pattern Wrapping)*

PATWRAP-EMPTY
$$\overline{wrap\ (\cdot; e \sim e)}$$

PATWRAP-VAR
$$\frac{wrap\ (\overline{\pi}; e_1 \sim e_2)}{wrap\ (x, \overline{\pi}; e_1 \sim \lambda x.e_2)}$$

PATWRAP-TYVAR
$$\frac{wrap\ (\overline{\pi}; e_1 \sim e_2)}{wrap\ (@a, \overline{\pi}; e_1 \sim \Lambda a.e_2)}$$

In addition to including the figures above, this appendix describes our treatment of let-declarations and patterns:

**Let Binders.** A let-expression let *decl* in *e* (rule ETm-InfLet and rule ETm-CheckLet) defines a single variable, with or without a type signature. The declaration typing judgement (Figure 3) produces a new context $\Gamma'$, extended with the binding from this declaration.

Rules Decl-NoAnnSingle and Decl-NoAnnMulti distinguish between a single equation without a type signature and multiple equations. In the former case, we synthesise the types of the patterns using the $\vdash^P$ judgement and then the type of the right-hand side. We assemble the complete type with *type*, and then generalise. The multiple-equation case is broadly similar, synthesising types for the patterns (note that each equation must yield the *same* types $\overline{\psi}$) and then synthesising types for the right-hand side. These types are then instantiated (only necessary under lazy instantiation—eager instantiation would have already done this step). This additional instantiation step is the only difference between the single-equation case and the multiple-equation case. The reason is that rule Decl-NoAnnMulti needs to construct a single type that subsumes the types of every branch. Following GHC, we simplify this process by first instantiating the types.

Rule Decl-Ann checks a declaration with a type signature. It works by first checking the patterns $\overline{\pi}_i$ on the left of the equals sign against the provided type $\sigma$. The right-hand sides $e_i$ are then checked against the remaining type $\sigma'_i$.

**Patterns.** The pattern synthesis relation $\Gamma \vdash^P \overline{\pi} \Rightarrow \overline{\psi}; \Delta$ and checking relation $\Gamma \vdash^P \overline{\pi} \Leftarrow \sigma \Rightarrow \sigma'; \Delta$ are presented in Figure 4. As the full type is not yet available, synthesis produces argument descriptors $\overline{\psi}$ and a typing context extension $\Delta$. When checking patterns, the type to check against $\sigma$ is available, and the relation produces a residual type $\sigma'$, along with the typing context extension $\Delta$.

Typing a variable pattern works similarly to expressions. Under inference (rule Pat-InfVar) we construct a monotype and place it in the context. When checking a variable (rule Pat-CheckVar), its type $\sigma_1$ is extracted from the known function type and placed in the context. Type abstraction $@a$ in both synthesis and checking mode (rule Pat-InfTyVar and rule Pat-CheckTyVar respectively) produces a type argument descriptor $@a$ and extends the typing environment.

Typing data constructor patterns (rule Pat-InfCon and rule Pat-CheckCon), works by looking up the type $\forall \overline{a}_0.\overline{\sigma}_0 \to T\,\overline{a}_0$ of the constructor $K$ in the typing context, and checking the applied patterns $\overline{\pi}$ against the instantiated type, and an extended context[13]. The remaining type should be the result type for the constructor, meaning that the constructor always needs to be fully applied. Note that full type schemes $\overline{\sigma}_1$ are allowed in patterns, where they are used to instantiate the variables $\overline{a}_0$ (possibly extended with guessed monotypes $\overline{\tau}_0$, if there are not enough $\overline{\sigma}_1$). Consider, for example, $f\,(Just\,@Int\,x) = x + 1$, where the $@Int$ refines the type of $Just$, which in turn assigns $x$ the type $Int$. Note that pattern checking allows skolemising bound type variables (rule Pat-CheckInfForall), but only when the patterns are not empty in order not to lose syntax-directedness of the rules. The same holds for rule Pat-CheckForall, which only applies when no other rules match.

## D  Core Language

The dynamic semantics of the languages in Section 4 are defined through a translation to System F. While the target language is largely standard, a few interesting remarks can be made. The language supports nested pattern matching through case lambdas $\overline{\text{case}\,\overline{\pi_{Fi}} : \overline{\psi_F} \to t_i}^i$, where patterns $\pi_F$ include both term and type variables, as well as nested constructor patterns. Note that while we reuse our type $\sigma$ grammar for the core language, System F does not distinguish between inferred and specified binders.

We also define two meta-language features to simplify the elaboration, and the proofs: Firstly, in order to support eta-expansion (for translating deep instantiation to System F), we define expression wrappers $\dot{t}$, essentially a limited form of expressions with a hole • in them. An expression $t$ can be filled in for the hole to get a new expression $\dot{t}[t]$. One especially noteworthy wrapper construct is $\lambda t_1.t_2$, explicitly abstracting over and handling the expression to be filled in. Note that, as expression wrappers are only designed to alter the type of expressions through eta-expansion, there is no need to support the full System F syntax.

---
[13]Extending the context for later patterns is not used in this system, but it would be required for extensions like view patterns.

Secondly, in order to define contextual equivalence, we introduce contexts $M$. These are again expressions with a hole • in them, but unlike expression wrappers, contexts do cover the entire System F syntax. Typing contexts is performed by the $M : \Gamma_1; \sigma_1 \mapsto \Gamma_2; \sigma_2$ relation: "Given an expression $t$ that has type $\sigma_1$ under typing environment $\Gamma_1$, then the resulting expression $M[t]$ has type $\sigma_2$ under typing environment $\Gamma_2$". We will elaborate further on contextual equivalence in Appendix E.2.

$$
\begin{array}{rcll}
t & ::= & x \mid K \mid t_1\, t_2 \mid \lambda x : \sigma.t \mid t\,\sigma \mid \Lambda a.t & \textit{Expression} \\
  & \mid & \textit{undefined} \mid \textit{seq} & \\
  & \mid & \text{case}\, \overline{\overline{\pi_{F_i}} : \overline{\psi_F} \to t_i}^{\,i} \mid \textit{true} \mid \textit{false} & \\
v & ::= & \lambda x : \sigma.t \mid \Lambda a.v \mid K\, \overline{t} & \textit{Value} \\
  & \mid & \text{case}\, \overline{\overline{\pi_{F_i}} : \overline{\psi_F} \to t_i}^{\,i} & \\
\dot{t} & ::= & \bullet \mid \lambda x : \sigma.\dot{t} \mid \dot{t}\,\sigma \mid \Lambda a.\dot{t} \mid \lambda t_1.t_2 & \textit{Expr. Wrapper} \\
M & ::= & \bullet \mid \lambda x : \sigma.M \mid M\, t \mid t\, M & \textit{Context} \\
  & \mid & \Lambda a.M \mid M\, \sigma & \\
\textit{arg}_F & ::= & t \mid \sigma & \textit{Argument} \\
\pi_F & ::= & x : \sigma \mid @a \mid K\, \overline{\pi_F} & \textit{Pattern} \\
\psi_F & ::= & \sigma \mid @a & \textit{Arg. descriptor}
\end{array}
$$

$\boxed{\Gamma \vdash t : \sigma}$  \hfill (System F Term Typing)

$$
\begin{array}{cccc}
\text{FTm-Var} & \text{FTm-Con} & \text{FTm-App} & \text{FTm-Abs} \\
\dfrac{x : \sigma \in \Gamma}{\Gamma \vdash x : \sigma} & \dfrac{K : \overline{a}; \overline{\sigma}; T \in \Gamma}{\Gamma \vdash K : \forall \overline{a}.\overline{\sigma} \to T\, \overline{a}} & \dfrac{\Gamma \vdash t_1 : \sigma_1 \to \sigma_2 \qquad \Gamma \vdash t_2 : \sigma_1}{\Gamma \vdash t_1\, t_2 : \sigma_2} & \dfrac{\Gamma, x : \sigma_1 \vdash t : \sigma_2}{\Gamma \vdash \lambda x : \sigma_1.t : \sigma_1 \to \sigma_2}
\end{array}
$$

$$
\begin{array}{ccccc}
\text{FTm-TyApp} & \text{FTm-TyAbs} & \text{FTm-Undef} & \text{FTm-True} & \text{FTm-False} \\
\dfrac{\Gamma \vdash t : \forall a.\sigma_1}{\Gamma \vdash t\, \sigma_2 : [\sigma_2/a]\, \sigma_1} & \dfrac{\Gamma, a \vdash t : \sigma}{\Gamma \vdash \Lambda a.t : \forall a.\sigma} & \dfrac{}{\Gamma \vdash \textit{undefined} : \forall a.a} & \dfrac{}{\Gamma \vdash \textit{true} : \textit{Bool}} & \dfrac{}{\Gamma \vdash \textit{false} : \textit{Bool}}
\end{array}
$$

$$
\begin{array}{cc}
 & \text{FTm-Case} \\
 & \overline{\Gamma \vdash^P \overline{\pi_{F_i}} : \overline{\psi_F}; \Delta}^{\,i} \\
 & \overline{\Gamma, \Delta \vdash t_i : \sigma_1}^{\,i} \\
\text{FTm-Seq} & \textit{type}\,(\overline{\psi_F}; \sigma_1 \sim \sigma_2) \\
\dfrac{}{\Gamma \vdash \textit{seq} : \forall a.\forall b.a \to b \to b} & \dfrac{}{\Gamma \vdash \text{case}\, \overline{\overline{\pi_{F_i}} : \overline{\psi_F} \to t_i}^{\,i} : \sigma_2}
\end{array}
$$

$\boxed{\Gamma \vdash^P \overline{\pi_F} : \overline{\psi_F}; \Delta}$ \hfill (System F Pattern Typing)

$$
\begin{array}{ccc}
\text{FPat-Empty} & \text{FPat-Var} & \text{FPat-TyVar} \\
 & \dfrac{\Gamma, x : \sigma \vdash^P \overline{\pi_F} : \overline{\psi_F}; \Delta}{\Gamma \vdash^P x : \sigma, \overline{\pi_F} : \sigma, \overline{\psi_F}; x : \sigma, \Delta} & \dfrac{\Gamma, a \vdash^P \overline{\pi_F} : \overline{\psi_F}; \Delta}{\Gamma \vdash^P @a, \overline{\pi_F} : @a, \overline{\psi_F}; a, \Delta} \\
\dfrac{}{\Gamma \vdash^P \cdot : \cdot; \cdot} & &
\end{array}
$$

$$
\text{FPat-Con} \\
\dfrac{\begin{array}{c} K : \overline{a}_0; \overline{\sigma}_0; T \in \Gamma \\ \Gamma \vdash^P \overline{\pi_F} : [\overline{\sigma}_1/\overline{a}_0]\, \overline{\sigma}_0; \Delta_1 \\ \Gamma, \Delta_1 \vdash^P \overline{\pi_F}' : \overline{\psi_F}; \Delta_2 \end{array}}{\Gamma \vdash^P (K\, \overline{\pi_F}), \overline{\pi_F}' : T\, \overline{\sigma}_1, \overline{\psi_F}; \Delta_1, \Delta_2}
$$

$$\boxed{M : \Gamma_1; \sigma_1 \mapsto \Gamma_2; \sigma_2}$$ *(System F Context Typing)*

**FCtx-Hole**
$$\bullet : \Gamma; \sigma \mapsto \Gamma; \sigma$$

**FCtx-Abs**
$$\frac{M : \Gamma_1; \sigma_2 \mapsto \Gamma_2, x : \sigma_1; \sigma_3}{\lambda x : \sigma_1.M : \Gamma_1; \sigma_2 \mapsto \Gamma_2; \sigma_1 \to \sigma_3}$$

**FCtx-AppR**
$$\frac{M_1 : \Gamma_1; \sigma_1 \mapsto \Gamma_2; \sigma_2 \to \sigma_3 \quad \Gamma_2 \vdash t_2 : \sigma_2}{M_1\ t_2 : \Gamma_1; \sigma_1 \mapsto \Gamma_2; \sigma_3}$$

**FCtx-AppL**
$$\frac{\Gamma_2 \vdash t_1 : \sigma_2 \to \sigma_3 \quad M_2 : \Gamma_1; \sigma_1 \mapsto \Gamma_2; \sigma_2}{t_1\ M_2 : \Gamma_1; \sigma_1 \mapsto \Gamma_2; \sigma_3}$$

**FCtx-TyAbs**
$$\frac{M : \Gamma_1; \sigma_1 \mapsto \Gamma_2, a; \sigma_2}{\Lambda a.M : \Gamma_1; \sigma_1 \mapsto \Gamma_2; \forall a.\sigma_2}$$

**FCtx-TyApp**
$$\frac{M : \Gamma_1; \sigma_1 \mapsto \Gamma_2; \forall a.\sigma_2}{M\ \sigma : \Gamma_1; \sigma_1 \mapsto \Gamma_2; [\sigma/a]\ \sigma_2}$$

**FCtx-Case**
$$\frac{\overline{\Gamma_1 \vdash^P \overline{\pi_{F_i}} : \overline{\psi_F}; \Delta}^i \quad \overline{M_i : \Gamma_1, \Delta; \sigma_1 \mapsto \Gamma_2; \sigma_2}^i \quad type\ (\overline{\psi_F}; \sigma_2 \sim \sigma_3)}{\text{case}\ \overline{\overline{\pi_{F_i}} : \overline{\psi_F} \to M_i}^i : \Gamma_1; \sigma_1 \mapsto \Gamma_2; \sigma_3}$$

Evaluation for our System F target language is largely standard and defined below. Note that, following GHC, our target language evaluates inside type abstractions (rule FEval-TyAbs). Because of this, a type abstraction $\Lambda a.t$ is a value if and only if $t$ is a value. A more extensive discussion can be found in Breitner et al. [2016, Appendix A.3].

$$\boxed{match\ (\overline{\pi_{F1}} \to t_1; t_2 : \sigma_2) \hookrightarrow \overline{\pi_{F2}} \to t_1'}$$ *(Core Pattern Matching)*

**FMatch-Var**
$$match\ (x : \sigma, \overline{\pi_F} \to t_1; t_2 : \sigma) \hookrightarrow \overline{\pi_F} \to [t_2/x]t_1$$

**FMatch-Con**
$$\frac{\sigma_2 = \overline{\psi_{F1}} \to \tau_2 \quad t_2 \hookrightarrow^{\Downarrow} K\ \overline{t} \quad (\text{case}\ \overline{\pi_{F1}} : \overline{\psi_{F1}} \to t_1)\ \overline{t} \hookrightarrow^{\Downarrow} v}{match\ ((K\ \overline{\pi_{F1}}), \overline{\pi_{F2}} \to t_1; t_2 : \sigma_2) \hookrightarrow \overline{\pi_{F2}} \to v}$$

$$\boxed{t_1 \hookrightarrow t_2}$$ *(Core Evaluation)*

**FEval-App**
$$\frac{t_1 \hookrightarrow t_1'}{t_1\ t_2 \hookrightarrow t_1'\ t_2}$$

**FEval-AppAbs**
$$(\lambda x : \sigma.t_1)\ t_2 \hookrightarrow [t_2/x]t_1$$

**FEval-Seq**
$$\frac{t_1 \hookrightarrow t_1'}{seq\ t_1\ t_2 \hookrightarrow seq\ t_1'\ t_2}$$

**FEval-SeqVal**
$$seq\ v_1\ t_2 \hookrightarrow t_2$$

**FEval-CaseEmpty**
$$\text{case}\ \overline{\cdot : \cdot \to t_i}^i \hookrightarrow t_0$$

**FEval-CaseMatch**
$$\frac{\forall j \in v\ \text{where}\ (match\ (\overline{\pi_{Fj}} \to t_j; t_2 : \sigma) \hookrightarrow \overline{\pi_{Fj}'} \to t_j')}{(\text{case}\ \overline{\overline{\pi_{Fi}} : \sigma, \overline{\psi_F} \to t_i}^{i<v})\ t_2 \hookrightarrow \text{case}\ \overline{\overline{\pi_{Fj}'} : \overline{\psi_F} \to t_j'}^{j<w}}$$

**FEval-TyAbs**
$$\frac{t \hookrightarrow t'}{\Lambda a.t \hookrightarrow \Lambda a.t'}$$

**FEval-TyApp**
$$\frac{t_1 \hookrightarrow t_1'}{t_1\ \sigma \hookrightarrow t_1'\ \sigma}$$

**FEval-TyAppAbs**
$$(\Lambda a.v_1)\ \sigma \hookrightarrow [\sigma/a]\ v_1$$

**FEval-Undef**
$$undefined \hookrightarrow undefined$$

**FEval-TyAbsCase**
$$(\text{case}\ \overline{@a_i, \overline{\pi_{Fi}} : @a, \overline{\psi_F} \to t_i}^i)\ \sigma \hookrightarrow \text{case}\ \overline{[\sigma/a]\overline{\pi_{Fi}} : [\sigma/a]\ \overline{\psi_F} \to [\sigma/a]\ t_i}^i$$

$$\boxed{t \hookrightarrow^{\Downarrow} v}$$ *(Big Step Evaluation)*

**FEvalBigStep-Step**
$$\frac{t \hookrightarrow t' \quad t' \hookrightarrow^{\Downarrow} v}{t \hookrightarrow^{\Downarrow} v}$$

**FEvalBigStep-Done**
$$v \hookrightarrow^{\Downarrow} v$$

## D.1 Translation from the Mixed Polymorphic $\lambda$-calculus

$\boxed{\Gamma \vdash^H h \Rightarrow \sigma \rightsquigarrow t}$ *(Head Type Synthesis)*

**H-Var**
$$\frac{x : \sigma \in \Gamma}{\Gamma \vdash^H x \Rightarrow \sigma \rightsquigarrow x}$$

**H-Con**
$$\frac{K : \overline{a}; \overline{\sigma}; T \in \Gamma}{\Gamma \vdash^H K \Rightarrow \forall \overline{a}.\overline{\sigma} \rightarrow T\,\overline{a} \rightsquigarrow K}$$

**H-Ann**
$$\frac{\Gamma \vdash e \Leftarrow \sigma \rightsquigarrow t}{\Gamma \vdash^H e : \sigma \Rightarrow \sigma \rightsquigarrow t}$$

**H-Undef**
$$\overline{\Gamma \vdash^H \text{undefined} \Rightarrow \forall a.a \rightsquigarrow \text{undefined}}$$

**H-Seq**
$$\overline{\Gamma \vdash^H \text{seq} \Rightarrow \forall a.\forall b.a \rightarrow b \rightarrow b \rightsquigarrow \text{seq}}$$

**H-Inf**
$$\frac{\Gamma \vdash e \Rightarrow \eta^\epsilon \rightsquigarrow t}{\Gamma \vdash^H e \Rightarrow \eta^\epsilon \rightsquigarrow t}$$

$\boxed{\Gamma \vdash e \Rightarrow \eta^\epsilon \rightsquigarrow t}$ *(Term Type Synthesis)*

**Tm-InfAbs**
$$\frac{\Gamma, x : \tau_1 \vdash e \Rightarrow \eta_2^\epsilon \rightsquigarrow t_1}{\Gamma \vdash \lambda x.e \Rightarrow \tau_1 \rightarrow \eta_2^\epsilon \rightsquigarrow \lambda x : \tau_1.t_1}$$

**Tm-InfTyAbs**
$$\frac{\Gamma, a \vdash e \Rightarrow \eta_1^\epsilon \rightsquigarrow t \quad \Gamma \vdash \forall a.\eta_1^\epsilon \xrightarrow{inst\,\delta} \eta_2^\epsilon \rightsquigarrow \dot{t}}{\Gamma \vdash \Lambda a.e \Rightarrow \eta_2^\epsilon \rightsquigarrow \dot{t}[\Lambda a.t]}$$

**Tm-InfApp**
$$\frac{\Gamma \vdash^H h \Rightarrow \sigma \rightsquigarrow t \quad \Gamma \vdash^A \overline{arg} \Leftarrow \sigma \Rightarrow \sigma' \rightsquigarrow \overline{arg_F} \quad \Gamma \vdash \sigma' \xrightarrow{inst\,\delta} \eta^\epsilon \rightsquigarrow \dot{t}}{\Gamma \vdash h\,\overline{arg} \Rightarrow \eta^\epsilon \rightsquigarrow \dot{t}[t\,\overline{arg_F}]}$$

**Tm-InfLet**
$$\frac{\Gamma \vdash decl \Rightarrow \Gamma' \rightsquigarrow x : \sigma = t_1 \quad \Gamma' \vdash e \Rightarrow \eta^\epsilon \rightsquigarrow t_2}{\Gamma \vdash \text{let } decl \text{ in } e \Rightarrow \eta^\epsilon \rightsquigarrow (\lambda x : \sigma.t_2)\,t_1}$$

**Tm-InfTrue**
$$\overline{\Gamma \vdash \text{true} \Rightarrow \text{Bool} \rightsquigarrow \text{true}}$$

**Tm-InfFalse**
$$\overline{\Gamma \vdash \text{false} \Rightarrow \text{Bool} \rightsquigarrow \text{false}}$$

$\boxed{\Gamma \vdash e \Leftarrow \sigma \rightsquigarrow t}$ *(Term Type Scheme Checking)*

**Tm-CheckAbs**
$$\frac{\Gamma \vdash \sigma \xdashrightarrow{skol\,\mathcal{S}} \sigma_1 \rightarrow \sigma_2; \Gamma_1 \rightsquigarrow \dot{t} \quad \Gamma_1, x : \sigma_1 \vdash e \Leftarrow \sigma_2 \rightsquigarrow t_1}{\Gamma \vdash \lambda x.e \Leftarrow \sigma \rightsquigarrow \dot{t}[\lambda x : \sigma_1.t_1]}$$

**Tm-CheckTyAbs**
$$\frac{\sigma = \forall \{\overline{a}\}.\forall a.\sigma' \quad \Gamma, \overline{a}, a \vdash e \Leftarrow \sigma' \rightsquigarrow t}{\Gamma \vdash \Lambda a.e \Leftarrow \sigma \rightsquigarrow \Lambda \overline{a}.\Lambda a.t}$$

**Tm-CheckLet**
$$\frac{\Gamma \vdash decl \Rightarrow \Gamma' \rightsquigarrow x : \sigma_1 = t_1 \quad \Gamma' \vdash e \Leftarrow \sigma \rightsquigarrow t_2}{\Gamma \vdash \text{let } decl \text{ in } e \Leftarrow \sigma \rightsquigarrow (\lambda x : \sigma_1.t_2)\,t_1}$$

**Tm-CheckInf**
$$\frac{\Gamma \vdash \sigma \xrightarrow{skol\,\delta} \rho; \Gamma_1 \rightsquigarrow \dot{t}_1 \quad \Gamma_1 \vdash e \Rightarrow \eta^\epsilon \rightsquigarrow t \quad \Gamma_1 \vdash \eta^\epsilon \xrightarrow{inst\,\delta} \rho \rightsquigarrow \dot{t}_2 \quad e \neq \lambda, \Lambda, \text{let}}{\Gamma \vdash e \Leftarrow \sigma \rightsquigarrow \dot{t}_1[\dot{t}_2[t]]}$$

$\boxed{\Gamma \vdash^A \overline{arg} \Leftarrow \sigma \Rightarrow \sigma' \rightsquigarrow \overline{arg_F}}$ *(Argument Type Checking)*

**Arg-Empty**
$$\overline{\Gamma \vdash^A \cdot \Leftarrow \sigma \Rightarrow \sigma \rightsquigarrow \cdot}$$

**Arg-App**
$$\frac{\Gamma \vdash e \Leftarrow \sigma_1 \rightsquigarrow t \quad \Gamma \vdash^A \overline{arg} \Leftarrow \sigma_2 \Rightarrow \sigma' \rightsquigarrow \overline{arg_F}}{\Gamma \vdash^A e, \overline{arg} \Leftarrow \sigma_1 \rightarrow \sigma_2 \Rightarrow \sigma' \rightsquigarrow t, \overline{arg_F}}$$

**Arg-Inst**
$$\frac{\Gamma \vdash^A e, \overline{arg} \Leftarrow [\tau_1/a]\,\sigma_2 \Rightarrow \sigma_3 \rightsquigarrow \overline{arg_F}}{\Gamma \vdash^A e, \overline{arg} \Leftarrow \forall a.\sigma_2 \Rightarrow \sigma_3 \rightsquigarrow \overline{arg_F}}$$

**Arg-TyApp**
$$\frac{\Gamma \vdash^A \overline{arg} \Leftarrow [\sigma_1/a]\,\sigma_2 \Rightarrow \sigma_3 \rightsquigarrow \overline{arg_F}}{\Gamma \vdash^A @\sigma_1, \overline{arg} \Leftarrow \forall a.\sigma_2 \Rightarrow \sigma_3 \rightsquigarrow \tau_1, \overline{arg_F}}$$

**Arg-InfInst**
$$\frac{\Gamma \vdash^A \overline{arg} \Leftarrow [\tau_1/a]\,\sigma_2 \Rightarrow \sigma_3 \rightsquigarrow \overline{arg_F}}{\Gamma \vdash^A \overline{arg} \Leftarrow \forall \{a\}.\sigma_2 \Rightarrow \sigma_3 \rightsquigarrow \overline{arg_F}}$$

$$\boxed{\Gamma \vdash \sigma \xrightarrow{inst\ \delta} \rho \rightsquigarrow \dot{t}}$$ *(Type Instantiation)*

**InstT-SInst**
$$\Gamma \vdash \rho \xrightarrow{inst\ \mathcal{S}} \rho \rightsquigarrow \bullet$$

**InstT-SForall**
$$\frac{\Gamma \vdash [\tau/a]\sigma \xrightarrow{inst\ \mathcal{S}} \rho \rightsquigarrow \dot{t}}{\Gamma \vdash \forall a.\sigma \xrightarrow{inst\ \mathcal{S}} \rho \rightsquigarrow \lambda t.(\dot{t}[t\ \tau])}$$

**InstT-SInfForall**
$$\frac{\Gamma \vdash [\tau/a]\sigma \xrightarrow{inst\ \mathcal{S}} \rho \rightsquigarrow \dot{t}}{\Gamma \vdash \forall \{a\}.\sigma \xrightarrow{inst\ \mathcal{S}} \rho \rightsquigarrow \lambda t.(\dot{t}[t\ \tau])}$$

**InstT-Mono**
$$\Gamma \vdash \tau \xrightarrow{inst\ \mathcal{D}} \tau \rightsquigarrow \bullet$$

**InstT-Function**
$$\frac{\Gamma \vdash \sigma_2 \xrightarrow{inst\ \mathcal{D}} \rho_2 \rightsquigarrow \dot{t}}{\Gamma \vdash \sigma_1 \rightarrow \sigma_2 \xrightarrow{inst\ \mathcal{D}} \sigma_1 \rightarrow \rho_2 \rightsquigarrow \lambda t.\lambda x:\sigma_1.(\dot{t}[t\ x])}$$

**InstT-Forall**
$$\frac{\Gamma \vdash [\tau/a]\sigma \xrightarrow{inst\ \mathcal{D}} \rho \rightsquigarrow \dot{t}}{\Gamma \vdash \forall a.\sigma \xrightarrow{inst\ \mathcal{D}} \rho \rightsquigarrow \lambda t.(\dot{t}[t\ \tau])}$$

**InstT-InfForall**
$$\frac{\Gamma \vdash [\tau/a]\sigma \xrightarrow{inst\ \mathcal{D}} \rho \rightsquigarrow \dot{t}}{\Gamma \vdash \forall \{a\}.\sigma \xrightarrow{inst\ \mathcal{D}} \rho \rightsquigarrow \lambda t.(\dot{t}[t\ \tau])}$$

$$\boxed{\Gamma \vdash \sigma \xrightarrow{skol\ \delta} \rho; \Gamma' \rightsquigarrow \dot{t}}$$ *(Type Skolemisation)*

**SkolT-SInst**
$$\Gamma \vdash \rho \xrightarrow{skol\ \mathcal{S}} \rho; \Gamma \rightsquigarrow \bullet$$

**SkolT-SForall**
$$\frac{\Gamma, a \vdash \sigma \xrightarrow{skol\ \mathcal{S}} \rho; \Gamma_1 \rightsquigarrow \dot{t}}{\Gamma \vdash \forall a.\sigma \xrightarrow{skol\ \mathcal{S}} \rho; \Gamma_1 \rightsquigarrow \Lambda a.\dot{t}}$$

**SkolT-SInfForall**
$$\frac{\Gamma, a \vdash \sigma \xrightarrow{skol\ \mathcal{S}} \rho; \Gamma_1 \rightsquigarrow \dot{t}}{\Gamma \vdash \forall \{a\}.\sigma \xrightarrow{skol\ \mathcal{S}} \rho; \Gamma_1 \rightsquigarrow \Lambda a.\dot{t}}$$

**SkolT-Mono**
$$\Gamma \vdash \tau \xrightarrow{skol\ \mathcal{D}} \tau; \Gamma \rightsquigarrow \bullet$$

**SkolT-Function**
$$\frac{\Gamma \vdash \sigma_2 \xrightarrow{skol\ \mathcal{D}} \rho_2; \Gamma_1 \rightsquigarrow \dot{t}}{\Gamma \vdash \sigma_1 \rightarrow \sigma_2 \xrightarrow{skol\ \mathcal{D}} \sigma_1 \rightarrow \rho_2; \Gamma_1 \rightsquigarrow \lambda t.\lambda x:\sigma_1.(\dot{t}[t\ x])}$$

**SkolT-Forall**
$$\frac{\Gamma, a \vdash \sigma \xrightarrow{skol\ \mathcal{D}} \rho; \Gamma_1 \rightsquigarrow \dot{t}}{\Gamma \vdash \forall a.\sigma \xrightarrow{skol\ \mathcal{D}} \rho; \Gamma_1 \rightsquigarrow \Lambda a.\dot{t}}$$

**SkolT-InfForall**
$$\frac{\Gamma, a \vdash \sigma \xrightarrow{skol\ \mathcal{D}} \rho; \Gamma_1 \rightsquigarrow \dot{t}}{\Gamma \vdash \forall \{a\}.\sigma \xrightarrow{skol\ \mathcal{D}} \rho; \Gamma_1 \rightsquigarrow \Lambda a.\dot{t}}$$

$\boxed{\Gamma \vdash \mathit{decl} \Rightarrow \Gamma' \leadsto x : \sigma = t}$ *(Declaration Checking)*

Decl-NoAnnSingle
$$\frac{\begin{array}{c}\Gamma \vdash^P \overline{\pi} \Rightarrow \overline{\psi}; \Delta \leadsto \overline{\pi_F} : \overline{\psi_F} \\ \Gamma, \Delta \vdash e \Rightarrow \eta^\epsilon \leadsto t \\ \mathit{type}\,(\overline{\psi}; \eta^\epsilon \sim \sigma) \quad \overline{a} = f_v(\sigma) \setminus \mathit{dom}\,(\Gamma)\end{array}}{\Gamma \vdash x\,\overline{\pi} = e \Rightarrow \Gamma, x : \forall \overline{\{a\}}.\sigma \leadsto x : \forall \overline{\{a\}}.\sigma = \mathrm{case}\,\overline{\pi_F} : \overline{\psi_F} \to t}$$

Decl-NoAnnMulti
$$\frac{\begin{array}{c}i > 1 \quad \overline{\Gamma \vdash^P \overline{\pi_i} \Rightarrow \overline{\psi}; \Delta_i \leadsto \overline{\pi_{F_i}} : \overline{\psi_F}}^i \\ \overline{\Gamma, \Delta_i \vdash e_i \Rightarrow \eta_i^\epsilon \leadsto t_i}^i \\ \overline{\Gamma, \Delta_i \vdash \eta_i^\epsilon \xrightarrow{\mathit{inst}\,\delta} \rho \leadsto \dot{t}_i}^i \quad \mathit{type}\,(\overline{\psi}; \rho \sim \sigma) \\ \overline{a} = f_v(\sigma) \setminus \mathit{dom}\,(\Gamma) \quad \sigma' = \forall \overline{\{a\}}.\sigma\end{array}}{\Gamma \vdash \overline{x\,\overline{\pi_i} = e_i}^i \Rightarrow \Gamma, x : \sigma' \leadsto x : \sigma' = \overline{\mathrm{case}\,\overline{\pi_{F_i}} : \overline{\psi_F} \to \dot{t}_i[t_i]}^i}$$

Decl-Ann
$$\frac{\overline{\Gamma \vdash^P \overline{\pi_i} \Leftarrow \sigma \Rightarrow \sigma_i'; \Delta_i \leadsto \overline{\pi_{F_i}} : \overline{\psi_F}}^i \quad \overline{\Gamma, \Delta_i \vdash e_i \Leftarrow \sigma_i' \leadsto t_i}^i}{\Gamma \vdash x : \sigma; \overline{x\,\overline{\pi_i} = e_i}^i \Rightarrow \Gamma, x : \sigma \leadsto x : \sigma = \overline{\mathrm{case}\,\overline{\pi_{F_i}} : \overline{\psi_F} \to t_i}^i}$$

$\boxed{\Gamma \vdash^P \overline{\pi} \Rightarrow \overline{\psi}; \Delta \leadsto \overline{\pi_F} : \overline{\psi_F}}$ *(Pattern Synthesis)*

Pat-InfEmpty
$$\Gamma \vdash^P \cdot \Rightarrow \cdot; \cdot \leadsto \cdot : \cdot$$

Pat-InfVar
$$\frac{\Gamma, x : \tau_1 \vdash^P \overline{\pi} \Rightarrow \overline{\psi}; \Delta \leadsto \overline{\pi_F} : \overline{\psi_F}}{\Gamma \vdash^P x, \overline{\pi} \Rightarrow \tau_1, \overline{\psi}; x : \tau_1, \Delta \leadsto x : \tau_1, \overline{\pi_F} : \tau_1, \overline{\psi_F}}$$

Pat-InfCon
$$\frac{\begin{array}{c}K : \overline{a}_0; \overline{\sigma}_0; T \in \Gamma \\ \Gamma \vdash^P \overline{\pi} \Leftarrow [\overline{\sigma}_1, \overline{\tau}_0/\overline{a}_0]\,(\overline{\sigma}_0 \to T\,\overline{a}_0) \Rightarrow T\,\overline{\tau}; \Delta_1 \leadsto \overline{\pi_{F_1}} : \overline{\psi_{F_1}} \\ \Gamma, \Delta_1 \vdash^P \overline{\pi}' \Rightarrow \overline{\psi}; \Delta_2 \leadsto \overline{\pi_{F_2}} : \overline{\psi_{F_2}}\end{array}}{\Gamma \vdash^P (K\,@\overline{\sigma}_1\,\overline{\pi}), \overline{\pi}' \Rightarrow T\,\overline{\tau}, \overline{\psi}; \Delta_1, \Delta_2 \leadsto (K\,\overline{\pi_{F_1}}), \overline{\pi_{F_2}} : T\,\overline{\tau}, \overline{\psi_{F_2}}}$$

Pat-InfTyVar
$$\frac{\Gamma, a \vdash^P \overline{\pi} \Rightarrow \overline{\psi}; \Delta \leadsto \overline{\pi_F} : \overline{\psi_F}}{\Gamma \vdash^P @a, \overline{\pi} \Rightarrow @a, \overline{\psi}; a, \Delta \leadsto @a, \overline{\pi_F} : @a, \overline{\psi_F}}$$

$\boxed{\Gamma \vdash^P \overline{\pi} \Leftarrow \sigma \Rightarrow \sigma'; \Delta \leadsto \overline{\pi_F} : \overline{\psi_F}}$ *(Pattern Checking)*

Pat-CheckEmpty
$$\Gamma \vdash^P \cdot \Leftarrow \sigma \Rightarrow \sigma; \cdot \leadsto \cdot : \cdot$$

Pat-CheckVar
$$\frac{\Gamma, x : \sigma_1 \vdash^P \overline{\pi} \Leftarrow \sigma_2 \Rightarrow \sigma'; \Delta \leadsto \overline{\pi_F} : \overline{\psi_F}}{\Gamma \vdash^P x, \overline{\pi} \Leftarrow \sigma_1 \to \sigma_2 \Rightarrow \sigma'; x : \sigma_1, \Delta \leadsto x : \sigma_1, \overline{\pi_F} : \sigma_1, \overline{\psi_F}}$$

Pat-CheckCon
$$\frac{\begin{array}{c}K : \overline{a}_0; \overline{\sigma}_0; T \in \Gamma \quad \Gamma \vdash \sigma_1 \xrightarrow{\mathit{inst}\,\delta} \rho_1 \leadsto \dot{t} \\ \Gamma \vdash^P \overline{\pi} \Leftarrow [\overline{\sigma}_1, \overline{\tau}_0/\overline{a}_0]\,(\overline{\sigma}_0 \to T\,\overline{a}_0) \Rightarrow \rho_1; \Delta_1 \leadsto \overline{\pi_{F_1}} : \overline{\psi_{F_1}} \\ \Gamma, \Delta_1 \vdash^P \overline{\pi}' \Leftarrow \sigma_2 \Rightarrow \sigma_2'; \Delta_2 \leadsto \overline{\pi_{F_2}} : \overline{\psi_{F_2}}\end{array}}{\Gamma \vdash^P (K\,@\overline{\sigma}_1\,\overline{\pi}), \overline{\pi}' \Leftarrow \sigma_1 \to \sigma_2 \Rightarrow \sigma_2'; \Delta_1, \Delta_2 \leadsto (K\,\overline{\pi_{F_1}}), \overline{\pi_{F_2}} : \sigma_1, \overline{\psi_{F_2}}}$$

Pat-CheckForall
$$\frac{\Gamma, a \vdash^P \overline{\pi} \Leftarrow \sigma \Rightarrow \sigma'; \Delta \rightsquigarrow \overline{\pi_F} : \overline{\psi_F} \quad \overline{\pi} \neq \cdot \text{ and } \overline{\pi} \neq @\sigma, \overline{\pi}'}{\Gamma \vdash^P \overline{\pi} \Leftarrow \forall a.\sigma \Rightarrow \sigma'; a, \Delta \rightsquigarrow \overline{\pi_F} : @a, \overline{\psi_F}}$$

Pat-CheckTyVar
$$\frac{\Gamma, a \vdash^P \overline{\pi} \Leftarrow [a/b]\,\sigma_1 \Rightarrow \sigma_2; \Delta \rightsquigarrow \overline{\pi_F} : \overline{\psi_F}}{\Gamma \vdash^P @a, \overline{\pi} \Leftarrow \forall b.\sigma_1 \Rightarrow \sigma_2; a, \Delta \rightsquigarrow \overline{\pi_F} : @a, \overline{\psi_F}}$$

Pat-CheckInfForall
$$\frac{\Gamma, a \vdash^P \overline{\pi} \Leftarrow \sigma \Rightarrow \sigma'; \Delta \rightsquigarrow \overline{\pi_F} : \overline{\psi_F} \quad \overline{\pi} \neq \cdot}{\Gamma \vdash^P \overline{\pi} \Leftarrow \forall \{a\}.\sigma \Rightarrow \sigma'; a, \Delta \rightsquigarrow \overline{\pi_F} : @\overline{a}, \overline{\psi_F}}$$

## E  Proofs

This section provides the proofs for the properties discussed in Section 5.

### E.1  Let-Inlining and Extraction

**Property 1** (Let Inlining is Type Preserving).
- *If* $\Gamma \vdash \text{let } x = e_1 \text{ in } e_2 \Rightarrow \eta^\epsilon$ *then* $\Gamma \vdash [e_1/x]\,e_2 \Rightarrow \eta^\epsilon$
- *If* $\Gamma \vdash \text{let } x = e_1 \text{ in } e_2 \Leftarrow \sigma$ *then* $\Gamma \vdash [e_1/x]\,e_2 \Leftarrow \sigma$

Before proving Property 1, we first introduce a number of helper lemmas:

**Lemma E.1** (Expression Inlining is Type Preserving (Synthesis)).
*If* $\Gamma_1 \vdash e_1 \Rightarrow \eta_1^\epsilon$ *and* $\Gamma_1, x : \forall \overline{\{a\}}.\eta_1^\epsilon, \Gamma_2 \vdash e_2 \Rightarrow \eta_2^\epsilon$ *where* $\overline{a} = f_v(\eta_1^\epsilon) \setminus dom(\Gamma_1)$
*then* $\Gamma_1, \Gamma_2 \vdash [e_1/x]\,e_2 \Rightarrow \eta_2^\epsilon$

**Lemma E.2** (Expression Inlining is Type Preserving (Checking)).
*If* $\Gamma_1 \vdash e_1 \Rightarrow \eta_1^\epsilon$ *and* $\Gamma_1, x : \forall \overline{\{a\}}.\eta_1^\epsilon, \Gamma_2 \vdash e_2 \Leftarrow \sigma_2$ *where* $\overline{a} = f_v(\eta_1^\epsilon) \setminus dom(\Gamma_1)$
*then* $\Gamma_1, \Gamma_2 \vdash [e_1/x]\,e_2 \Leftarrow \sigma_2$

**Lemma E.3** (Head Inlining is Type Preserving).
*If* $\Gamma_1 \vdash e_1 \Rightarrow \eta_1^\epsilon$ *and* $\Gamma_1, x : \forall \overline{\{a\}}.\eta_1^\epsilon, \Gamma_2 \vdash^H h \Rightarrow \sigma_2$ *where* $\overline{a} = f_v(\eta_1^\epsilon) \setminus dom(\Gamma_1)$
*then* $\Gamma_1, \Gamma_2 \vdash^H [e_1/x]\,h \Rightarrow \sigma_2$

**Lemma E.4** (Argument Inlining is Type Preserving).
*If* $\Gamma_1 \vdash e_1 \Rightarrow \eta_1^\epsilon$ *and* $\Gamma_1, x : \forall \overline{\{a\}}.\eta_1^\epsilon, \Gamma_2 \vdash^A \overline{arg} \Leftarrow \sigma_1 \Rightarrow \sigma_2$ *where* $\overline{a} = f_v(\eta_1^\epsilon) \setminus dom(\Gamma_1)$
*then* $\Gamma_1, \Gamma_2 \vdash^A [e_1/x]\,\overline{arg} \Leftarrow \sigma_1 \Rightarrow \sigma_2$

**Lemma E.5** (Declaration Inlining is Type Preserving).
*If* $\Gamma_1 \vdash e_1 \Rightarrow \eta_1^\epsilon$ *and* $\Gamma_1, x : \forall \overline{\{a\}}.\eta_1^\epsilon, \Gamma_2 \vdash decl \Rightarrow \Gamma_3$ *where* $\overline{a} = f_v(\eta_1^\epsilon) \setminus dom(\Gamma_1)$
*then* $\Gamma_1, \Gamma_2 \vdash [e_1/x]\,decl \Rightarrow \Gamma_3$

Figure 6 shows the dependencies between the different relations, and by extension the different helper lemmas. An arrow from $A$ to $B$ denotes that $B$ depends on $A$. Note that these 5 lemmas need to be proven through mutual induction. The proof proceeds by structural induction on the second typing derivation. While the number of cases gets quite large, each case is entirely trivial.

Using these additional lemmas, we then continue proving Property 1. By case analysis on the premise (rule Tm-InfLet or rule Tm-CheckLet, followed by rule Decl-NoAnnSingle), we learn that $\Gamma \vdash x = e_1 \Rightarrow \Gamma, x : \forall \overline{\{a\}}.\eta_1^\epsilon$, $\Gamma \vdash e_1 \Rightarrow \eta_1^\epsilon$, and either $\Gamma, x : \forall \overline{\{a\}}.\eta_1^\epsilon \vdash e_2 \Rightarrow \eta^\epsilon$ or $\Gamma, x : \forall \overline{\{a\}}.\eta_1^\epsilon \vdash e_2 \Leftarrow \sigma$. Both parts of the goal now follow trivially from Lemma E.1 and E.2 respectively. □

**Property 2** (Let Extraction is Type Preserving).
- *If* $\Gamma \vdash [e_1/x]\,e_2 \Rightarrow \eta_2^\epsilon$ *and* $\Gamma \vdash e_1 \Rightarrow \eta_1^\epsilon$ *then* $\Gamma \vdash \text{let } x = e_1 \text{ in } e_2 \Rightarrow \eta_2^\epsilon$
- *If* $\Gamma \vdash [e_1/x]\,e_2 \Leftarrow \sigma_2$ *and* $\Gamma \vdash e_1 \Rightarrow \eta_1^\epsilon$ *then* $\Gamma \vdash \text{let } x = e_1 \text{ in } e_2 \Leftarrow \sigma_2$

Similarly to before, we start by introducing a number of helper lemmas:

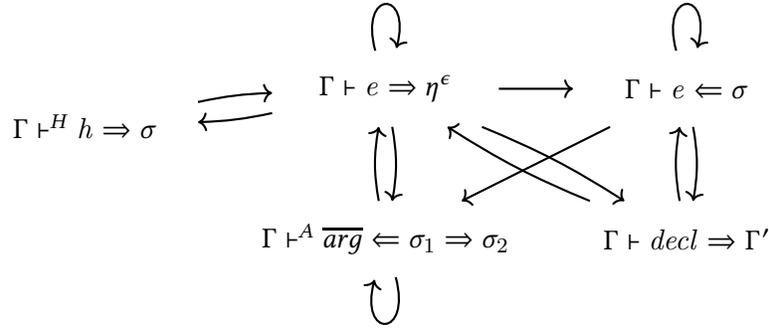

**Figure 6.** Relation dependencies

**Lemma E.6** (Expression Extraction is Type Preserving (Synthesis)).
If $\Gamma \vdash e_1 \Rightarrow \eta_1^\epsilon$ and $\Gamma \vdash [e_1/x]\, e_2 \Rightarrow \eta_2^\epsilon$
then $\Gamma, x : \forall \overline{\{a\}}.\eta_1^\epsilon \vdash e_2 \Rightarrow \eta_2^\epsilon$ where $\overline{a} = f_v(\eta_1^\epsilon) \setminus dom\,(\Gamma)$

**Lemma E.7** (Expression Extraction is Type Preserving (Checking)).
If $\Gamma \vdash e_1 \Rightarrow \eta_1^\epsilon$ and $\Gamma \vdash [e_1/x]\, e_2 \Leftarrow \sigma_2$
then $\Gamma, x : \forall \overline{\{a\}}.\eta_1^\epsilon \vdash e_2 \Leftarrow \sigma_2$ where $\overline{a} = f_v(\eta_1^\epsilon) \setminus dom\,(\Gamma)$

**Lemma E.8** (Head Extraction is Type Preserving).
If $\Gamma \vdash e_1 \Rightarrow \eta_1^\epsilon$ and $\Gamma \vdash^H [e_1/x]\, h \Rightarrow \sigma_2$
then $\Gamma, x : \forall \overline{\{a\}}.\eta_1^\epsilon \vdash^H h \Rightarrow \sigma_2$ where $\overline{a} = f_v(\eta_1^\epsilon) \setminus dom\,(\Gamma)$

**Lemma E.9** (Argument Extraction is Type Preserving).
If $\Gamma \vdash e_1 \Rightarrow \eta_1^\epsilon$ and $\Gamma \vdash^A [e_1/x]\, \overline{arg} \Leftarrow \sigma_1 \Rightarrow \sigma_2$
then $\Gamma, x : \forall \overline{\{a\}}.\eta_1^\epsilon \vdash^A \overline{arg} \Leftarrow \sigma_1 \Rightarrow \sigma_2$ where $\overline{a} = f_v(\eta_1^\epsilon) \setminus dom\,(\Gamma)$

**Lemma E.10** (Declaration Extraction is Type Preserving).
If $\Gamma \vdash e_1 \Rightarrow \eta_1^\epsilon$ and $\Gamma \vdash [e_1/x]\, decl \Rightarrow \Gamma, \Gamma'$
then $\Gamma, x : \forall \overline{\{a\}}.\eta_1^\epsilon \vdash decl \Rightarrow \Gamma, x : \forall \overline{\{a\}}.\eta_1^\epsilon, \Gamma'$ where $\overline{a} = f_v(\eta_1^\epsilon) \setminus dom\,(\Gamma)$

In addition to these helper lemmas, we also introduce two typing context lemmas:

**Lemma E.11** (Environment Variable Shifting is Type Preserving).
- If $\Gamma_1, x_1 : \sigma_1, x_2 : \sigma_2, \Gamma_2 \vdash e \Rightarrow \eta^\epsilon$ then $\Gamma_1, x_2 : \sigma_2, x_1 : \sigma_1, \Gamma_2 \vdash e \Rightarrow \eta^\epsilon$
- If $\Gamma_1, x_1 : \sigma_1, x_2 : \sigma_2, \Gamma_2 \vdash e \Leftarrow \sigma$ then $\Gamma_1, x_2 : \sigma_2, x_1 : \sigma_1, \Gamma_2 \vdash e \Leftarrow \sigma$

**Lemma E.12** (Environment Type Variable Shifting is Type Preserving).
- If $\Gamma_1, a, x : \sigma, \Gamma_2 \vdash e \Rightarrow \eta^\epsilon$ and $\cdot = f_v(\sigma) \setminus dom\,(\Gamma_1)$ then $\Gamma_1, x : \sigma, a, \Gamma_2 \vdash e \Rightarrow \eta^\epsilon$
- If $\Gamma_1, a, x : \sigma, \Gamma_2 \vdash e \Leftarrow \sigma$ and $\cdot = f_v(\sigma) \setminus dom\,(\Gamma_1)$ then $\Gamma_1, x : \sigma, a, \Gamma_2 \vdash e \Leftarrow \sigma$
- If $\Gamma_1, x : \sigma, a, \Gamma_2 \vdash e \Rightarrow \eta^\epsilon$ then $\Gamma_1, a, x : \sigma, \Gamma_2 \vdash e \Rightarrow \eta^\epsilon$
- If $\Gamma_1, x : \sigma, a, \Gamma_2 \vdash e \Leftarrow \sigma$ then $\Gamma_1, a, x : \sigma, \Gamma_2 \vdash e \Leftarrow \sigma$

Lemmas E.11 and E.12 are folklore, and can be proven through straightforward induction.

Now we can go about proving Lemmas E.6 till E.10. Similarly to the Property 1 helper lemmas, they have to be proven using mutual induction. Most cases are quite straightforward, and we will focus only on Lemma E.8. We start by performing case analysis on $h$:

**Case** $h = y$ **where** $y = x$

By evaluating the substitution, we know from the premise that $\Gamma \vdash e_1 \Rightarrow \eta_1^\epsilon$ and $\Gamma \vdash^H e_1 \Rightarrow \sigma_2$, while the goal remains $\Gamma, x : \forall \overline{\{a\}}.\eta_1^\epsilon \vdash^H x \Rightarrow \sigma_2$. It is clear from rule H-Var that in order for the goal to hold, $\sigma_2 = \forall \overline{\{a\}}.\eta_1^\epsilon$. We proceed by case analysis on the second derivation:

**case rule** H-Var $e_1 = x'$ : The rule premise tells us that $x' : \sigma_2 \in \Gamma$. The goal follows directly under lazy instantiation. However, under eager instantiation, rule Tm-InfApp instantiates the type $\Gamma \vdash \sigma_2 \xrightarrow{inst\ \delta} \eta_1^\epsilon$ making the goal invalid.

**case rule** H-Con $e_1 = K$, **rule** H-Ann $e_1 = e_3 : \sigma_3$, **rule** H-Inf $e_1 = e_1$, **rule** H-Undef $e_1 = undefined$, **or rule** H-Seq $e_1 = seq$ :
Similarly to the previous case, the goal is only valid under eager instantiation.

**Case** $h = y$ **where** $y \neq x$
This case is trivial, as the substitution $[e_1/x]$ does not alter $h$. The result thus follows from weakening.

**Case** $h = K$, $h = undefined$, **or** $h = seq$
Similarly to the previous case, as the substitution does not alter $h$, the result thus follows from weakening.

**Case** $h = e : \sigma$
The result follows by applying Lemma E.7.

**Case** $h = e$
The result follows by applying Lemma E.6. □

Using these lemmas, both Property 2 goals follow straightforwardly using rule Decl-NoAnnSingle, in combination with rule Tm-InfLet and Lemma E.6 or rule Tm-CheckLet and Lemma E.7, respectively. □

### E.2 Contextual Equivalence

As we've now arrived at properties involving the runtime semantics of the language, we first need to formalise our definition of contextual equivalence, and introduce a number of useful lemmas.

**Definition 2** (Contextual Equavalence).

$$\begin{aligned}
t_1 \simeq t_2 \equiv\ & \Gamma \vdash t_1 : \sigma_1 \quad \wedge \quad \Gamma \vdash \sigma_1 \xrightarrow{inst\ \delta} \rho_3 \rightsquigarrow \dot{t}_1 \\
& \wedge\ \Gamma \vdash t_2 : \sigma_2 \quad \wedge \quad \Gamma \vdash \sigma_2 \xrightarrow{inst\ \delta} \rho_3 \rightsquigarrow \dot{t}_2 \\
& \wedge\ \forall M : \Gamma; \rho_3 \mapsto \cdot; Bool, \\
& \exists v : M[\dot{t}_1[t_1]] \hookrightarrow^{\Downarrow} v \quad \wedge \quad M[\dot{t}_2[t_2]] \hookrightarrow^{\Downarrow} v
\end{aligned}$$

This definition for contextual equivalence is modified from Harper [2016, Chapter 46]. Two core expressions are thus contextually equivalent, if a common type exists to which both their types instantiate, and if no (closed) context can distinguish between them. This can either mean that both applied expressions evaluate to the same value $v$ or both diverge. Note that while we require the context to map to a closed, Boolean expression, other base types, like $Int$, would have been valid alternatives as well.

We first introduce reflexivity, commutativity and transitivity lemmas:

**Lemma E.13** (Contextual Equivalence Reflexivity).
*If* $\Gamma \vdash t : \sigma$ *then* $t \simeq t$

The proof follows directly from the definition of contextual equivalence, along with the determinism of System F evaluation.

**Lemma E.14** (Contextual Equivalence Commutativity).
*If* $t_1 \simeq t_2$ *then* $t_2 \simeq t_1$

Trivial proof by unfolding the definition of contextual equivalence.

**Lemma E.15** (Contextual Equivalence Transitivity).
*If* $t_1 \simeq t_2$ *and* $t_2 \simeq t_3$ *then* $t_1 \simeq t_3$

Trivial proof by unfolding the definition of contextual equivalence.

Furthermore, we also introduce a number of compatibility lemmas for the contextual equivalence relation, along with two helper lemmas:

**Lemma E.16** (Compatibility Term Abstraction).
*If $t_1 \simeq t_2$ then $\lambda x : \sigma.t_1 \simeq \lambda x : \sigma.t_2$*

**Lemma E.17** (Compatibility Term Application).
*If $t_1 \simeq t_2$ and $t_1' \simeq t_2'$ then $t_1\ t_1' \simeq t_2\ t_2'$*

**Lemma E.18** (Compatibility Type Abstraction).
*If $t_1 \simeq t_2$ then $\Lambda a.t_1 \simeq \Lambda a.t_2$*

**Lemma E.19** (Compatibility Type Application).
*If $t_1 \simeq t_2$ then $t_1\ \sigma \simeq t_2\ \sigma$*

**Lemma E.20** (Compatibility Case Abstraction).
*If $\forall\ i : t_{1\ i} \simeq t_{2\ i}$ then $\mathsf{case}\ \overline{\pi_{F\ i} : \overline{\psi_F} \to t_{1\ i}}^{\ i} \simeq \mathsf{case}\ \overline{\pi_{F\ i} : \overline{\psi_F} \to t_{2\ i}}^{\ i}$*

**Lemma E.21** (Compatibility Expression Wrapper).
*If $t_1 \simeq t_2$ then $\dot{t}[t_1] \simeq \dot{t}[t_2]$*

**Lemma E.22** (Compatibility Helper Forwards).
*If $M[t_1] \hookrightarrow^{\Downarrow} v$ and $t_1 \hookrightarrow t_2$ then $M[t_2] \hookrightarrow^{\Downarrow} v$*

**Lemma E.23** (Compatibility Helper Backwards).
*If $M[t_2] \hookrightarrow^{\Downarrow} v$ and $t_1 \hookrightarrow t_2$ then $M[t_1] \hookrightarrow^{\Downarrow} v$*

The helper lemmas are proven by straightforward induction on the evaluation step derivation. We will prove Lemma E.18 as an example, as it is non-trivial. The other compatibility lemmas are proven similarly.

We start by unfolding the definition of contextual equivalence in both the premise: $\Gamma \vdash t_1 : \sigma_1, \Gamma \vdash \sigma_1 \xrightarrow{inst\ \delta} \rho_3 \rightsquigarrow \dot{t}_1, \Gamma \vdash t_2 : \sigma_2, \Gamma \vdash \sigma_2 \xrightarrow{inst\ \delta} \rho_3 \rightsquigarrow \dot{t}_2, \forall M : \Gamma; \rho_3 \mapsto \cdot; Bool, \exists v : M[\dot{t}_1[t_1]] \hookrightarrow^{\Downarrow} v$ and $M[\dot{t}_2[t_2]] \hookrightarrow^{\Downarrow} v$. Unfolding the definition reduces the goal to be proven to $\Gamma' \vdash \Lambda a.t_1 : \sigma_1', \Gamma' \vdash \sigma_1' \xrightarrow{inst\ \delta} \rho_3' \rightsquigarrow \dot{t}_1', \Gamma' \vdash \Lambda a.t_2 : \sigma_2'$, $\Gamma' \vdash \sigma_2' \xrightarrow{inst\ \delta} \rho_3' \rightsquigarrow \dot{t}_2', \forall M' : \Gamma'; \rho_3' \mapsto \cdot; Bool, \exists v' : M'[\dot{t}_1'[\Lambda a.t_1]] \hookrightarrow^{\Downarrow} v'$ and $M'[\dot{t}_2'[\Lambda a.t_2]] \hookrightarrow^{\Downarrow} v'$.

The typing judgement goals follow directly from rule FTm-TyAbs, where we take $\sigma_1' = \forall\ a.\sigma_1, \sigma_2' = \forall a.\sigma_2$ and $\Gamma' = [\tau/a]\ \Gamma$ for some $\tau$.

As we know $\Gamma \vdash \sigma_1 \xrightarrow{inst\ \delta} \rho_3 \rightsquigarrow \dot{t}_1$, it is easy to see that $[\tau/a]\ \Gamma \vdash [\tau/a]\ \sigma_1 \xrightarrow{inst\ \delta} [\tau/a]\ \rho_3 \rightsquigarrow [\tau/a]\ \dot{t}_1$, and similarly for $[\tau/a]\ \sigma_2$. Using this, the instantiation goals follow from rule InstT-SForall and rule InstT-Forall with $\rho_3' = [\tau/a]\ \rho_3, \dot{t}_1' = \lambda t.([\tau/a]\ \dot{t}_1[t\ \tau])$ and $\dot{t}_2' = \lambda t.([\tau/a]\ \dot{t}_2[t\ \tau])$.

Finally, by inlining the definitions, the first halve of the third goal becomes $M'[(\lambda t.([\tau/a]\ \dot{t}_1[t\ \tau]))[\Lambda a.t_1]] \hookrightarrow^{\Downarrow} v'$. This reduces to $M'[[\tau/a]\ \dot{t}_1[(\Lambda a.t_1)\ \tau]] \hookrightarrow^{\Downarrow} v'$. By lemma E.22 (note that we can consider the combination of a context and an expression wrapper as a new context): $M'[[\tau/a]\ \dot{t}_1[[\tau/a]\ t_1]] \hookrightarrow^{\Downarrow} v'$. We can now bring the substitutions to the front, and reduce the goal (by Lemma E.23) $M''[\dot{t}_1[t_1]] \hookrightarrow^{\Downarrow} v'$ where we define $M'' = \lambda t.M'[(\Lambda a.t)\ \tau]$ (note that we use $\lambda t$ as meta-notation here, to simplify our definition of $M''$). We perform the same derivation for the second halve of the goal: $M''[\dot{t}_2[t_2]] \hookrightarrow^{\Downarrow} v'$. As $M'' : \Gamma; \rho_3 \mapsto \cdot; Bool$, the goal follows directly from the unfolded premise, where $v' = v$. □

We introduce an additional lemma stating that instantiating the type of expressions does not alter their behaviour:

**Lemma E.24** (Type Instantiation is Runtime Semantics Preserving).
*If $\Gamma \vdash t : \sigma$ and $\Gamma \vdash \sigma \xrightarrow{inst\ \delta} \rho \rightsquigarrow \dot{t}$ then $t \simeq \dot{t}[t]$*

The proof proceeds by induction on the instantiation relation:
**Case rule** InstT-SInst $\dot{t} = \bullet$ :
Trivial case, as $\dot{t}[t] = t$, the goal follows directly from Lemma E.13.
**Case rule** InstT-SForall $\dot{t} = \lambda t_1.(\dot{t}'[t_1\ \tau])$ :
We know from the first premise, along with rule FTm-TyApp that $\Gamma \vdash t\ \tau : [\tau/a]\ \sigma'$ where $\sigma = \forall\ a.\sigma'$. By applying the induction hypothesis we get $t\ \tau \simeq \dot{t}'[t\ \tau]$. The goal to be proven is $t \simeq (\lambda t_1.(\dot{t}'[t_1\ \tau]))[t]$, which reduces to $t \simeq \dot{t}'[t\,\tau]$. By unfolding the definition of contextual equivalence in both the goal and the induction hypothesis result (using Lemma E.15), the remaining goals are:

- $\Gamma \vdash t : \sigma_1$ : follows directly from the first premise.
- $\Gamma \vdash \forall a.\sigma' \xrightarrow{inst\ S} \rho' \rightsquigarrow \dot{t}_1$ and $\Gamma \vdash \rho' \xrightarrow{inst\ S} \rho \rightsquigarrow \dot{t}_2$ : follows directly from the premise if we take $\rho' = \rho$, $\dot{t}_1 = \dot{t}$ and $\dot{t}_2 = \bullet$.
- $M[\dot{t}_1[t]] \hookrightarrow^{\Downarrow} v$ and $M[\dot{t}[t]] \hookrightarrow^{\Downarrow} v$ : trivial as both sides are identical and evaluation is deterministic.

**Case rule** INSTT-SINFFORALL $\dot{t} = \lambda t_1.(\dot{t}'[t_1\,\tau])$ :
The proof follows analogously to the previous case. We have thus proven Lemma E.24 under shallow instantiation.

**Case rule** INSTT-MONO $\dot{t} = \bullet$ :
Trivial case, as $\dot{t}[t] = t$, the goal follows directly from Lemma E.13.

**Case rule** INSTT-FUNCTION $\dot{t} = \lambda t_1.\lambda x : \sigma_1.(\dot{t}'[t_1\,x])$ :
It is clear that the goal does not hold in this case. Under deep instantiation, full eta expansion is performed, which alters the evaluation behaviour. Consider for example $undefined$ and its expansion $\lambda x : \sigma.undefined\ x$. □

Finally, we introduce a lemma stating that evaluation preserves contextual equivalence. However, in order to prove it, we first need to introduce the common preservation lemma:

**Lemma E.25** (Preservation).
*If $\Gamma \vdash t : \sigma$ and $t \hookrightarrow t'$ then $\Gamma \vdash t' : \sigma$*

The preservation proof for System F is folklore, and proceeds by straightforward induction on the evaluation relation.

**Lemma E.26** (Evaluation is Contextual Equivalence Preserving).
*If $t_1 \simeq t_2$ and $t_2 \hookrightarrow t_2'$ then $t_1 \simeq t_2'$*

The proof follows by Lemma E.25 (to cover type preservation) and Lemma E.22 (to cover the evaluation aspect).

### E.3 Let-Inlining and Extraction, Continued

**Property 3** (Let Inlining is Runtime Semantics Preserving).
- *If $\Gamma \vdash \text{let } x = e_1 \text{ in } e_2 \Rightarrow \eta^\epsilon \rightsquigarrow t_1$ and $\Gamma \vdash [e_1/x]\,e_2 \Rightarrow \eta^\epsilon \rightsquigarrow t_2$ then $t_1 \simeq t_2$*
- *If $\Gamma \vdash \text{let } x = e_1 \text{ in } e_2 \Leftarrow \sigma \rightsquigarrow t_1$ and $\Gamma \vdash [e_1/x]\,e_2 \Leftarrow \sigma \rightsquigarrow t_2$ then $t_1 \simeq t_2$*

We first need typing preservation lemmas before we can prove Property 3.

**Lemma E.27** (Expression Typing Preservation (Synthesis)).
*If $\Gamma \vdash e \Rightarrow \eta \rightsquigarrow t$ then $\Gamma \vdash t : \eta$*

**Lemma E.28** (Expression Typing Preservation (Checking)).
*If $\Gamma \vdash e \Leftarrow \sigma \rightsquigarrow t$ then $\Gamma \vdash t : \sigma$*

**Lemma E.29** (Head Typing Preservation).
*If $\Gamma \vdash^H h \Rightarrow \sigma \rightsquigarrow t$ then $\Gamma \vdash t : \sigma$*

**Lemma E.30** (Argument Typing Preservation).
*If $\Gamma \vdash^A \overline{arg} \Leftarrow \sigma \Rightarrow \sigma' \rightsquigarrow \overline{arg_F}$ then $\forall t_i \in \overline{arg_F} : \Gamma \vdash t_i : \sigma_i$*

**Lemma E.31** (Declaration Typing Preservation).
*If $\Gamma \vdash decl \Rightarrow \Gamma' \rightsquigarrow x : \sigma = t$ then $\Gamma \vdash t : \sigma$*

Similarly to the helper lemmas for Property 1, these lemmas need to be proven using mutual induction. The proofs follow through straightforward induction on the typing derivation.

We continue by introducing another set of helper lemmas:

**Lemma E.32** (Expression Inlining is Runtime Semantics Preserving (Synthesis)).
*If $\Gamma_1, x : \forall \overline{\{a\}}.\eta_1^\epsilon, \Gamma_2 \vdash e_2 \Rightarrow \eta_2^\epsilon \rightsquigarrow t_2$, $\Gamma_1 \vdash e_1 \Rightarrow \eta_1^\epsilon \rightsquigarrow t_1$ and $\Gamma_1, \Gamma_2 \vdash [e_1/x]\,e_2 \Rightarrow \eta_2^\epsilon \rightsquigarrow t_3$ where $\overline{a} = f_v(\eta_1^\epsilon) \setminus dom\,(\Gamma_1)$ then $t_3 \simeq (\lambda x : \forall \overline{a}.\eta_1^\epsilon.t_2)\,t_1$*

**Lemma E.33** (Expression Inlining is Runtime Semantics Preserving (Checking)).
*If* $\Gamma_1, x : \forall \overline{\{a\}}.\eta_1^\epsilon, \Gamma_2 \vdash e_2 \Leftarrow \sigma_2 \leadsto t_2$, $\Gamma_1 \vdash e_1 \Rightarrow \eta_1^\epsilon \leadsto t_1$ *and* $\Gamma_1, \Gamma_2 \vdash [e_1/x] e_2 \Leftarrow \sigma_2 \leadsto t_3$ *where* $\overline{a} = f_v(\eta_1^\epsilon) \setminus dom(\Gamma_1)$ *then* $t_3 \simeq (\lambda x : \forall \overline{a}.\eta_1^\epsilon.t_2) t_1$

**Lemma E.34** (Head Inlining is Runtime Semantics Preserving).
*If* $\Gamma_1, x : \forall \overline{\{a\}}.\eta_1^\epsilon, \Gamma_2 \vdash^H h \Rightarrow \sigma \leadsto t_2$, $\Gamma_1 \vdash e_1 \Rightarrow \eta_1^\epsilon \leadsto t_1$ *and* $\Gamma_1, \Gamma_2 \vdash^H [e_1/x] h \Rightarrow \sigma \leadsto t_3$ *where* $\overline{a} = f_v(\eta_1^\epsilon) \setminus dom(\Gamma_1)$ *then* $t_3 \simeq (\lambda x : \forall \overline{a}.\eta_1^\epsilon.t_2) t_1$

**Lemma E.35** (Argument Inlining is Runtime Semantics Preserving).
*If* $\Gamma_1, x : \forall \overline{\{a\}}.\eta_1^\epsilon, \Gamma_2 \vdash^A \overline{arg} \Leftarrow \sigma_1 \Rightarrow \sigma_2 \leadsto \overline{arg_F}_1$, $\Gamma_1 \vdash e_1 \Rightarrow \eta_1^\epsilon \leadsto t_1$
*and* $\Gamma_1, \Gamma_2 \vdash^A [e_1/x] \overline{arg} \Leftarrow \sigma_1 \Rightarrow \sigma_2 \leadsto \overline{arg_F}_2$ *where* $\overline{a} = f_v(\eta_1^\epsilon) \setminus dom(\Gamma_1)$
*then* $\forall t_i \in \overline{arg_F}_1, t'_i \in \overline{arg_F}_2 : t'_i \simeq (\lambda x : \forall \overline{a}.\eta_1^\epsilon.t_i) t_1$

**Lemma E.36** (Declaration Inlining is Runtime Semantics Preserving).
*If* $\Gamma_1, x : \forall \overline{\{a\}}.\eta_1^\epsilon, \Gamma_2 \vdash decl \Rightarrow \Gamma_3 \leadsto y : \sigma_2 = t_2$, $\Gamma_1 \vdash e_1 \Rightarrow \eta_1^\epsilon \leadsto t_1$ *and* $\Gamma_1, \Gamma_2 \vdash [e_1/x] decl \Rightarrow \Gamma_3 \leadsto y : \sigma_2 = t_3$
*where* $\overline{a} = f_v(\eta_1^\epsilon) \setminus dom(\Gamma_1)$ *then* $t_3 \simeq (\lambda x : \forall \overline{a}.\eta_1^\epsilon.t_2) t_1$

As is probably clear by now, these lemmas are proven through mutual induction. The proof proceeds by structural induction on the first typing derivation. We will focus on the non-trivial cases:

**Case rule** H-Var $h = y$ **where** $y = x$ :
The goal reduces to $t_1 \simeq (\lambda x : \forall \overline{a}.\eta_1^\epsilon.x) t_1$, which follows directly from Lemmas E.13 and E.26.

**Case rule** H-Var $h = y$ **where** $y \neq x$ :
The goal reduces to $y \simeq (\lambda x : \forall \overline{a}.\eta_1^\epsilon.y) t_1$. Since $(\lambda x : \forall \overline{a}.\eta_1^\epsilon.y) t_1 \hookrightarrow y$, the goal follows directly from Lemmas E.13 and E.26.

**Case rule** Tm-InfAbs $e_2 = \lambda y.e_4$ :
The premise tells us $\Gamma_1, x : \forall \overline{\{a\}}.\eta_1^\epsilon, \Gamma_2, y : \tau_1 \vdash e_4 \Rightarrow \eta_4^\epsilon \leadsto t_4$ and $\Gamma_1, \Gamma_2, y : \tau_1 \vdash [e_1/x] e_4 \Rightarrow \eta_4^\epsilon \leadsto t_5$. Applying the induction hypothesis gives us $t_5 \simeq (\lambda x : \forall \overline{a}.\eta_1^\epsilon.t_4) t_1$. The goal reduces to $\lambda y : \tau_1.t_5 \simeq (\lambda x : \forall \overline{a}.\eta_1^\epsilon.\lambda y : \tau_1.t_4) t_1$. In order not to clutter the proof too much, we introduce an additional helper lemma E.37. The goal then follows from Lemmas E.16 and E.37.

**Case rule** Tm-InfTyAbs $e_2 = \Lambda a.e_4$ :
The premise tells us $\Gamma_1, x : \forall \overline{\{a\}}.\eta_1^\epsilon, \Gamma_2, a \vdash e_4 \Rightarrow \eta_4^\epsilon \leadsto t_4$, $\Gamma_1, \Gamma_2, a \vdash [e_1/x] e_4 \Rightarrow \eta_4^\epsilon \leadsto t_5$ and $\Gamma_1, \Gamma_2 \vdash \forall a.\eta_4^\epsilon \xrightarrow{inst\ \delta} \eta_5^\epsilon \leadsto \dot{t}$. Applying the induction hypothesis gives us $t_5 \simeq (\lambda x : \forall \overline{a}.\eta_1^\epsilon.t_4) t_1$. The goal reduces to $\dot{t}[\Lambda a.t_5] \simeq (\lambda x : \forall \overline{a}.\eta_1^\epsilon.\dot{t}[\Lambda a.t_4]) t_1$. Similarly to before, we avoid cluttering the proof by introducing an additional helper lemma E.38. The goal then follows from Lemmas E.18, E.24 and E.38. □

**Lemma E.37** (Property 3 Term Abstraction Helper).
*If* $\Gamma \vdash \lambda x : \sigma_2.((\lambda y : \sigma_1.t_2) t_1) : \sigma_3$ *and* $\Gamma \vdash t_1 : \sigma_1$ *then* $\lambda x : \sigma_2.((\lambda y : \sigma_1.t_2) t_1) \simeq (\lambda y : \sigma_1.\lambda x : \sigma_2.t_2) t_1$

**Lemma E.38** (Property 3 Type Abstraction Helper).
*If* $\Gamma \vdash \Lambda a.((\lambda x : \sigma_1.t_2) t_1) : \sigma_2$ *and* $a \notin f_v(\sigma_1)$ *then* $\Lambda a.((\lambda x : \sigma_1.t_2) t_1) \simeq (\lambda x : \sigma_1.\Lambda a.t_2) t_1$

Both lemmas follow from the definition of contextual equivalence.

We now return to proving Property 3. By case analysis (Either rule Tm-InfLet or rule Tm-CheckLet, followed by rule Decl-NoAnnSingle) we know $\Gamma, x : \forall \overline{\{a\}}.\eta_1^\epsilon \vdash e_2 \Rightarrow \eta^\epsilon \leadsto t_3$ or $\Gamma, x : \forall \overline{\{a\}}.\eta_1^\epsilon \vdash e_2 \Leftarrow \sigma \leadsto t_3$ where $t_1 = (\lambda x : \forall \overline{a}.\eta_1^\epsilon.t_3) t_4$, $\Gamma \vdash e_1 \Rightarrow \eta_1^\epsilon \leadsto t_4$ and $\overline{a} = f_v(\eta_1^\epsilon) \setminus dom(\Gamma)$. The goal thus follows directly from Lemma E.32 or E.33. However, as Lemma E.24 only holds under shallow instantiation, we cannot prove Property 3 under deep instantiation. □

### E.4 Type Signatures

**Property 4b** (Signature Property is Type Preserving).
*If* $\Gamma \vdash x \overline{\pi} = e \Rightarrow \Gamma'$ *and* $x : \sigma \in \Gamma'$ *then* $\Gamma \vdash x : \sigma; x \overline{\pi} = e \Rightarrow \Gamma'$

Before proving Property 4b, we first introduce a number of helper lemmas:

**Lemma E.39** (Skolemisation Exists).
If $f_v(\sigma) \in \Gamma$ then $\exists \rho, \Gamma'$ such that $\Gamma \vdash \sigma \xrightarrow{skol\ \delta} \rho; \Gamma'$

The proof follows through careful examination of the skolemisation relation.

**Lemma E.40** (Skolemisation Implies Instantiation).
If $\Gamma \vdash \sigma \xrightarrow{skol\ \delta} \rho; \Gamma'$ then $\Gamma' \vdash \sigma \xrightarrow{inst\ \delta} \rho$

The proof follows by straightforward induction on the skolemisation relation. Note that as skolemisation binds all type variables in $\Gamma'$, they can then be used for instantiation.

**Lemma E.41** (Inferred Type Binders Preserve Expression Checking).
If $\Gamma \vdash e \Leftarrow \sigma$ then $\Gamma \vdash e \Leftarrow \forall \overline{\{a\}}.\sigma$

The proof follows by straightforward induction on the typing derivation.

**Lemma E.42** (Pattern Synthesis Implies Checking).
If $\Gamma \vdash^P \overline{\pi} \Rightarrow \overline{\psi}; \Delta$ then $\forall \sigma', \exists \sigma : \Gamma \vdash^P \overline{\pi} \Leftarrow \sigma \Rightarrow \sigma'; \Delta$ where $type\ (\overline{\psi}; \sigma' \sim \sigma)$

The proof follows by straightforward induction on the pattern typing derivation.

**Lemma E.43** (Expression Synthesis Implies Checking).
If $\Gamma \vdash e \Rightarrow \eta^\epsilon$ then $\Gamma \vdash e \Leftarrow \eta^\epsilon$

The proof follows by induction on the typing derivation. We will focus on the non-trivial cases below:

**Case rule** Tm-InfAbs $e = \lambda x.e'$:
We know from the premise of the typing rule that $\Gamma, x : \tau_1 \vdash e' \Rightarrow \eta_2^\epsilon$ where $\eta^\epsilon = \tau_1 \rightarrow \eta_2^\epsilon$. By rule Tm-CheckAbs, the goal reduces to $\Gamma \vdash \tau_1 \rightarrow \eta_2^\epsilon \xrightarrow{skol\ S} \tau_1 \rightarrow \eta_2^\epsilon; \Gamma$ (which follows directly by rule SkolT-SInst) and $\Gamma, x : \tau_1 \vdash e' \Leftarrow \eta_2^\epsilon$ (which follows by the induction hypothesis).

**Case rule** Tm-InfTyAbs $e = \Lambda a.e'$:
The typing rule premise tells us that $\Gamma, a \vdash e' \Rightarrow \eta_1^\epsilon$ and $\Gamma \vdash \forall a.\eta_1^\epsilon \xrightarrow{inst\ \delta} \eta_2^\epsilon$. By rule Tm-CheckTyAbs, the goal reduces to $\eta_2^\epsilon = \forall \overline{\{a\}}.\forall a.\sigma'$ and $\Gamma, \overline{\{a\}}, a \vdash e' \Leftarrow \sigma'$. It is now clear that this property can never hold under eager instantiation, as the forall type in $\forall a.\eta_1^\epsilon$ would always be instantiated away. We will thus focus solely on lazy instantiation from here on out, where $\eta_2^\epsilon = \forall a.\eta_1^\epsilon$. In this case, the goal follows directly from the induction hypothesis.

**Case rule** Tm-InfApp $e = h\ \overline{arg}$:
We know from the typing rule premise that $\Gamma \vdash^H h \Rightarrow \sigma$, $\Gamma \vdash^A \overline{arg} \Leftarrow \sigma \Rightarrow \sigma'$ and $\Gamma \vdash \sigma' \xrightarrow{inst\ \delta} \eta^\epsilon$. Note that as we assume lazy instantiation, $\eta^\epsilon = \sigma'$. By rule Tm-CheckInf, the goal reduces to $\Gamma \vdash \eta^\epsilon \xrightarrow{skol\ \delta} \rho; \Gamma'$ (follows by Lemma E.39), $\Gamma' \vdash h\ \overline{arg} \Rightarrow \eta_1^\epsilon$ (follows by performing environment weakening on the premise, with $\eta_1^\epsilon = \eta^\epsilon$) and $\Gamma' \vdash \eta_1^\epsilon \xrightarrow{inst\ \delta} \rho$ (given that $\eta_1^\epsilon = \eta^\epsilon$, this follows by Lemma E.40). □

We now proceed with proving Property 4b, through case analysis on the declaration typing derivation (rule Decl-NoAnnSing

We know from the typing rule premise that $\Gamma \vdash^P \overline{\pi} \Rightarrow \overline{\psi}; \Delta$, $\Gamma, \Delta \vdash e \Rightarrow \eta^\epsilon$, $type\ (\overline{\psi}; \eta^\epsilon \sim \sigma_1)$ and $\sigma = \forall \overline{\{a\}}.\sigma_1$ where $\overline{a} = f_v(\sigma_1) \setminus dom\ (\Gamma)$. By rule Decl-Ann, the goal reduces to $\Gamma \vdash^P \overline{\pi} \Leftarrow \forall \overline{\{a\}}.\sigma_1 \Rightarrow \sigma_2; \Delta_2$ and $\Gamma, \Delta_2 \vdash e \Leftarrow \sigma_2$. We know from Lemma E.42 that $\Gamma \vdash^P \overline{\pi} \Leftarrow \sigma_1 \Rightarrow \sigma_3; \Delta$ where $type\ (\overline{\psi}; \sigma_3 \sim \sigma_1)$. Furthermore, from Lemma E.43 we get $\Gamma, \Delta \vdash e \Leftarrow \eta^\epsilon$. Note that we thus only prove Property 4b under lazy instantiation. We now proceed by case analysis on $\overline{\pi}$:

**Case** $\overline{\pi} = \cdot$:
The first goal now follows trivially by rule Pat-CheckEmpty with $\sigma_2 = \forall \overline{\{a\}}.\sigma_1$, $\sigma_1 = \eta^\epsilon$ and $\Delta = \Delta_2 = \cdot$. The second goal follows by Lemma E.41.

**Case** $\overline{\pi} \neq \cdot$:
The first goal follows by repeated application of rule Pat-CheckInfForall with $\sigma_2 = \sigma_3 = \eta^\epsilon$. The second goal then follows directly from Lemma E.43. □

**Property 5** (Signature Property is Runtime Semantics Preserving).
*If* $\Gamma \vdash \overline{x\,\overline{\pi}_i = e_i}^{\,i} \Rightarrow \Gamma' \rightsquigarrow x : \sigma = t_1$ *and* $\Gamma \vdash x : \sigma; \overline{x\,\overline{\pi}_i = e_i}^{\,i} \Rightarrow \Gamma' \rightsquigarrow x : \sigma = t_2$ *then* $t_1 \simeq t_2$

We start by introducing a number of helper lemmas:

**Lemma E.44** (Pattern Typing Mode Preserves Translation).
*If* $\Gamma \vdash^P \overline{\pi} \Rightarrow \overline{\psi}; \Delta \rightsquigarrow \overline{\pi_{F1}} : \overline{\psi_{F1}}$ *and* $\Gamma \vdash^P \overline{\pi} \Leftarrow \sigma \Rightarrow \sigma'; \Delta \rightsquigarrow \overline{\pi_{F2}} : \overline{\psi_{F2}}$ *where* $type\,(\overline{\psi}; \sigma' \sim \sigma)$
*then* $\overline{\pi_{F1}} = \overline{\pi_{F2}}$ *and* $\overline{\psi_{F1}} = \overline{\psi_{F2}}$

The proof follows by straightforward induction on the pattern type inference derivation.

**Lemma E.45** (Compatibility One-Sided Type Abstraction).
*If* $t_1 \simeq t_2$ *then* $t_1 \simeq \Lambda a.t_2$

The proof follows by the definition of contextual equivalence. Note that while the left and right hand sides have different types, they still instantiate to a single common type.

**Lemma E.46** (Partial Skolemisation Preserves Type Checking and Runtime Semantics).
*If* $\Gamma \vdash e \Leftarrow \forall \overline{\{a\}}.\sigma \rightsquigarrow t_1$ *then* $\Gamma, \overline{a} \vdash e \Leftarrow \sigma \rightsquigarrow t_2$ *where* $t_1 \simeq t_2$.

The proof proceeds by induction on the type checking derivation. Note that every case performs a (limited) form of skolemisation. Every case proceeds by applying the induction hypothesis, followed by Lemma E.45.

**Lemma E.47** (Typing Mode Preserves Runtime Semantics).
*If* $\Gamma \vdash e \Rightarrow \eta^\epsilon \rightsquigarrow t_1$ *and* $\Gamma \vdash e \Leftarrow \sigma \rightsquigarrow t_2$ *where* $\Gamma \vdash \eta^\epsilon \xrightarrow{inst\,\delta} \rho \rightsquigarrow \dot{t}_1$ *and* $\Gamma \vdash \sigma \xrightarrow{inst\,\delta} \rho \rightsquigarrow \dot{t}_2$
*then* $t_1 \simeq t_2$

The proof proceeds by induction on the first typing derivation. Each case follows straightforwardly by applying the induction hypothesis, along with the corresponding compatibility lemma (Lemmas E.16 till E.20).

We now turn to proving property 5, through case analysis on the first declaration typing derivation:

**Case rule** DECL-NOANNSINGLE:

We know from the premise of the first derivation that $\Gamma \vdash^P \overline{\pi} \Rightarrow \overline{\psi}; \Delta \rightsquigarrow \overline{\pi_{F1}} : \overline{\psi_{F1}}$, $\Gamma, \Delta \vdash e \Rightarrow \eta^\epsilon \rightsquigarrow t'_1$, $type\,(\overline{\psi}; \eta^\epsilon \sim \sigma_1)$, $t_1 = \text{case}\,\overline{\pi_{F1}} : \overline{\psi_{F1}} \to t'_1$ and $\sigma = \forall \overline{\{a\}}.\sigma_1$ where $\overline{a} = f_v(\sigma_1) \setminus dom\,(\Gamma)$. By case analysis on the second derivation (rule DECL-ANN), we get $\Gamma \vdash^P \overline{\pi} \Leftarrow \forall \overline{\{a\}}.\sigma_1 \Rightarrow \sigma_2; \Delta \rightsquigarrow \overline{\pi_{F2}} : \overline{\psi_{F2}}$, $\Gamma, \Delta \vdash e \Leftarrow \sigma_2 \rightsquigarrow t'_2$ and $t_2 = \text{case}\,\overline{\pi_{F2}} : \overline{\psi_{F2}} \to t'_2$.

We proceed by case analysis on the patterns $\overline{\pi}$:

**case** $\overline{\pi} = \cdot$: We know from rule PAT-INFEMPTY, rule PAT-CHECKEMPTY and rule TYPE-EMPTY that $\sigma_2 = \forall \overline{\{a\}}.\sigma_1 = \forall \overline{\{a\}}.\eta^\epsilon$. By applying Lemma E.46, we get $\Gamma, \overline{a} \vdash e \Leftarrow \eta^\epsilon \rightsquigarrow t_3$ where $t'_2 \simeq t_3$. The goal now follows by Lemma E.47 (after environment weakening, where $\sigma = \rho = \eta^\epsilon$), and Lemma E.15.

**case** $\overline{\pi} \neq \cdot$: By case analysis on the pattern checking derivation (rule PAT-CHECKINFFORALL), we know that $\Gamma, \overline{a} \vdash^P \overline{\pi} \Leftarrow \sigma_1 \Rightarrow \sigma_2; \Delta' \rightsquigarrow \overline{\pi_{F2}} : \overline{\psi_{F2}}'$ where $\Delta = \overline{a}, \Delta'$ and $\overline{\psi_{F2}} = @\overline{a}, \overline{\psi_{F2}}'$. By Lemma E.42 (where we take $\sigma = \sigma_1$), we know that $type\,(\overline{\psi}; \sigma_2 \sim \sigma_1)$. This thus means that $\sigma_2 = \eta^\epsilon$. By Lemma E.44, the goal reduces to $\text{case}\,\overline{\pi_{F1}} : \overline{\psi_{F1}} \to t'_1 \simeq \text{case}\,\overline{\pi_{F1}} : \overline{\psi_{F1}} \to t'_2$. Applying Lemma E.20 reduces this goal further to $t'_1 \simeq t'_2$. This follows directly from Lemma E.47 (where $\sigma = \rho = \eta^\epsilon$).

**Case rule** DECL-NOANNMULTI:

We know from the premise of the first derivation that $\forall i : \Gamma \vdash^P \overline{\pi}_i \Rightarrow \overline{\psi}; \Delta_i \rightsquigarrow \overline{\pi_{Fi}} : \overline{\psi_F}$, $\Gamma, \Delta_i \vdash e_i \Rightarrow \eta^\epsilon_i \rightsquigarrow t_i$ and $\Gamma, \Delta_i \vdash \eta^\epsilon_i \xrightarrow{inst\,\delta} \rho' \rightsquigarrow \dot{t}_i$. Furthermore, $t_1 = \text{case}\,\overline{\overline{\pi_{Fi}} : \overline{\psi_F} \to \dot{t}_i[t_i]}^{\,i}$, $type\,(\overline{\psi}; \rho' \sim \sigma')$ and $\sigma = \forall \overline{\{a\}}.\sigma'$ where $\overline{a} = f_v(\sigma') \setminus dom\,(\Gamma)$. By case analysis on the second derivation (rule DECL-ANN), we know that $\forall i : \Gamma \vdash^P \overline{\pi}_i \Leftarrow \forall \overline{\{a\}}.\sigma' \Rightarrow \sigma_i; \Delta_i \rightsquigarrow \overline{\pi'_{Fi}} : \overline{\psi_F}'$, $\Gamma, \Delta_i \vdash e_i \Leftarrow \sigma_i \rightsquigarrow t'_i$ and $t_2 = \text{case}\,\overline{\overline{\pi'_{Fi}} : \overline{\psi_F}' \to t'_i}^{\,i}$.

We again perform case analysis on the patterns $\overline{\pi}$:

**case** $\overline{\pi} = \cdot$: Similarly to last time, we know that $\sigma' = \rho'$ and $\forall i : \sigma_i = \forall \overline{\{a\}}.\rho'$. We know by Lemma E.46 that $\forall i : \Gamma, \overline{a} \vdash e_i \Leftarrow \rho' \rightsquigarrow t''_i$ where $t'_i \simeq t''_i$. The goal now follows by Lemma E.47 (where we take $\sigma = \rho = \rho'$) and Lemma E.15.

**case** $\overline{\pi} \neq \cdot$: Similarly to the previous case, we can derive that $\forall i : \Gamma, \overline{a} \vdash^P \overline{\pi} \Leftarrow \sigma' \Rightarrow \sigma_i; \Delta'_i \rightsquigarrow \overline{\pi'_{F\,i}} : \overline{\psi''_F}^i$ where $\Delta_i = \overline{a}, \Delta'_i$ and $\overline{\psi'_F} = @\overline{a}, \overline{\psi''_F}$. We again derive by Lemma E.42 that $type\,(\overline{\psi}; \sigma_i \sim \sigma')$ and thus that $\sigma_i = \rho'$.
By Lemma E.44, the goal reduces to case $\overline{\pi_{F\,i} : \overline{\psi_F} \rightarrow \dot{t}_i[t_i]}^i \simeq$ case $\overline{\pi_{F\,i} : \overline{\psi_F} \rightarrow t'_i}^i$. We reduce this goal further by applying Lemma E.20 to $\forall i : \dot{t}_i[t_i] \simeq t'_i$. This follows directly from Lemma E.47 (where $\sigma = \rho = \rho'$).

Note however, that as Lemma E.47 only holds under shallow instantiation, that the same holds true for Property 5. □

**Property 6** (Type Signatures are Runtime Semantics Preserving).
$If\,\Gamma \vdash x : \sigma_1; \overline{x\,\overline{\pi}_i = e_i}^i \Rightarrow \Gamma_1 \rightsquigarrow x : \sigma_1 = t_1\, and\, \Gamma \vdash x : \sigma_2; \overline{x\,\overline{\pi}_i = e_i}^i \Rightarrow \Gamma_1 \rightsquigarrow x : \sigma_2 = t_2\, where\, \Gamma \vdash \sigma_1 \xrightarrow{inst\,\delta} \rho \rightsquigarrow \dot{t}_1\, and\, \Gamma \vdash \sigma_2 \xrightarrow{inst\,\delta} \rho \rightsquigarrow \dot{t}_2\, then\, \dot{t}_1[t_1] \simeq \dot{t}_2[t_2]$

We start by introducing a number of helper lemmas:

**Lemma E.48** (Substitution in Expressions is Type Preserving (Synthesis)).
$If\,\Gamma, a \vdash e \Rightarrow \eta^\epsilon \rightsquigarrow t\, then\, \Gamma \vdash [\tau/a]\,e \Rightarrow [\tau/a]\,\eta^\epsilon \rightsquigarrow [\tau/a]\,t$

**Lemma E.49** (Substitution in Expressions is Type Preserving (Checking)).
$If\,\Gamma, a \vdash e \Leftarrow \sigma \rightsquigarrow t\, then\, \Gamma \vdash [\tau/a]\,e \Leftarrow [\tau/a]\,\sigma \rightsquigarrow [\tau/a]\,t$

**Lemma E.50** (Substitution in Heads is Type Preserving).
$If\,\Gamma, a \vdash^H h \Rightarrow \sigma \rightsquigarrow t\, then\, \Gamma \vdash^H [\tau/a]\,h \Rightarrow [\tau/a]\,\sigma \rightsquigarrow [\tau/a]\,t$

**Lemma E.51** (Substitution in Arguments is Type Preserving).
$If\,\Gamma, a \vdash^A \overline{arg} \Leftarrow \sigma \Rightarrow \sigma' \rightsquigarrow \overline{arg_F}\, then\, \Gamma \vdash^A [\tau/a]\,\overline{arg} \Leftarrow [\tau/a]\,\sigma \Rightarrow [\tau/a]\,\sigma' \rightsquigarrow [\tau/a]\,\overline{arg_F}$

**Lemma E.52** (Substitution in Declarations is Type Preserving).
$If\,\Gamma, a \vdash decl \Rightarrow \Gamma, a, x : \sigma \rightsquigarrow x : \sigma = t\, then\, \Gamma \vdash [\tau/a]\,decl \Rightarrow \Gamma, x : [\tau/a]\,\sigma \rightsquigarrow x : \sigma = [\tau/a]\,t$

The proof proceeds by mutual induction on the typing derivation. While the number of cases gets pretty large, each is quite straightforward.

**Lemma E.53** (Type Instantiation Produces Equivalent Expressions (Synthesis)).
$If\,\Gamma_1 \vdash e \Rightarrow \eta^\epsilon_1 \rightsquigarrow t_1,\, \Gamma_2 \vdash e \Rightarrow \eta^\epsilon_2 \rightsquigarrow t_2\, and\, \exists\,\overline{a} \subseteq f_v(\eta^\epsilon_1) \cup f_v(\eta^\epsilon_2)$
such that $\Gamma' = [\overline{\tau}/\overline{a}]\,\Gamma_1 = [\overline{\tau}/\overline{a}]\,\Gamma_2\, and\, \Gamma' \vdash \forall \overline{a}.\eta^\epsilon_1 \xrightarrow{inst\,\delta} \rho \rightsquigarrow \dot{t}_1\, and\, \Gamma' \vdash \forall \overline{a}.\eta^\epsilon_2 \xrightarrow{inst\,\delta} \rho \rightsquigarrow \dot{t}_2$
then $\dot{t}_1[\Lambda\overline{a}.t_1] \simeq \dot{t}_2[\Lambda\overline{a}.t_2]$

**Lemma E.54** (Type Instantiation Produces Equivalent Expressions (Checking)).
$If\,\Gamma_1 \vdash e \Leftarrow \sigma_1 \rightsquigarrow t_1\, and\, \Gamma_2 \vdash e \Leftarrow \sigma_2 \rightsquigarrow t_2\, and\, \exists\,\overline{a} \subseteq f_v(\sigma_1) \cup f_v(\sigma_2)$
such that $\Gamma' = [\overline{\tau}/\overline{a}]\,\Gamma_1 = [\overline{\tau}/\overline{a}]\,\Gamma_2\, and\, \Gamma' \vdash \forall \overline{a}.\sigma_1 \xrightarrow{inst\,\delta} \rho \rightsquigarrow \dot{t}_1\, and\, \Gamma' \vdash \forall \overline{a}.\sigma_2 \xrightarrow{inst\,\delta} \rho \rightsquigarrow \dot{t}_2$
then $\dot{t}_1[\Lambda\overline{a}.t_1] \simeq \dot{t}_2[\Lambda\overline{a}.t_2]$

**Lemma E.55** (Type Instantiation Produces Equivalent Expressions (Head Judgement)).
$If\,\Gamma_1 \vdash^H h \Rightarrow \sigma_1 \rightsquigarrow t_1,\, \Gamma_2 \vdash^H h \Rightarrow \sigma_2 \rightsquigarrow t_2\, and\, \exists\,\overline{a} \subseteq f_v(\eta^\epsilon_1) \cup f_v(\eta^\epsilon_2)$
such that $\Gamma' = [\overline{\tau}/\overline{a}]\,\Gamma_1 = [\overline{\tau}/\overline{a}]\,\Gamma_2\, and\, \Gamma' \vdash \forall \overline{a}.\sigma_1 \xrightarrow{inst\,\delta} \rho \rightsquigarrow \dot{t}_1\, and\, \Gamma' \vdash \forall \overline{a}.\sigma_2 \xrightarrow{inst\,\delta} \rho \rightsquigarrow \dot{t}_2$
then $\dot{t}_1[\Lambda\overline{a}.t_1] \simeq \dot{t}_2[\Lambda\overline{a}.t_2]$

Note that we define $[\tau/a]\,\Gamma$ as removing $a$ from the environment $\Gamma$ and substituting any occurrence of $a$ in types bound to term variables. Furthermore, we use $\overline{a}_1 \cup \overline{a}_2$ as a shorthand for list concatenation, removing duplicates. The proof proceeds by induction on the first typing derivation. Note that Lemmas E.53, E.54 and E.55 have to be proven using mutual induction. However, the proof for Lemma E.55 is trivial, as every case besides rule H-Inf is deterministic. As usual, we will focus on the non-trivial cases:

**Case rule** Tm-CheckAbs $e = \lambda x.e'$:
We know from the premise of the first and second (as the relation is syntax directed) typing derivation that $\Gamma_1 \vdash \sigma_1 \xrightarrow{skol\,S} \sigma_4 \rightarrow \sigma_5; \Gamma'_1 \rightsquigarrow \dot{t}'_1,\, \Gamma_2 \vdash \sigma_2 \xrightarrow{skol\,S} \sigma'_4 \rightarrow \sigma'_5; \Gamma'_2 \rightsquigarrow \dot{t}'_2,\, \Gamma'_1, x : \sigma_4 \vdash e' \Leftarrow \sigma_5 \rightsquigarrow t_3$ and $\Gamma'_2, x : \sigma'_4 \vdash e' \Leftarrow \sigma'_5 \rightsquigarrow t_4$, where $t_1 = \dot{t}'_1[\lambda x : \sigma_4.t_3]$ and $t_2 = \dot{t}'_2[\lambda x : \sigma'_4.t_4]$.

At this point, it is already clear that Lemma E.54 can not hold under deep instantiation, as instantiation performs full eta expansion. We will thus focus on shallow instantiation from here on out.

By case analysis on the skolemisation and instantiation premises, it is clear that $\Gamma'_1 = \Gamma_1, \overline{a}_1$, $\Gamma'_2 = \Gamma_2, \overline{a}_2$ and $\rho = [\overline{\tau}_1/\overline{a}_1](\sigma_4 \rightarrow \sigma_5) = [\overline{\tau}_2/\overline{a}_2](\sigma'_4 \rightarrow \sigma'_5) = \sigma_3 \rightarrow \sigma'_3$. In order to apply the induction hypothesis, we take $\overline{a}'$ as $\overline{a} \cup \overline{a}_1 \cup \overline{a}_2$. Note that this does not alter the instantiation to $\rho$ in any way, as these variables would already have been instantiated. We apply the induction hypothesis with $\Gamma_1 \vdash \forall \overline{a}'.\sigma_5 \xrightarrow{inst\ \delta} \sigma'_3 \rightsquigarrow \dot{t}_3$ and $\Gamma_2 \vdash \forall \overline{a}'.\sigma'_5 \xrightarrow{inst\ \delta} \sigma'_3 \rightsquigarrow \dot{t}_4$ (after weakening), producing $\dot{t}_3[\Lambda\overline{a}'.t_3] \simeq \dot{t}_4[\Lambda\overline{a}'.t_4]$. Under shallow instantiation, these two instantiations follow directly from the premise with $\dot{t}_3 = \dot{t}_1$ and $\dot{t}_4 = \dot{t}_2$.

The goal reduces to $\dot{t}_1[\Lambda\overline{a}.\dot{t}'_1[\lambda x : \sigma_4.t_3]] \simeq \dot{t}_2[\Lambda\overline{a}.\dot{t}'_2[\lambda x : \sigma'_4.t_4]]$. By the definition of skolemisation, this further reduces to $\dot{t}_1[\Lambda\overline{a}.\Lambda\overline{a}_1.\lambda x : \sigma_4.t_3] \simeq \dot{t}_2[\Lambda\overline{a}.\Lambda\overline{a}_2.\lambda x : \sigma'_4.t_4]$. Finally, the goal follows by the induction hypothesis and compatibility Lemmas E.18, E.16 and E.21, along with transitivity Lemma E.15.

**Case rule Tm-CheckTyAbs** $e = \Lambda a.e'$:

We know the premise of the typing derivation that $\sigma_1 = \forall \overline{\{a\}}_1.\forall a.\sigma'_1$, $\sigma_2 = \forall \overline{\{a\}}.\forall a.\sigma'_2$, $\Gamma_1, \overline{a}_1, a \vdash e' \Leftarrow \sigma'_1 \rightsquigarrow t'_1$, $\Gamma_2, \overline{a}_2, a \vdash e' \Leftarrow \sigma'_2 \rightsquigarrow t'_2$, $t_1 = \Lambda\overline{a}_1.\Lambda a.t'_1$ and $t_2 = \Lambda\overline{a}_2.\Lambda a.t'_2$. By case analysis on the type instantiation (rule InstT-SForall and rule InstT-SInfForall), we get $\Gamma' \vdash [\overline{\tau}_1/\overline{a}][\overline{\tau}'_1/\overline{a}_1][\tau_1/a]\sigma'_1 \xrightarrow{inst\ \delta} \rho \rightsquigarrow \dot{t}'_1$ and $\Gamma' \vdash [\overline{\tau}_2/\overline{a}][\overline{\tau}'_2/\overline{a}_2][\tau_2/a]\sigma'_2 \xrightarrow{inst\ \delta} \rho \rightsquigarrow \dot{t}'_2$ where $\dot{t}_1 = \lambda t.(\dot{t}'_1[t\ \overline{\tau}_1\ \overline{\tau}'_1\ \tau_1])$ and $\dot{t}_2 = \lambda t.(\dot{t}'_2[t\ \overline{\tau}_2\ \overline{\tau}'_2\ \tau_2])$.

The goal to be proven is $\dot{t}_1[\Lambda\overline{a}.\Lambda\overline{a}_1.\Lambda a.t'_1] \simeq \dot{t}_2[\Lambda\overline{a}.\Lambda\overline{a}_2.\Lambda a.t'_2]$. This reduces to $\dot{t}'_1[(\Lambda\overline{a}.\Lambda\overline{a}_1.\Lambda a.t'_1)\ \overline{\tau}_1\ \overline{\tau}'_1\ \tau_1] \simeq \dot{t}'_2[(\Lambda\overline{a}.\Lambda\overline{a}_2.\Lambda a.t'_2)\ \overline{\tau}_2\ \overline{\tau}'_2\ \tau_2]$.

We now define a substitution $\theta = [\overline{\tau}_1/\overline{a}].[\overline{\tau}_2/\overline{a}].[\overline{\tau}'_1/\overline{a}_1].[\overline{\tau}'_2/\overline{a}_2].[\tau_1/a].[\tau_2/a]$. From the instantiation relation (and the fact that both types instantiate to the same type $\rho$, we conclude that if $[\tau_i/a] \in \theta$ and $[\tau_j/a] \in \theta$ that $\tau_i = \tau_j$. By applying Lemma E.49, we transform the premise to $[\overline{\tau}_1/\overline{a}]\Gamma_1 \vdash \theta\ e' \Leftarrow \theta\ \sigma'_1 \rightsquigarrow \theta\ t'_1$ and $[\overline{\tau}_2/\overline{a}]\Gamma_2 \vdash \theta\ e' \Leftarrow \theta\ \sigma'_2 \rightsquigarrow \theta\ t'_2$.

By applying the induction hypothesis, we get that $\dot{t}'_1[\theta\ t'_1] \simeq \dot{t}'_2[\theta\ t'_2]$. The goal follows directly from the definition of $\theta$.

**Case rule Tm-CheckInf** :

We know from the premise of the typing derivation that $\Gamma_1 \vdash \sigma_1 \xrightarrow{skol\ \delta} \rho_1; \Gamma'_1 \rightsquigarrow \dot{t}'_1$, $\Gamma_2 \vdash \sigma_2 \xrightarrow{skol\ \delta} \rho_2; \Gamma'_2 \rightsquigarrow \dot{t}'_2$, $\Gamma'_1 \vdash e \Rightarrow \eta^\epsilon_1 \rightsquigarrow t'_1$, $\Gamma'_2 \vdash e \Rightarrow \eta^\epsilon_2 \rightsquigarrow t'_2$, $\Gamma'_1 \vdash \eta^\epsilon_1 \xrightarrow{inst\ \delta} \rho_1 \rightsquigarrow \dot{t}''_1$, $\Gamma'_2 \vdash \eta^\epsilon_2 \xrightarrow{inst\ \delta} \rho_2 \rightsquigarrow \dot{t}''_2$, $t_1 = \dot{t}'_1[\dot{t}''_1[t'_1]]$ and $t_2 = \dot{t}''_2[\dot{t}'_2[t'_2]]$. The goal to be proven is thus $\dot{t}_1[\Lambda\overline{a}.\dot{t}'_1[\dot{t}''_1[t'_1]]] \simeq \dot{t}_2[\Lambda\overline{a}.\dot{t}'_2[\dot{t}''_2[t'_2]]]$.

From the definition of shallow skolemisation, we know that $\Gamma'_1 = \Gamma_1, \overline{a}_1$, $\Gamma'_2 = \Gamma_2, \overline{a}_2$, $\dot{t}'_1 = \lambda t.\Lambda\overline{a}_1.t$ and $\dot{t}'_2 = \lambda t.\Lambda\overline{a}_2.t$. We now take $\overline{a}' = \overline{a} \cup \overline{a}_1 \cup \overline{a}_2$. As $\sigma_1$ and $\sigma_2$ instantiate to the same type $\rho$, it is not hard to see from the definition of skolemisation that $\Gamma'_1 \vdash \forall \overline{a}'.\eta^\epsilon_1 \xrightarrow{inst\ \delta} \rho \rightsquigarrow \dot{t}_3$ and $\Gamma'_2 \vdash \forall \overline{a}'.\eta^\epsilon_2 \xrightarrow{inst\ \delta} \rho \rightsquigarrow \dot{t}_4$. By applying Lemma E.53, we thus get $\dot{t}_3[\Lambda\overline{a}'.t'_1] \simeq \dot{t}_4[\Lambda\overline{a}'.t'_2]$. The goal follows through careful examination of the skolemisation and instantiation premises. □

**Lemma E.56** (Pattern Checking Implies Synthesis).
*If* $\Gamma \vdash^P \overline{\pi} \Leftarrow \sigma \Rightarrow \sigma'; \Delta \rightsquigarrow \overline{\pi_F} : \overline{\psi_F}$ *then* $\exists \overline{\psi} : \Gamma \vdash^P \overline{\pi} \Rightarrow \overline{\psi}; \Delta \rightsquigarrow \overline{\pi_F} : \overline{\psi_F}$ *where type* $(\overline{\psi}; \sigma' \sim \sigma)$

The proof follows by straightforward induction on the pattern typing derivation.

We now go back to proving Property 6, and proceed by case analysis on both typing derivations (rule Decl-Ann). We know from the premise that $\Gamma \vdash^P \overline{\pi}_i \Leftarrow \sigma_1 \Rightarrow \sigma_{i1}; \Delta_{i1} \rightsquigarrow \overline{\pi_{Fi}} : \overline{\psi_{F1}}$, $\Gamma \vdash^P \overline{\pi}_i \Leftarrow \sigma_2 \Rightarrow \sigma_{i2}; \Delta_{i2} \rightsquigarrow \overline{\pi_{Fi}} : \overline{\psi_{F2}}$, $\Gamma, \Delta_{i1} \vdash e_i \Leftarrow \sigma_{i1} \rightsquigarrow t_{i1}$, $\Gamma, \Delta_{i2} \vdash e_i \Leftarrow \sigma_{i2} \rightsquigarrow t_{i2}$, $t_1 = \text{case}\ \overline{\overline{\pi_{Fi}} : \overline{\psi_{F1}} \rightarrow t_{i1}}^i$ and $t_2 = \text{case}\ \overline{\overline{\pi_{Fi}} : \overline{\psi_{F2}} \rightarrow t_{i2}}^i$.
The goal to be proven is $\dot{t}_1[\text{case}\ \overline{\overline{\pi_{Fi}} : \overline{\psi_{F1}} \rightarrow t_{i1}}^i] \simeq \dot{t}_2[\text{case}\ \overline{\overline{\pi_{Fi}} : \overline{\psi_{F2}} \rightarrow t_{i2}}^i]$. Lemma E.20 reduces this to $\forall i : \dot{t}_1[t_{i1}] \simeq \dot{t}_2[t_{i2}]$.

We take $\overline{a}_i = dom\ (\Delta_{i1}) \cup dom\ (\Delta_{i2}) \setminus dom\ (\Gamma)$, and apply weakening to get $\Gamma, \overline{a}_i \vdash e_i \Leftarrow \sigma_{i1} \rightsquigarrow t_{i1}$ and $\Gamma, \overline{a}_i \vdash e_i \Leftarrow \sigma_{i2} \rightsquigarrow t_{i2}$. The goal now follows directly from Lemma E.54 with $\overline{a}_i = \cdot$, if we can show that $\Gamma, \overline{a}_i \vdash \sigma_{i1} \xrightarrow{inst\ \delta} \rho' \rightsquigarrow \dot{t}_1$ and $\Gamma, \overline{a}_i \vdash \sigma_{i2} \xrightarrow{inst\ \delta} \rho' \rightsquigarrow \dot{t}_2$ for some $\rho'$ (Note that Lemma E.54 only holds under shallow instantiation).

We know from Lemma E.56 that $\exists \overline{\psi_F} : \Gamma \vdash^P \overline{\pi}_i \Rightarrow \overline{\psi}; \Delta_i \rightsquigarrow \overline{\pi_{F\,i}} : \overline{\psi_F}$ such that $type\,(\overline{\psi}; \sigma_{i\,1} \sim \sigma_1)$ and $type\,(\overline{\psi}; \sigma_{i\,2} \sim \sigma_2)$. The remaining goal follows from the definition of the type relation, and shallow instantiation. □

## E.5 Pattern Inlining and Extraction

**Property 7** (Pattern Inlining is Type Preserving).
*If* $\Gamma \vdash x\,\overline{\pi} = e_1 \Rightarrow \Gamma'$ *and* $wrap\,(\overline{\pi}; e_1 \sim e_2)$ *then* $\Gamma \vdash x = e_2 \Rightarrow \Gamma'$

We first introduce a helper lemma to prove pattern inlining in expressions preserves the type:

**Lemma E.57** (Pattern Inlining in Expressions is Type Preserving (Synthesis)).
*If* $\Gamma \vdash^P \overline{\pi} \Rightarrow \overline{\psi}; \Delta$ *and* $\Gamma, \Delta \vdash e_1 \Rightarrow \eta_1^\epsilon$ *where* $wrap\,(\overline{\pi}; e_1 \sim e_2)$
*then* $\Gamma \vdash e_2 \Rightarrow \eta_2^\epsilon$ *and* $type\,(\overline{\psi}; \eta_1^\epsilon \sim \eta_2^\epsilon)$

The proof proceeds by induction on the pattern typing derivation. We will focus on the non-trivial cases below. Note that the rule PAT-INFCON is an impossible case as $wrap\,(K\,\overline{\pi}; e_1 \sim e_2)$ is undefined.

**Case rule** PAT-INFVAR $\overline{\pi} = x, \overline{\pi}', \overline{\psi} = \tau_1, \overline{\psi}'$ **and** $\Delta = x : \tau_1, \Delta'$:
We know from the rule premise that $\Gamma, x : \tau_1 \vdash^P \overline{\pi}' \Rightarrow \overline{\psi}'; \Delta'$. Furthermore, by inlining the definitions of $\Delta$ and $\overline{\pi}$ in the lemma premise, we get $\Gamma, x : \tau_1, \Delta' \vdash e_1 \Rightarrow \eta_1^\epsilon$ and $wrap\,(x, \overline{\pi}'; e_1 \sim \lambda x.e_2')$ and thus (by rule PATWRAP-VAR) $wrap\,(\overline{\pi}'; e_1 \sim e_2')$. By the induction hypothesis, we get $\Gamma, x : \tau_1 \vdash e_2' \Rightarrow \eta_3^\epsilon$ and $type\,(\overline{\psi}'; \eta_1^\epsilon \sim \eta_3^\epsilon)$. The goal follows by rule TM-INFABS and rule TYPE-VAR.

**Case rule** PAT-INFTYVAR $\overline{\pi} = @a, \overline{\pi}', \overline{\psi} = @a, \overline{\psi}'$ **and** $\Delta = a, \Delta'$:
We know from the rule premise that $\Gamma, a \vdash^P \overline{\pi}' \Rightarrow \overline{\psi}'; \Delta'$. Again, by inlining the definitions in the lemma premise, we get $\Gamma, a, \Delta' \vdash e_1 \Rightarrow \eta_1^\epsilon$ and $wrap\,(@a, \overline{\pi}'; e_1 \sim \Lambda a.e_2')$ and thus (by rule PATWRAP-TYVAR) $wrap\,(\overline{\pi}'; e_1 \sim e_2')$. By the induction hypothesis, we get $\Gamma, a \vdash e_2' \Rightarrow \eta_3^\epsilon$ and $type\,(\overline{\psi}'; \eta_1^\epsilon \sim \eta_3^\epsilon)$.

The goal to be proven is $\Gamma \vdash \Lambda a.e_2' \Rightarrow \forall a.\eta_3^\epsilon$ where $type\,(@a, \overline{\psi}'; \eta_1^\epsilon \sim \forall a.\eta_3^\epsilon)$ (follows by rule TYPE-TYVAR). However, under eager instantiation, this goal can never hold as rule TM-INFTYABS would instantiate the forall binder away. We can thus only prove this lemma under lazy instantiation, where the goal follows trivially from rule TM-INFTYABS. □

We now proceed with proving Property 7, through case analysis on the declaration typing relation (rule DECL-NOANNSINGLE). We know from the premise of the first derivation that $\Gamma \vdash^P \overline{\pi} \Rightarrow \overline{\psi}; \Delta$, $\Gamma, \Delta \vdash e_1 \Rightarrow \eta_1^\epsilon$, $type\,(\overline{\psi}; \eta_1^\epsilon \sim \sigma)$ and $\Gamma' = \Gamma, x : \forall \overline{\{a\}}.\sigma$ where $\overline{a} = f_v(\sigma) \setminus dom\,(\Gamma)$. The goal to be proven thus becomes $\Gamma \vdash^P \cdot \Rightarrow \cdot; \cdot$ (follows directly from rule PAT-INFEMPTY) and $\Gamma \vdash e_2 \Rightarrow \eta_2^\epsilon$ where $\eta_2^\epsilon = \sigma$ (follows from Lemma E.57). Note that as we require Lemma E.57, we can only prove Property 7 under lazy instantiation. □

**Property 9** (Pattern Extraction is Type Preserving).
*If* $\Gamma \vdash x = e_2 \Rightarrow \Gamma'$ *and* $wrap\,(\overline{\pi}; e_1 \sim e_2)$ *then* $\Gamma \vdash x\,\overline{\pi} = e_1 \Rightarrow \Gamma'$

We first introduce another helper lemma to prove that pattern extraction from expressions preserves the typing:

**Lemma E.58** (Pattern Extraction from Expressions is Type Preserving (Synthesis)).
*If* $\Gamma \vdash e_2 \Rightarrow \eta_2^\epsilon$ *and* $\exists\,e_1, \overline{\pi}$ *such that* $wrap\,(\overline{\pi}; e_1 \sim e_2)$
*then* $\Gamma \vdash^P \overline{\pi} \Rightarrow \overline{\psi}; \Delta$ *and* $\Gamma, \Delta \vdash e_1 \Rightarrow \eta_1^\epsilon$ *where* $type\,(\overline{\psi}; \eta_1^\epsilon \sim \eta_2^\epsilon)$

The proof proceeds by induction on the $e_2$ typing derivation. As usual, we will focus on the non-trivial cases:

**Case rule** TM-INFABS $e_2 = \lambda x.e_2'$ **and** $\eta_2^\epsilon = \tau_2 \rightarrow \eta_3^\epsilon$:
We know from the rule premise that $\Gamma, x : \tau_2 \vdash e_2' \Rightarrow \eta_3^\epsilon$. It is clear by case analysis on $wrap\,(\overline{\pi}; e_1 \sim \lambda x.e_2')$ that $\overline{\pi} = x, \overline{\pi}'$ and $wrap\,(\overline{\pi}'; e_1 \sim e_2')$. By applying the induction hypothesis, we get $\Gamma, x : \tau_2 \vdash^P \overline{\pi}' \Rightarrow \overline{\psi}'; \Delta'$, $\Gamma, x : \tau_2, \Delta' \vdash e_1 \Rightarrow \eta_1^\epsilon$ and $type\,(\overline{\psi}'; \eta_1^\epsilon \sim \eta_3^\epsilon)$. The goal thus follows straightforwardly by rule PAT-INFVAR and rule TYPE-VAR.

**Case rule** TM-INFTYABS $e_2 = \Lambda a.e_2'$:

We know from the rule premise that $\Gamma, a \vdash e_2' \Rightarrow \eta_3^\epsilon$ and $\Gamma \vdash \forall a.\eta_3^\epsilon \xrightarrow{inst\ \delta} \eta_2^\epsilon$. Furthermore, it is clear by case analysis on $wrap\ (\overline{\pi}; e_1 \sim \Lambda a.e_2')$ that $\overline{\pi} = @a, \overline{\pi}'$ and $wrap\ (\overline{\pi}'; e_1 \sim e_2')$. By the induction hypothesis, we get $\Gamma, a \vdash^P \overline{\pi}' \Rightarrow \overline{\psi}'; \Delta', \Gamma, a, \Delta' \vdash e_1 \Rightarrow \eta_1^\epsilon$ and $type\ (\overline{\psi}'; \eta_1^\epsilon \sim \eta_3^\epsilon)$.

The goal to be proven is $\Gamma \vdash^P @a, \overline{\pi}' \Rightarrow @a, \overline{\psi}'; a, \Delta'$ (follows by rule Pat-InfTyVar), $\Gamma, a, \Delta' \vdash e_1 \Rightarrow \eta_1^\epsilon$ (follows by the induction hypothesis) and $type\ (@a, \overline{\psi}'; \eta_1^\epsilon \sim \eta_2^\epsilon)$. However, it is clear that this final goal does not hold under eager instantiation, as rule Tm-InfTyAbs instantiates the forall binder away. Under lazy instantiation, the remaining goal follows directly from the premise.

**Case rule** Tm-InfApp $e_2 = h\ \overline{arg}$ **and** $\overline{arg} = \cdot$ **and** $h = e$ **:**
The goal follows directly by the induction hypothesis.

**Case rule** Tm-InfApp $e_2 = h\ \overline{arg}$ **and** $\overline{arg} \neq \cdot$ **or** $h \neq e$ **:**
It is clear from the definition of $wrap\ (\overline{\pi}; e_1 \sim h\ \overline{arg})$ that $\overline{\pi} = \cdot$. The goal thus follows trivially. □

We now return to prove Property 9 by case analysis on the declaration typing derivation (rule Decl-NoAnnSingle). We know from the derivation premise that $\Gamma \vdash e_2 \Rightarrow \eta_2^\epsilon$ and $\sigma = \forall \overline{\{a\}}.\eta_2^\epsilon$ where $\overline{a} = f_v(\eta_2^\epsilon) \setminus dom\ (\Gamma)$. The goal follows directly from Lemma E.58. Note that as Lemma E.58 only holds under lazy instantiation, the same holds true for Property 9. □

**Property 8** (Pattern Inlining / Extraction is Runtime Semantics Preserving).
*If $\Gamma \vdash x\ \overline{\pi} = e_1 \Rightarrow \Gamma' \rightsquigarrow x : \sigma = t_1$, $wrap\ (\overline{\pi}; e_1 \sim e_2)$, and $\Gamma \vdash x = e_2 \Rightarrow \Gamma' \rightsquigarrow x : \sigma = t_2$ then $t_1 \simeq t_2$*

We start by introducing a helper lemma, proving pattern inlining preserves the runtime semantics for expressions.

**Lemma E.59** (Pattern Inlining in Expressions is Runtime Semantics Preserving).
*If $\Gamma \vdash^P \overline{\pi} \Rightarrow \overline{\psi}; \Delta \rightsquigarrow \overline{\pi_F} : \overline{\psi_F}$ and $\Gamma, \Delta \vdash e_1 \Rightarrow \eta_1^\epsilon \rightsquigarrow t_1$ and $\Gamma \vdash e_2 \Rightarrow \eta_2^\epsilon \rightsquigarrow t_2$ where $wrap\ (\overline{\pi}; e_1 \sim e_2)$ then $\text{case } \overline{\pi_F} : \overline{\psi_F} \to t_1 \simeq t_2$*

The proof proceeds by induction on the pattern typing derivation. We will focus on the non-trivial cases. Note that, as $wrap\ (K\ \overline{\pi}; e_1 \sim e_2)$ is undefined, rule Pat-InfCon is an impossible case.

**Case rule** Pat-InfVar $\overline{\pi} = x, \overline{\pi}', \overline{\psi} = \tau_1, \overline{\psi}', \Delta = x : \tau_1, \Delta', \overline{\pi_F} = x : \tau_1, \overline{\pi_F}'$ **and** $\overline{\psi_F} = \tau_1, \overline{\psi_F}'$ **:**
We know from the pattern typing derivation premise that $\Gamma, x : \tau_1 \vdash^P \overline{\pi}' \Rightarrow \overline{\psi}'; \Delta' \rightsquigarrow \overline{\pi_F}' : \overline{\psi_F}'$. By inlining the definitions and rule PatWrap-Var, we get $e_2 = \lambda x.e_2'$ and $wrap\ (\overline{\pi}'; e_1 \sim e_2')$. By case analysis on the $e_2$ typing derivation (rule Tm-InfAbs), we know $\Gamma, x : \tau_1 \vdash e_2' \Rightarrow \eta_3^\epsilon \rightsquigarrow t_2'$ where $\eta_2^\epsilon = \tau_1 \to \eta_3^\epsilon$ and $t_2 = \lambda x : \tau_1.t_2'$. By applying the induction hypothesis, we get $\text{case } \overline{\pi_F}' : \overline{\psi_F}' \to t_1 \simeq t_2'$. The goal to be proven is $\lambda x : \tau_1.\text{case } \overline{\pi_F}' : \overline{\psi_F}' \to t_1 = \lambda x : \tau_1.t_2'$, and follows directly from Lemma E.16.

**Case rule** Pat-InfTyVar $\overline{\pi} = @a, \overline{\pi}', \overline{\psi} = @a, \overline{\psi}', \Delta = a, \Delta', \overline{\pi_F} = @a, \overline{\pi_F}'$ **and** $\overline{\psi_F} = @a, \overline{\psi_F}'$ **:**
We know from the pattern typing derivation premise that $\Gamma, a \vdash^P \overline{\pi}' \Rightarrow \overline{\psi}'; \Delta' \rightsquigarrow \overline{\pi_F}' : \overline{\psi_F}'$. Similarly to the previous case, by inlining and rule PatWrap-TyVar, we get $e_2 = \Lambda a.e_2'$ and $wrap\ (\overline{\pi}'; e_1 \sim e_2')$. By case analysis on the $e_2$ typing derivation (rule Tm-InfTyAbs), we get $\Gamma, a \vdash e_2' \Rightarrow \eta_3^\epsilon \rightsquigarrow t_2', \Gamma \vdash \forall a.\eta_3^\epsilon \xrightarrow{inst\ \delta} \eta_2^\epsilon \rightsquigarrow \dot{t}$ and $t_2 = \dot{t}[\Lambda a.t_2']$. Applying the induction hypothesis tells us that $\text{case } \overline{\pi_F}' : \overline{\psi_F}' \to t_1 \simeq t_2'$.

The goal to be proven is $\Lambda a.\text{case } \overline{\pi_F}' : \overline{\psi_F}' \to t_1 \simeq \dot{t}[\Lambda a.t_2']$. By applying Lemma E.18 to the result of the induction hypothesis, we get $\Lambda a.\text{case } \overline{\pi_F}' : \overline{\psi_F}' \to t_1 \simeq \Lambda a.t_2'$. Under lazy instantiation, the goal follows directly from this result, as $\dot{t} = \bullet$. Under eager deep instantiation, it is clear that the goal does not hold, as $\dot{t}$ might perform eta expansion, thus altering the runtime semantics. Under eager shallow instantiation, the goal follows straightforwardly, as $\dot{t}$ can only perform type applications. Note that this implies that $\Lambda a.\text{case } \overline{\pi_F}' : \overline{\psi_F}' \to t_1$ and $\dot{t}[\Lambda a.t_2']$ could thus have different types, but can always instantiate to the same type. □

We now return to proving Property 8, by case analysis on the first declaration typing relation (rule Decl-NoAnnSingle). We know from the derivation premise that $\Gamma \vdash^P \overline{\pi} \Rightarrow \overline{\psi}; \Delta \rightsquigarrow \overline{\pi_F} : \overline{\psi_F}, \Gamma, \Delta \vdash e_1 \Rightarrow \eta_1^\epsilon \rightsquigarrow t_1', t_1 = \text{case } \overline{\pi_F} : \overline{\psi_F} \to t_1'$,

$$\boxed{numargs\,(\sigma) = m} \qquad \text{(Explicit Argument Counting)}$$

NUMARGS-TYVAR
$$numargs\,(a) = 0$$

NUMARGS-CON
$$numargs\,(T\,\overline{\tau}) = 0$$

NUMARGS-ARROW
$$\frac{numargs\,(\sigma_2) = m}{numargs\,(\sigma_1 \to \sigma_2) = m + 1}$$

NUMARGS-FORALL
$$\frac{numargs\,(\sigma) = m}{numargs\,(\forall\,a.\sigma) = m}$$

NUMARGS-INFFORALL
$$\frac{numargs\,(\sigma) = m}{numargs\,(\forall\,\{a\}.\sigma) = m}$$

**Figure 7.** Counting Explicit Arguments

$type\,(\overline{\psi}; \eta_1^\epsilon \sim \sigma')$, $\sigma = \forall\,\overline{\{a\}}.\sigma'$ where $\overline{a} = f_v(\sigma') \setminus dom\,(\Gamma)$. The premise of the second declaration typing derivations tells us that $\Gamma \vdash e_2 \Rightarrow \eta_2^\epsilon \leadsto t_2$. The goal now follows directly from Lemma E.59. Note that as Lemma E.59 does not hold under eager deep instantiation, the same is true for Property 8. □

### E.6 Single vs. Multiple Equations

**Property 10** (Single/multiple Equations is Type Preserving).
*If* $\Gamma \vdash x\,\overline{\pi} = e \Rightarrow \Gamma, x : \sigma$ *then* $\Gamma \vdash x\,\overline{\pi} = e, x\,\overline{\pi} = e \Rightarrow \Gamma'$

The proof proceeds by case analysis on the declaration typing derivation (rule DECL-NOANNSINGLE). From the derivation premise, we get $\Gamma \vdash^P \overline{\pi} \Rightarrow \overline{\psi}; \Delta$, $\Gamma, \Delta \vdash e \Rightarrow \eta^\epsilon$, $type\,(\overline{\psi}; \eta^\epsilon \sim \sigma_1)$ and $\sigma = \forall\,\overline{\{a\}}_1.\sigma_1$ where $\overline{a}_1 = f_v(\sigma_1) \setminus dom\,(\Gamma)$. The goal to be proven thus reduces to $\Gamma, \Delta \vdash \eta^\epsilon \xrightarrow{inst\,\delta} \rho$, $type\,(\overline{\psi}; \rho \sim \sigma_2)$ and $\sigma = \forall\,\overline{\{a\}}_2.\sigma_2$ where $\overline{a}_2 = f_v(\sigma_2) \setminus dom\,(\Gamma)$. It is clear that the property can not hold under lazy instantiation, as rule DECL-NOANNMULTI performs an additional instantiation step, thus altering the type. Under eager instantiation, $\eta^\epsilon$ is already an instantiated type by the type inference relation, making the instantiation in the goal a no-op (by definition). The goal is thus trivially true. □

### E.7 $\eta$-expansion

**Property 11b** ($\eta$-expansion is Type Preserving).

- *If* $\Gamma \vdash e \Rightarrow \eta^\epsilon$ *where* $numargs(\eta^\epsilon) = n$ *and* $\Gamma \vdash \eta^\epsilon \xrightarrow{inst\,\delta} \tau$ *then* $\Gamma \vdash \lambda\overline{x}^n.e\,\overline{x}^n \Rightarrow \eta^\epsilon$
- *If* $\Gamma \vdash e \Leftarrow \sigma$ *where* $numargs(\rho) = n$ *then* $\Gamma \vdash \lambda\overline{x}^n.e\,\overline{x}^n \Leftarrow \sigma$

A formal definition of *numargs* is shown in Figure 7. We prove Property 11b by first introducing a slightly more general lemma:

**Lemma E.60** ($\eta$-expansion is Type Preserving - Generalised).

- *If* $\Gamma \vdash e \Rightarrow \eta^\epsilon$ *where* $0 \leqslant n \leqslant numargs(\eta^\epsilon)$ *and* $\Gamma \vdash \eta^\epsilon \xrightarrow{inst\,\delta} \tau$ *then* $\Gamma \vdash \lambda\overline{x}^n.e\,\overline{x}^n \Rightarrow \eta^\epsilon$
- *If* $\Gamma \vdash e \Leftarrow \sigma$ *where* $0 \leqslant n \leqslant numargs(\rho)$ *then* $\Gamma \vdash \lambda\overline{x}^n.e\,\overline{x}^n \Leftarrow \sigma$

The proof proceeds by induction on the integer $n$.
**Case** $n = 0$:
This case is trivial, as it follows directly from the premise.
**Case** $n = m + 1 \leqslant numargs(\eta^\epsilon)$:
  **case synthesis mode:** We know from the induction hypothesis that $\Gamma \vdash \lambda\overline{x}^m.e\,\overline{x}^m \Rightarrow \eta^\epsilon$. We perform case analysis on this result ($m$ repeated applications of rule TM-INFABS) to get $\Gamma, \overline{x_i : \tau_i}^{i<m} \vdash e\,\overline{x}^m \Rightarrow \eta_1^\epsilon$ where $\eta^\epsilon = \overline{\tau_i}^{i<m} \to \eta_1^\epsilon$. Performing case analysis again on this result (rule TM-INFAPP), gives us $\Gamma, \overline{x_i : \tau_i}^{i<m} \vdash^H e \Rightarrow \sigma_1$, $\Gamma, \overline{x_i : \tau_i}^{i<m} \vdash^A \overline{x}^m \Leftarrow \sigma_1 \Rightarrow \sigma_2$ and $\Gamma, \overline{x_i : \tau_i}^{i<m} \vdash \sigma_2 \xrightarrow{inst\,\delta} \eta_1^\epsilon$.
  The goal to be proven is $\Gamma \vdash \lambda\overline{x}^{m+1}.e\,\overline{x}^{m+1} \Rightarrow \eta^\epsilon$, which (by rule TMINFABS) reduces to $\Gamma, \overline{x_i : \tau_i}^{i<m}, x : \tau \vdash e\,\overline{x}^{m+1} \Rightarrow \eta_2^\epsilon$, where $\eta^\epsilon = \overline{\tau_i}^{i<m} \to \tau \to \eta_2^\epsilon$.

Note that this requires proving that $\eta_1^\epsilon = \tau \to \eta_2^\epsilon$. While we know that $m < numargs(\eta^\epsilon)$, we can only prove this under eager deep instantiation. Under lazy instantiation, type inference does not instantiate the result type at all. Under eager shallow, it is instantiated, but only up to the first function type. From here on out, we will thus assume eager deep instantiation. Furthermore, note that as even deep instantiation does not instantiate argument types, we need the additional premise that $\eta^\epsilon$ instantiates into a monotype, in order to prove this goal.

This result in turn (by rule TM-INFAPP) reduces to $\Gamma, \overline{x_i : \tau_i}^{i<m}, x : \tau \vdash^H e \Rightarrow \sigma_1$ (follows by weakening), $\Gamma, \overline{x_i : \tau_i}^{i<m}, x : \tau \vdash^A x, \overline{x}^m \Leftarrow \sigma_1 \Rightarrow \sigma_3$ (follows by rule ARG-INST, rule ARG-APP and the fact that $\eta_1^\epsilon = \tau \to \eta_2^\epsilon$) and $\Gamma, \overline{x_i : \tau_i}^{i<m}, x : \tau \vdash \sigma_3 \xrightarrow{inst\ \delta} \eta_2^\epsilon$ (follows by the definition of instantiation).

**case checking mode :** We know from the induction hypothesis that $\Gamma \vdash \lambda \overline{x}^m . e\, \overline{x}^m \Leftarrow \sigma$. The proof proceeds similarly to the synthesis mode case, by case analysis on this result (rule TM-CHECKABS). One additional step is that rule TM-CHECKINF is applied to type $e\, \overline{x}^m$. The derivation switches to synthesis mode at this point, and becomes completely identical to the previous case. □

The proof for Property 11b now follows directly by Lemma E.60, by taking $n = numargs(\eta^\epsilon)$. □